\documentclass[structabstract]{aa}  
\usepackage{graphicx}
\usepackage{txfonts}
\begin{document}
   \title{Population gradients and photometric metallicities in early- and transition-type dwarf galaxies: Clues from the Sculptor Group \thanks{Based on observations made with the NASA/ESA Hubble Space Telescope, obtained from the data archive at the Space Telescope Institute. STScI is operated by the association of Universities for Research in Astronomy, Inc. under the NASA contract NAS 5-26555.}}

   \author{S.~Lianou 
              \inst{1} 
          \and
           E.~K.~Grebel 
              \inst{2} 
          \and 
           G.~S.~Da Costa
              \inst{3}
          \and
           M.~Rejkuba
              \inst{4}
          \and
           H.~Jerjen
              \inst{3}
          \and 
           A.~Koch 
              \inst{5}
          }

           \institute{Institute for Astronomy, Astrophysics, Space Applications \& Remote Sensing, National Observatory of Athens, I.~Metaxa and Vas.~Pavlou, GR-15236 Palaia Penteli, Greece
           \and
                  Astronomisches Rechen-Institut, Zentrum f\"ur Astronomie der Universit\"at Heidelberg, M\"onchhofstrasse 12-14, D-69120 Heidelberg, Germany
           \and
                  Research School of Astronomy \& Astrophysics, Australian National University, Mt Stromlo Observatory, via Cotter Rd, Weston, ACT 2611, Australia
           \and
                  European Southern Observatory, Karl-Schwarzschild-Strasse 2, 85748 Garching bei M\"unchen, Germany
           \and
                  Landessternwarte, Zentrum f\"ur Astronomie der Universit\"at Heidelberg, K\"onigstuhl 12, D-69117 Heidelberg, Germany
                  }

   \date{Received July 6, 2012; accepted November 8, 2012}

 
  \abstract
   {}
   {We focus on the resolved stellar populations of one early-type and four transition-type dwarf galaxies in the Sculptor group, with the aim to examine the potential presence of population gradients and place constraints on their $\it{mean}$ metallicities. }
   {We use deep Hubble Space Telescope images to construct color-magnitude diagrams, from which we select stellar populations that trace different evolutionary phases in order to constrain their range of ages and metallicities, as well as to examine their spatial distribution. In addition, we use the resolved stars in the red giant branch in order to derive photometric metallicities.}
   {All studied dwarfs contain intermediate-age stars with ages of $\sim$1~Gyr and older as traced by the luminous asymptotic giant branch and red clump stars, while the transition-type dwarfs contain also stars younger than $\sim$1~Gyr as traced by a young main sequence and vertical red clump stars. Moreover, the spatial distribution of the stars that trace different evolutionary phases shows a population gradient in all transition-type dwarfs. The derived error-weighted mean metallicities, assuming purely old stellar populations, range from $-$1.5~dex for ESO294-G010 to $-$1.9~dex for Scl-dE1, and should be considered as lower limits to their true metallicities. Assuming intermediate-age stellar populations to dominate the dwarfs, we derive upper limits for the metallicities that are 0.3 to 0.2~dex higher than the metallicities derived assuming purely old populations. We discuss how photometric metallicity gradients are affected by the age-metallicity degeneracy, which prevents strong conclusions regarding their actual presence. Finally, the transition-type dwarfs lie beyond the virial radius of their closest bright galaxy, as also observed for the Local Group transition-type dwarfs. Scl-dE1 is the only dwarf spheroidal in our sample and is an outlier in a potential morphology-distance relation, similar as the two isolated dwarf spheroidals of the Local Group, Tucana and Cetus.}
   {}

   \keywords{Galaxies: dwarf --
             Galaxies: stellar content -- 
             Galaxies: evolution --
             Galaxies: groups: individual: Sculptor group}

\titlerunning{Scl Group dwarfs: gradients and metallicities}
   \maketitle
%

\section{Introduction}

   The evolution of dwarf galaxies may be influenced by several environmental mechanisms, such as ram pressure stripping and tidal interactions (for a review see Boselli \& Gavazzi \cite{sl_boselli06}). Such mechanisms are important drivers of galaxy evolution within a dense environment, such as clusters of galaxies, as advocated by observations which show, for example, that bright early-type dwarf galaxies in the Virgo or Fornax clusters are rotationally supported (Pedraz et al.~\cite{sl_pedraz02}; de Rijcke et al.~\cite{sl_derijcke03}; Geha, Guhathakurta \& van der Marel \cite{sl_geha03}; van Zee, Skillman \& Haynes \cite{sl_vanzee04}; Toloba et al.~\cite{sl_toloba09}) or that there are sub-classes of Virgo cluster dwarf ellipticals (dEs) with spiral features (Jerjen, Kalnajs \& Binggeli \cite{sl_jerjen00b}; Barazza, Binggeli \& Jerjen \cite{sl_barazza02}; Lisker, Grebel \& Binggeli \cite{sl_lisker06}). The situation is similar within a group environment, where tidal interactions may affect the evolution of dwarf galaxies as observed in the Local Group (LG; e.g., Ibata, Gilmore \& Irwin \cite{sl_ibata95}; Putman et al.~\cite{sl_putman03}; McConnachie et al.~\cite{sl_mcconnachie09}), or these may even lead to the formation of tidal dwarf galaxies, as is the case for instance in the M81 group (e.g., Yun, Ho \& Lo \cite{sl_yun94}; Sakai \& Madore \cite{sl_sakai01}; Makarova et al.~\cite{sl_makarova02}; Sabbi et al.~\cite{sl_sabbi08}; Croxall et al.~\cite{sl_croxall09}). Other effects include ram pressure stripping, ionization, and evaporation, all of which may contribute to the observed relation between the morphology of dwarf galaxies in groups and their distance from the nearest giant galaxy (e.g., van den Bergh \cite{sl_vandenbergh94}; Grebel et al.~\cite{sl_grebel03}). We will refer to the latter relation as the ``morphology-distance'' relation. The environment plays an important role in shaping the properties and evolution of dwarf galaxies, while there are some properties that to a first degree seem to remain unaffected, such as color or metallicity gradients (den Brok et al.~\cite{sl_denbrok11}; Koleva et al.~\cite{sl_koleva11}). 

   The LG has provided a benchmark environment in which dwarf galaxies have been studied with the best means available in progressively more detail. These studies have revealed that LG dwarf galaxies experienced a variety of star formation histories (SFHs; e.g., Tolstoy, Tosi \& Hill \cite{sl_tolstoy09}; and references therein). The variety of SFHs observed in LG dwarf galaxies can be explained by invoking the combination of ram pressure stripping, tidal stirring and cosmic ultra-violet background radiation on gas-dominated dwarf galaxies (e.g., Mayer \cite{sl_mayer10}; Kazantzidis et al.~\cite{sl_kazantzidis11}; Nichols \& Bland-Hawthorn \cite{sl_nichols11}; and references therein). Furthermore, Grcevich \& Putman (\cite{sl_grcevich09}) confirm that LG dwarfs within $\sim$270\,kpc from the Galaxy or Andromeda are deficient in detectable HI gas, as well as that their HI mass correlates with galactocentric distance, suggestive of an environmental influence. Likewise, Weisz et al.~(\cite{sl_weisz11b}) find a pronounced morphology-distance relation. Outliers to the morphology-distance relation for LG dwarfs do exist, as the cases of the isolated Cetus and Tucana dwarf spheroidals (dSphs) demonstrate (e.g., Lewis et al.~\cite{sl_lewis07}; Fraternali et al.~\cite{sl_fraternali09}; Monelli et al.~\cite{sl_monelli10a, sl_monelli10b}; and references therein).

   Bouchard, Da Costa \& Jerjen (\cite{sl_bouchard09}) analyze H$\alpha$ emission of dwarf galaxies in nearby groups. They find strong correlations between the local density of a larger sample of dwarfs (also including LG dwarfs) and other physical properties, such as current star formation rates and HI masses. They conclude that the local galaxy density, representing a measure of the environment, influences the properties of dwarf galaxies. To that end, the comparative analysis of the cumulative SFHs of nearby dwarf galaxies ($D<4$~Mpc) shows that the environment in which dwarf galaxies live plays an important role in shaping their current morphological type (Weisz et al.~\cite{sl_weisz11a}).

   In the present study, we focus on a sample of dwarf galaxies in the Sculptor group ($1.5<D<4.5$~Mpc). The Sculptor group is at an early stage of its evolution during which galaxies are still falling in along extended filamentary structures, forming a "cigar"-like structure (Jerjen, Freeman \& Binggeli \cite{sl_jerjen98}; Karachentsev et al.~\cite{sl_karachentsev03}). Embedded in this filamentary structure are several sub-groups of relatively bright galaxies surrounded by the fainter dwarf galaxies of both early- and late-type such as dSphs and dwarf irregulars (dIrrs) (C\^ot\'e et al.~\cite{sl_cote97}; Jerjen, Freeman \& Binggeli \cite{sl_jerjen98}; Jerjen, Binggeli \& Freeman \cite{sl_jerjen00}; Karachentsev et al.~\cite{sl_karachentsev03}). 

  The dwarf galaxy population in the Sculptor group includes two early-type dwarfs, namely Scl-dE1 (dSph) and NGC59 (dSO), and four transition-type dwarfs (dIrrs/dSphs), namely ESO410-G005, ESO294-G010, ESO540-G030, and ESO540-G032 (Bouchard et al.~\cite{sl_bouchard05}; Beaulieu et al.~\cite{sl_beaulieu06}). The transition-type dwarfs are galaxies with intermediate properties between those of the gas-poor dSphs and the gas-rich dIrrs, and their seemingly intermediate stage between dSphs and dIrrs has brought attention to them as their potential evolutionary links (e.g., Grebel, Gallagher \& Harbeck \cite{sl_grebel03}; Knezek, Sembach \& Gallagher \cite{sl_knezek99}; Mateo \cite{sl_mateo98}; Heisler et al.~\cite{sl_heisler97}). The four transition-type dwarfs and NGC59 have been detected in HI, with masses ranging from 3$\times$10$^{5}$ to 10$^{6}$\,M$_{\sun}$ for the transition-type dwarfs (Bouchard et al.~\cite{sl_bouchard05}), and 1.4$\times$10$^{7}$\,M$_{\sun}$ for NGC59 (Beaulieu et al.~\cite{sl_beaulieu06}), while for a few cases H$\alpha$ emission has also been detected (ESO540-G032 and ESO294-G010; Bouchard, Jerjen \& Da Costa \cite{sl_bouchard09}). Scl-dE1 is the sole dSph in the Sculptor group with so far no HI or H$\alpha$ detections (Bouchard et al.~\cite{sl_bouchard05,sl_bouchard09}). In addition to these types, there is a large number of late-type dwarfs in the Sculptor group (e.g., C\^ot\'e et al.~\cite{sl_cote97}; Karachentsev et al.~\cite{sl_karachentsev04}; Bouchard et al.~\cite{sl_bouchard09}).

   For our stellar population study we selected Scl-dE1 and all four transition-type dwarfs. This sample of five dwarfs represents $\sim$80\% of the early- and transition-type dwarfs currently known in the Sculptor group. With our study, we aim at placing constraints on the early chemical enrichment of our galaxy sample. Within the context of population gradients, we further examine the spatial distribution of stars that mark different evolutionary phases. Population or metallicity gradients, in the sense of the young or metal-rich populations being more centrally concentrated, have been uncovered in some LG dwarfs (e.g., Harbeck et al.~\cite{sl_harbeck01}; Tolstoy et al.~\cite{sl_tolstoy04}; Kirby et al.~\cite{sl_kirby09}; Battaglia et al.~\cite{sl_battaglia11}; de Boer et al.~\cite{sl_deboer11}), but not all (e.g., Harbeck et al.~\cite{sl_harbeck01}; Koch et al.~\cite{sl_koch07a}).

   This paper is structured as follows. In \S2 we present our observational data set and photometry, including photometric errors and completeness estimates analyses. In \S3 we present the color-magnitude diagrams and discuss the stellar content of our studied dwarfs, with a further focus in placing constraints on their ages and metallicities. In \S4 we investigate the stellar spatial distribution and the cumulative distribution of stars in different evolutionary phases, in order to probe the potential presence of population gradients. In \S5 we focus on the red giant branch and present the photometric metallicity distribution functions. In addition, we discuss biases due to the presence of intermediate-age stars, as well as due to an old single age assumption. In \S6 we discuss the morphology-distance relation. In \S7 we summarise our findings and conclude. 

\section{Observations and photometry}

   We use deep observations obtained with the Advanced Camera for Surveys (ACS) on board the Hubble Space Telescope (HST) through
%
\begin{table}
     \begin{minipage}[t]{\columnwidth}
      \caption{Log of Observations.}
      \label{table1} 
      \centering
      \renewcommand{\footnoterule}{}
      \begin{tabular}{l c c c c }
\hline\hline
    Galaxy      	&RA               &Dec                 &\multicolumn{2}{c}{Exposure time (s)}   \\ 
\cline{4-5}
                        &(J2\,000.0)      &(J2\,000.0)          &F606W        &F814W                    \\ 
    (1)                 &(2)              &(3)                  &(4)          &(5)                      \\ 
\hline 
    ESO540-G030         &00~49~21.10      &$-$18~04~34.0        &8\,960       &7\,840                   \\
    ESO540-G032         &00~50~24.50	  &$-$19~54~23.0        &8\,960       &6\,708                   \\
    ESO294-G010   	&00~26~33.40	  &$-$41~51~19.0        &13\,920      &27\,840                  \\
    ESO410-G005  	&00~15~31.40	  &$-$32~10~47.0        &13\,440      &26\,880                  \\
    Scl-dE1	   	&00~23~51.70	  &$-$24~42~18.0        &17\,920      &17\,920                  \\
\hline
\end{tabular} 
\footnotetext{Note.-- Units of right ascension are hours, minutes, and seconds, and units of declination are degrees, arcminutes and arcseconds.}%
\end{minipage}
\end{table}
%
Program GO\,10503 (PI: Da Costa). A summary of the observations is given in Table~\ref{table1}, where the columns show: (1) the galaxy name; (2) and (3) equatorial coordinates of the field centers (J2000.0); (4) and (5) the total exposure time in the F606W and F814W filters, respectively. Additional information on the data sets can be found in Da Costa et al.~(\cite{sl_dacosta09,sl_dacosta10}).  

   We used the ST-ECF Hubble Science Archive in order to download the scientific images, which were already pre-reduced through the HST pipeline. We perform stellar point source photometry using the ACS module of DOLPHOT, a modified version of the HSTphot photometry package (Dolphin \cite{sl_dolphin00}). We conduct the point source photometry simultaneously on all the individual, calibrated and flat-fielded ``FLT'' images, while we use as a reference image the deepest drizzled image available. We follow the photometry reduction steps as described in the manual of DOLPHOT for the ACS module. In the final photometric catalogues we allow only objects with $S/N>$~9 and ``type'' equal to 1, i.e., ``good stars''. The ``type'' is a DOLPHOT parameter that is used to distinguish objects that are classified, amongst others, as ``good stars'', ``elongated objects'', ``too sharp objects''. We further apply photometric quality cuts based on the distributions of the sharpness and crowding parameters, as suggested in the DOLPHOT manual and in Williams et al.~(\cite{sl_williams09}). We use for the sharpness parameter the restriction of $|$Sharp$_{F606W}+$Sharp$_{F814W}$$|<$~1, and for the crowding parameter the requirement (Crowd$_{F606W}+$~Crowd$_{F814W})<$~1. With these selection criteria, the number of likely stellar objects retrieved in the final photometric catalog 
\begin{table*}
\begin{minipage}[t]{\textwidth}
\caption[]{Global and derived photometric properties (see text for references).}
\label{table2} 
\centering
\renewcommand{\footnoterule}{}
\begin{tabular}{l c c c c c c c c c c}

\hline\hline
  Galaxy      &Type          &$N_{\star}$ &F814W(50\%)  &F606W(50\%)     &E(B-V)   &$A_{F814W}$   &$A_{F606W}$    &$F814W_{TRGB}$   &$(m-M)_{O}$      &R$_{\rm disturber}$   \\
              &              &           &(mag)        &(mag)           &(mag)    &(mag)       &(mag)          &(mag)           &(mag)            &kpc  \\
   (1)        &(2)           &(3)        &(4)          &(5)             &(6)      &(7)         &(8)            &(9)             &(10)             &(11) \\
\hline                                                                                                                         
 ESO294-G010  &dIrr/dSph     &27\,605     &26.6         &28.1           &0.006     &0.011       &0.017          &22.36$\pm$0.07  &26.43$\pm$0.05   &209$\pm$92  \\
 ESO410-G005  &dIrr/dSph     &34\,284     &27.6         &28.1           &0.014     &0.025       &0.039          &22.39$\pm$0.06  &26.45$\pm$0.05   &290$\pm$61  \\
 ESO540-G030  &dIrr/dSph     &13\,439     &26.7         &27.2           &0.023     &0.042       &0.065          &23.63$\pm$0.07  &27.68$\pm$0.06   &434$\pm$17 \\
 ESO540-G032  &dIrr/dSph     &16\,504     &26.6         &27.1           &0.020     &0.036       &0.056          &23.68$\pm$0.09  &27.72$\pm$0.06   &335$\pm$41  \\
 Scl-dE1      &dSph          &8\,310      &26.9         &27.4           &0.012     &0.027       &0.042          &24.07$\pm$0.11  &28.13$\pm$0.06   &870$\pm$260  \\
\hline
\end{tabular}
\end{minipage}
\end{table*}
%
is listed in column 3 of Table~\ref{table2}. 

   For the current analysis we choose to work in the ACS\,/\,WFC filter system, although DOLPHOT provides magnitudes in both the ACS\,/\,WFC and the Landolt UBVRI photometric systems. Therefore, we transform the Galactic foreground extinction from the V-band and I-band, $A_{I}$ and $A_{V}$ (Schlegel, Finkbeiner \& Davis \cite{sl_schlegel98}), into the ACS\,/\,WFC system, using the corresponding extinction ratios $A(P)\,/\,E(B-V)$ for a G2 star. $A(P)$ corresponds to the extinctions in the filters F814W and F606W, which are provided by Sirianni et al.~(\cite{sl_sirianni05}; their Table 14). These extinction ratios are multiplied with the corresponding $E(B-V)$ value in order to finally get the extinctions in the ACS filters. We list the  $E(B-V)$ values in column (6) of Table~\ref{table2}, adopted from Schlegel et al.~(\cite{sl_schlegel98}). The transformed values of extinction in the F814W and F606W filters are listed in Table~\ref{table2}, columns (7) and (8), respectively.  

   Table~\ref{table2} includes information on the global and derived photometric properties of the studied dwarf galaxy sample. The columns contain: (1) the galaxy name; (2) the galaxy type; (3) the number of stellar objects after all photometric selection criteria have been applied; (4) and (5) the F814W and F606W magnitudes, respectively, that correspond to the 50\% completeness limit, discussed in the following paragraph; (6) the reddening adopted from Schlegel et al.~(\cite{sl_schlegel98}); (7) and (8) the extinction in the F814W and F606W filters, as detailed in the preceding paragraph; (9) the F814W magnitude of the tip of the red giant branch (TRGB), derived in \S3; (10) the true distance modulus, derived in \S3; (11) the deprojected distance of each dwarf from its main disturber galaxy, defined in \S6 (Karachentsev et al.~\cite{sl_karachentsev04}). 

  \subsection{Photometric errors and completeness}

   In order to quantify the completeness factors and the photometric errors, we perform artificial star tests using standard routines that are included in the DOLPHOT software package. For each galaxy, we use 10$^{5}$ stars per ACS/WFC chip, distributed within a color and magnitude range similar to that of the observed stars, but extended by $\sim$0.5~mag to fainter and brighter magnitudes, as well as to bluer and redder colors. The artificial stars are added and measured one at a time (see Dolphin \cite{sl_dolphin00}; Perina et al.~\cite{sl_perina09}) to avoid self-induced crowding. 

 \begin{figure}
   \centering
      \includegraphics[width=4cm,clip]{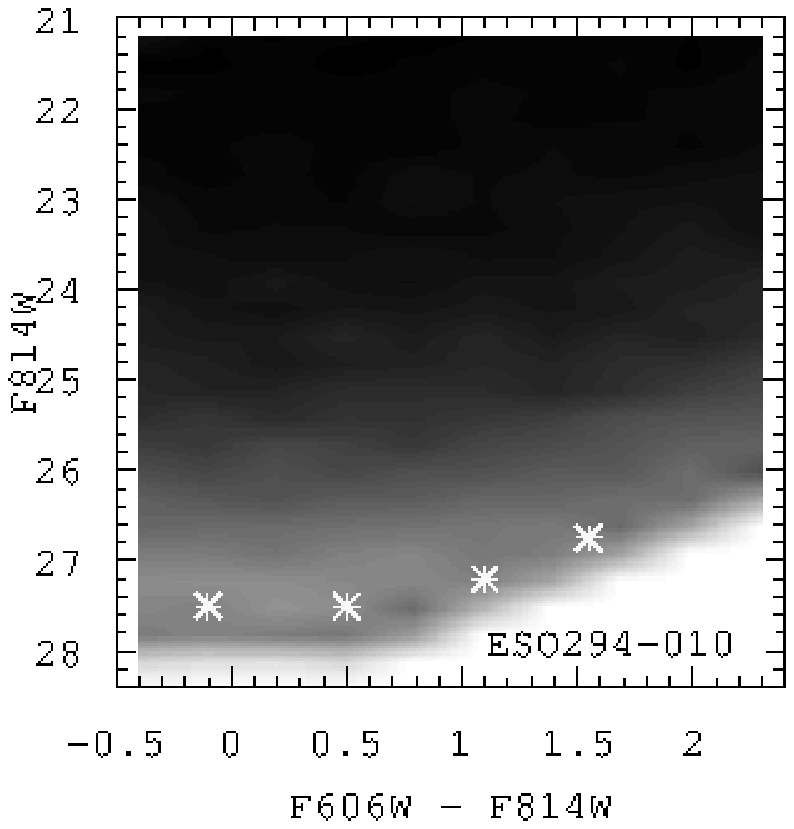}
      \includegraphics[width=4cm,clip]{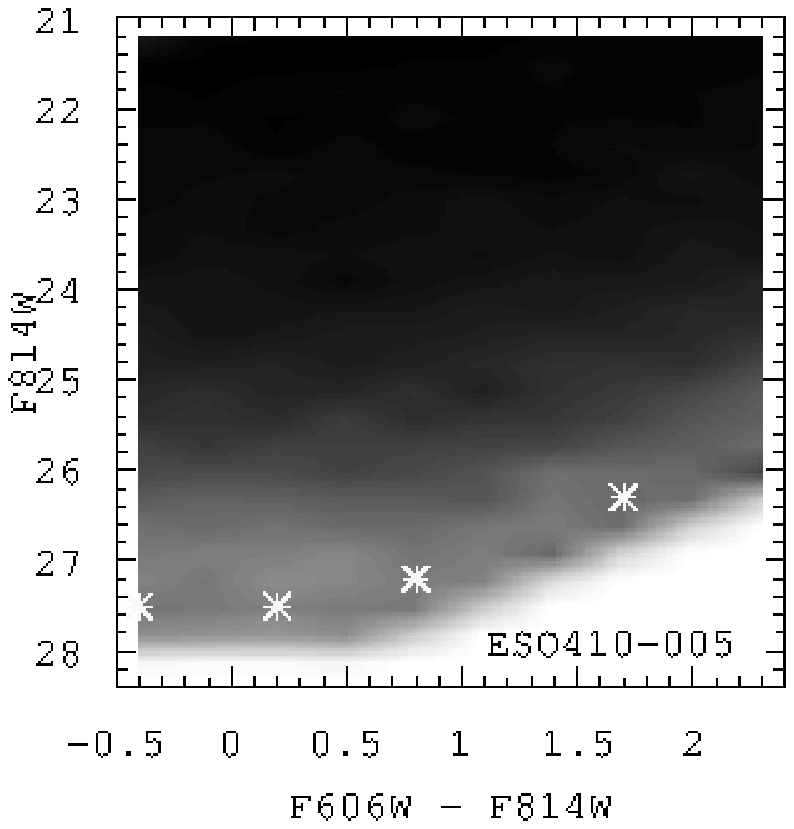}
      \includegraphics[width=4cm,clip]{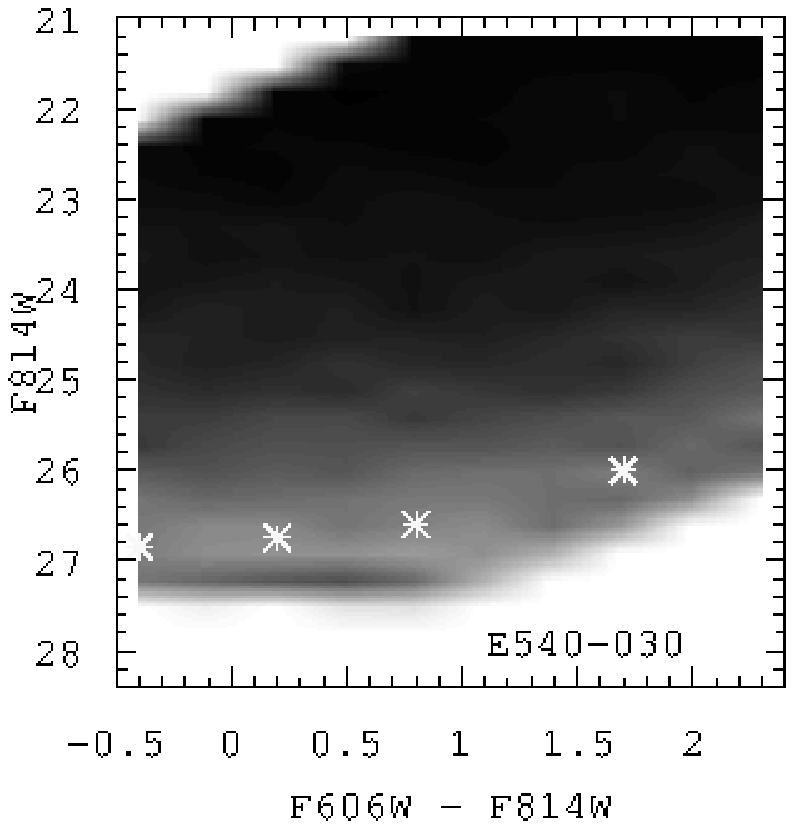}
      \includegraphics[width=4cm,clip]{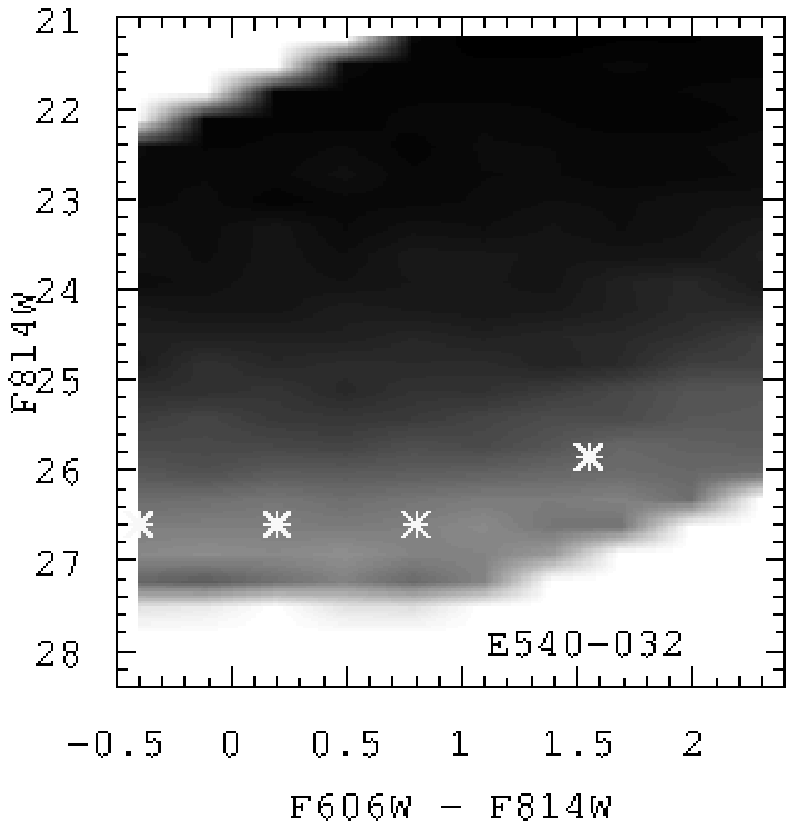}
      \includegraphics[width=4cm,clip]{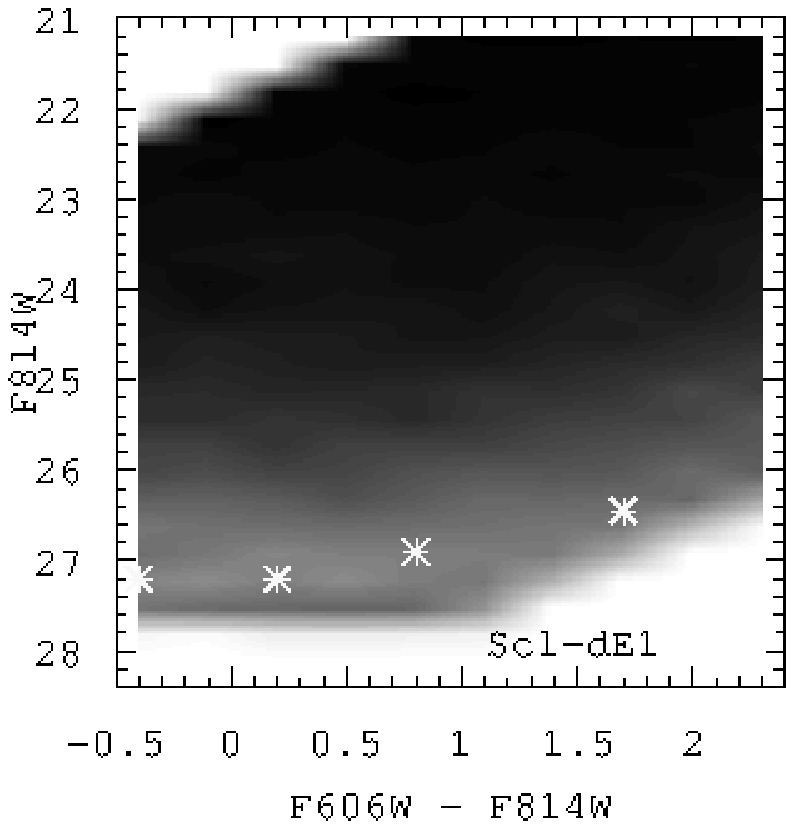}
      \includegraphics[scale=0.51]{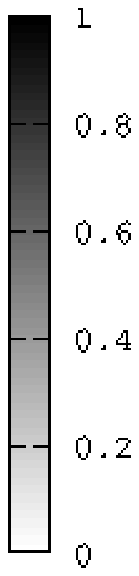}
      \caption{Photometric completeness as a function of color and magnitude for each studied galaxy. Areas with a 100\% completeness factor are shown in black, while areas with 0\% completeness factors are shown with the lightest gray, and the colorbar next to Scl-dE1 indicates the completeness factors valid for all dwarfs. The gray asterisks indicate the 50\% completeness factors as a function of color and magnitude. Note that in the middle and lower panels, the upper left corners are shown in white due to the non-sampling of the completeness at this color and magnitude region.}
      \label{sl_figure01}%
\end{figure}
%
   In Fig.~\ref{sl_figure01} we show the completeness factors as a function of color and magnitude. These were computed by counting the input and output artificial stars with a binning in color and magnitude of 0.3~mag. The F814W and F606W magnitudes that correspond to the 50\% completeness factors, and at a color $($F606W$-$F814W$)=$0.5~mag, are listed in Table~\ref{table2}, columns (4) and (5), respectively, while they are also shown in Fig.~\ref{sl_figure01} with the gray asterisks.

 \begin{figure*}
   \centering
      \includegraphics[width=8cm,clip]{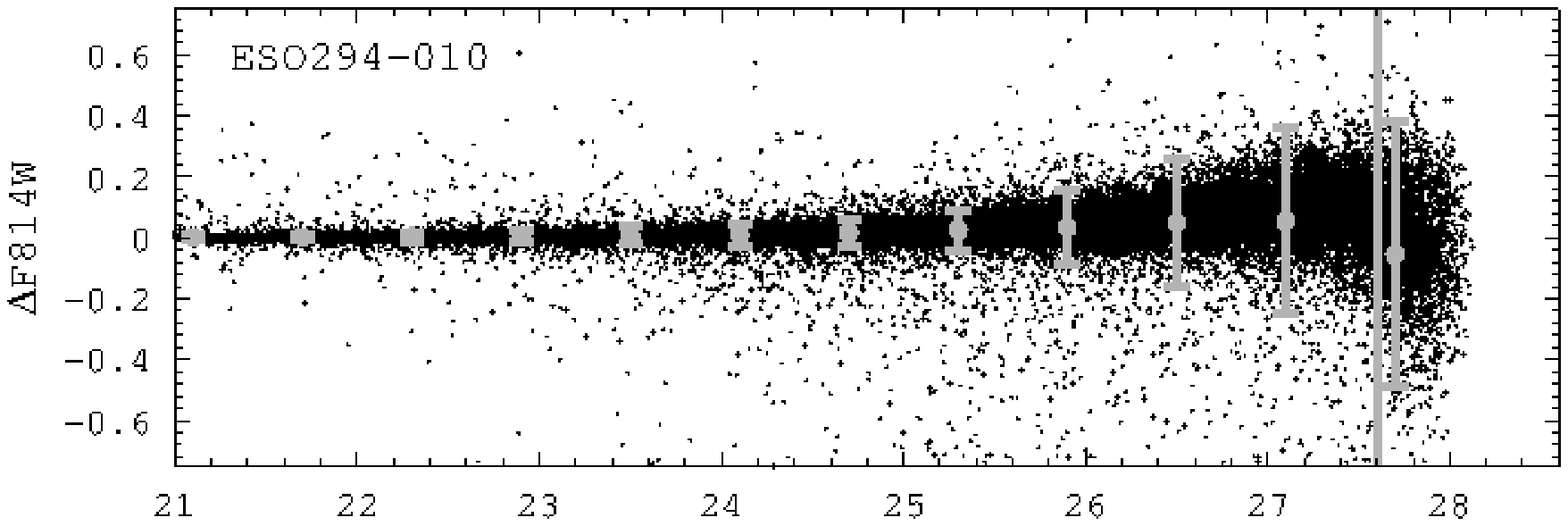}
      \includegraphics[width=8cm,clip]{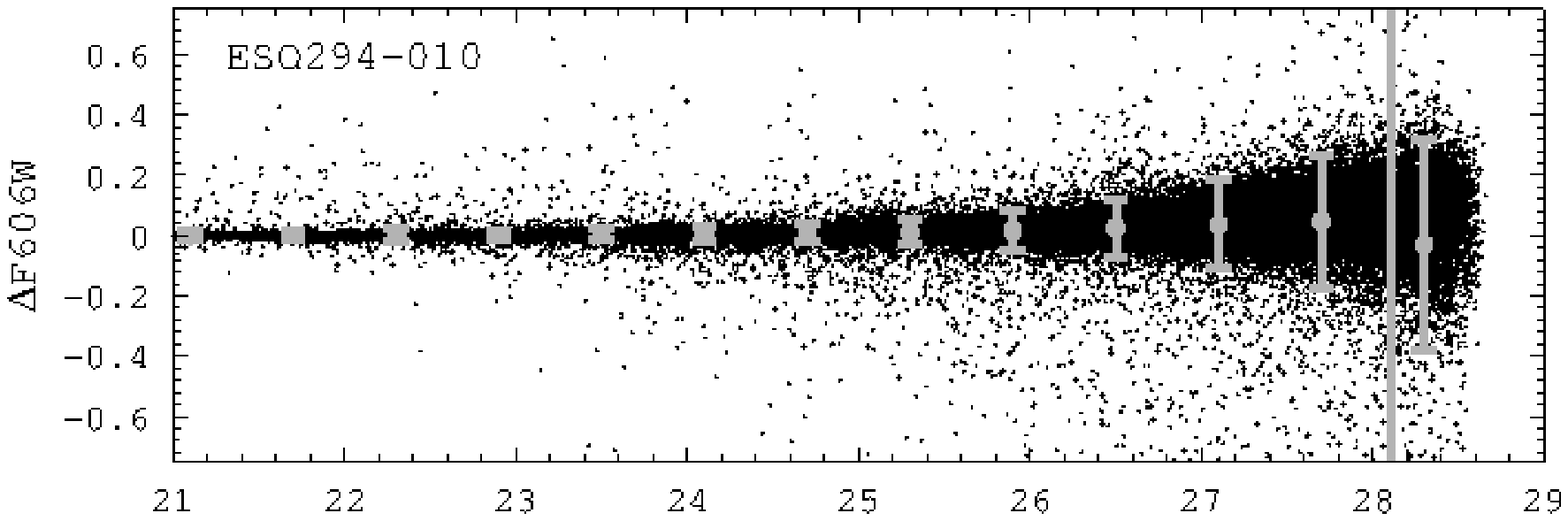}
      \includegraphics[width=8cm,clip]{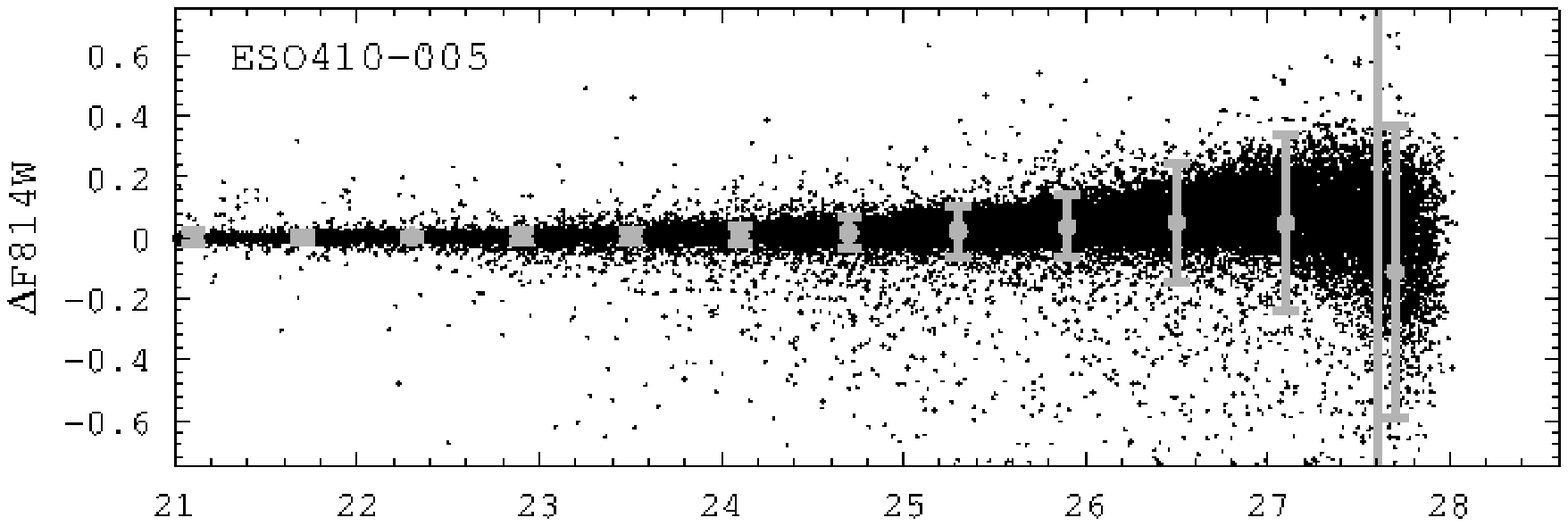}
      \includegraphics[width=8cm,clip]{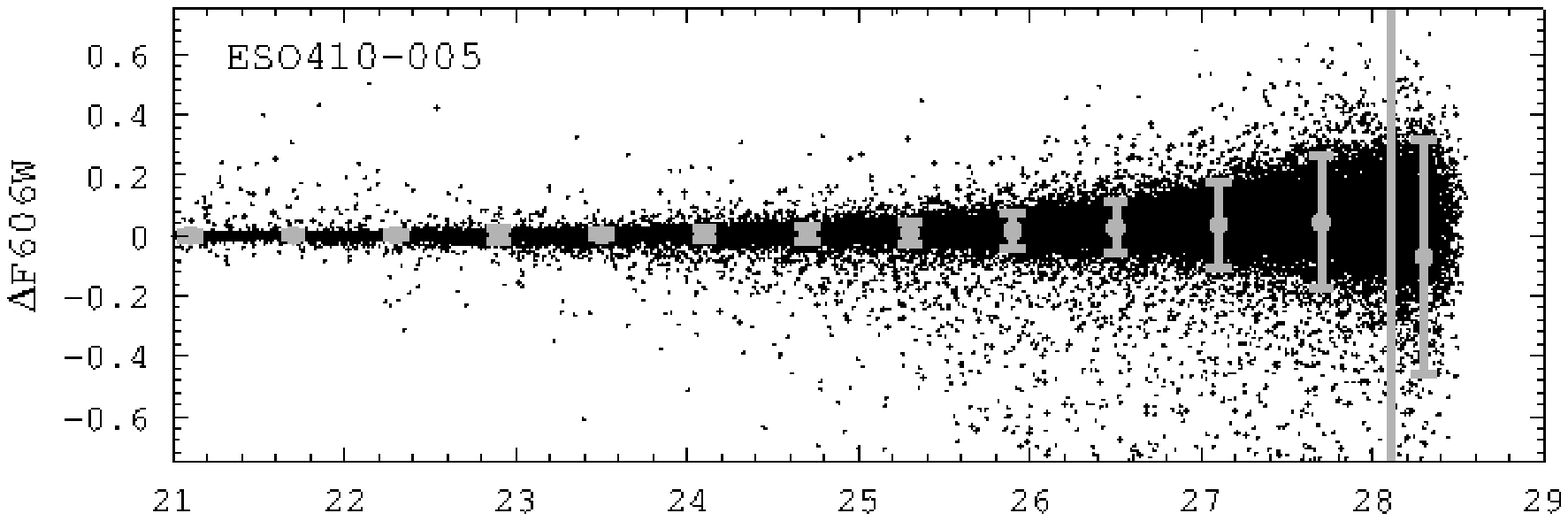}
      \includegraphics[width=8cm,clip]{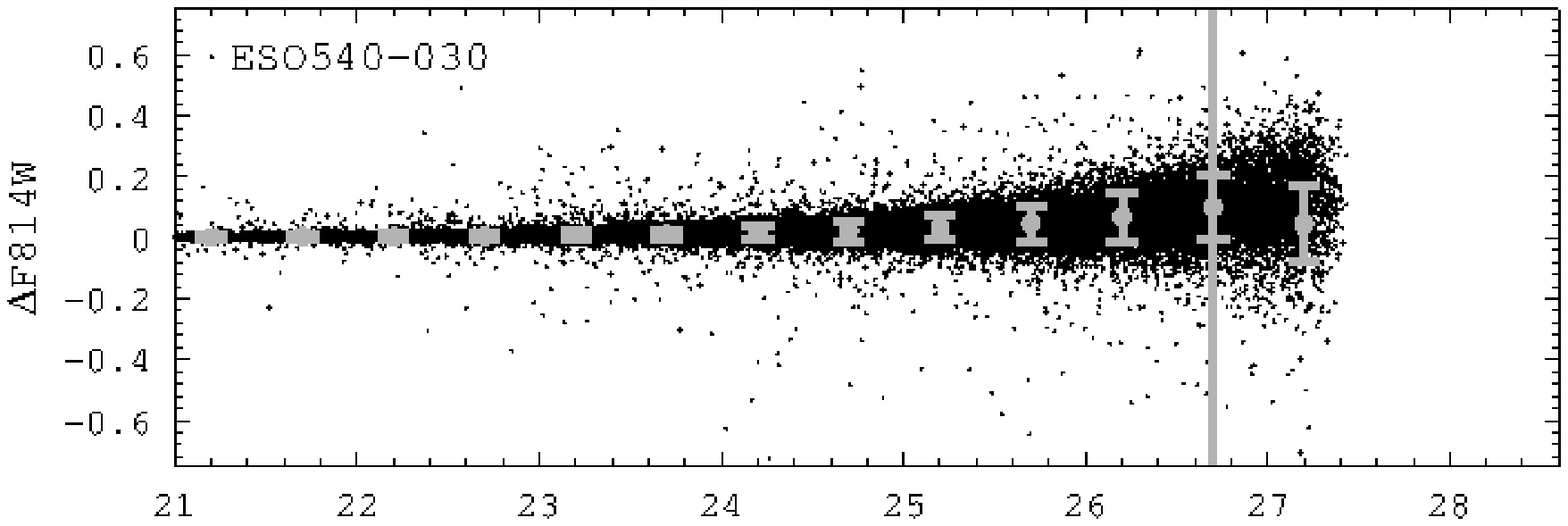}
      \includegraphics[width=8cm,clip]{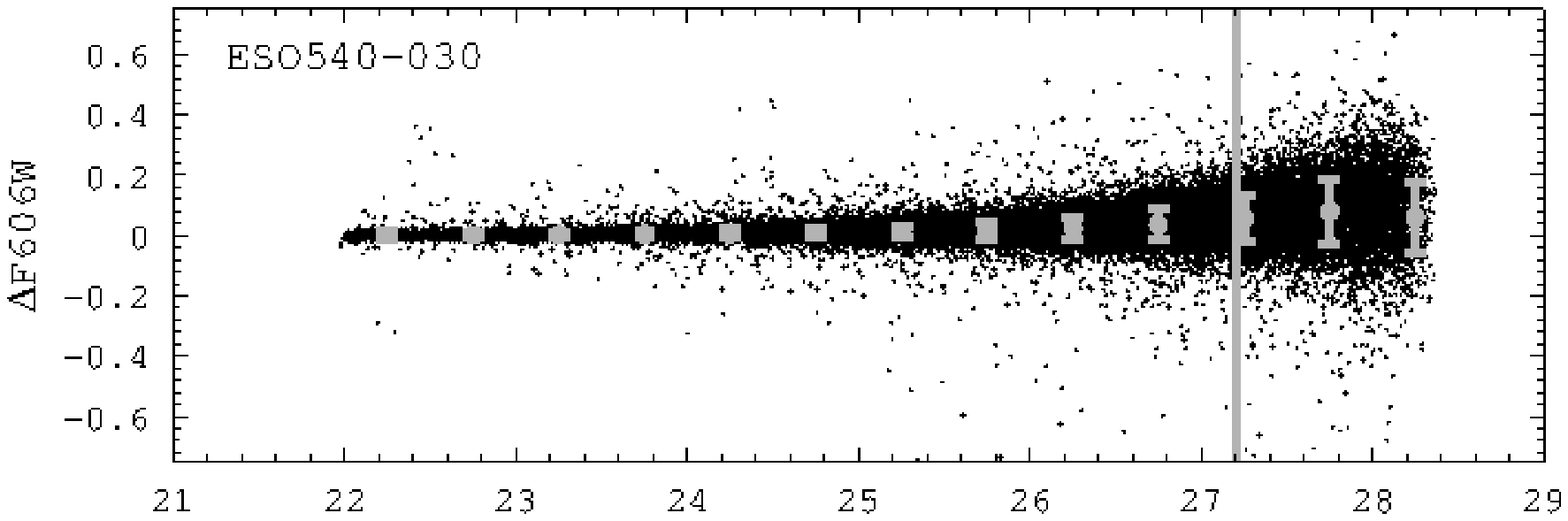}
      \includegraphics[width=8cm,clip]{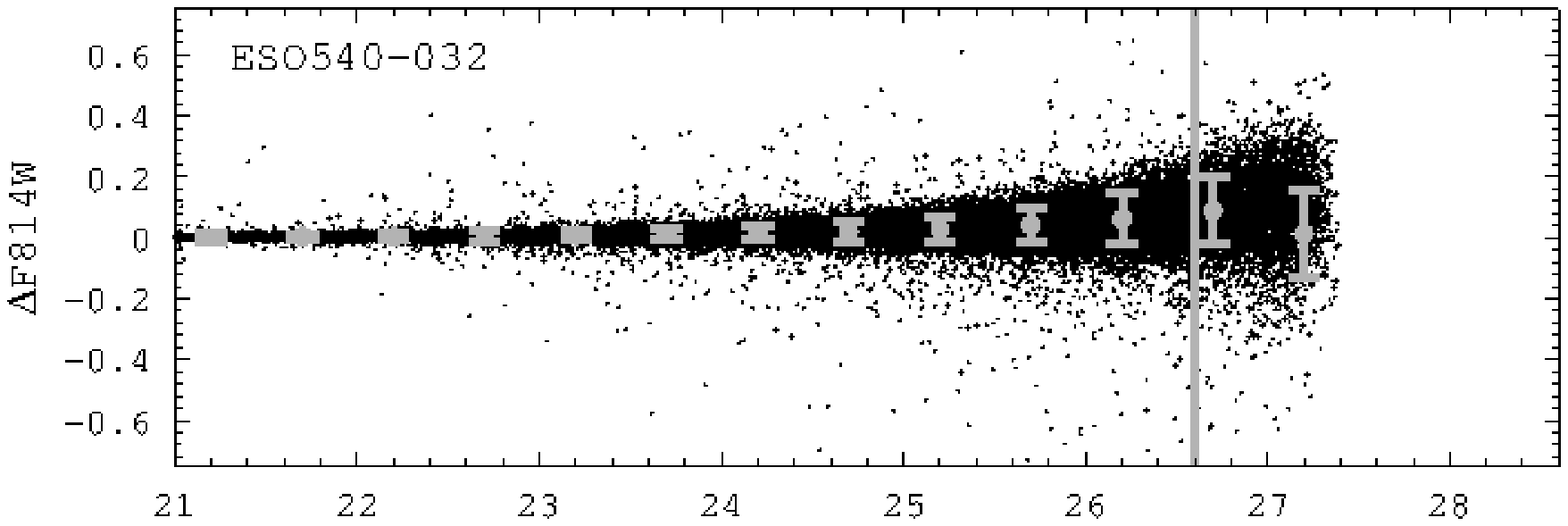}
      \includegraphics[width=8cm,clip]{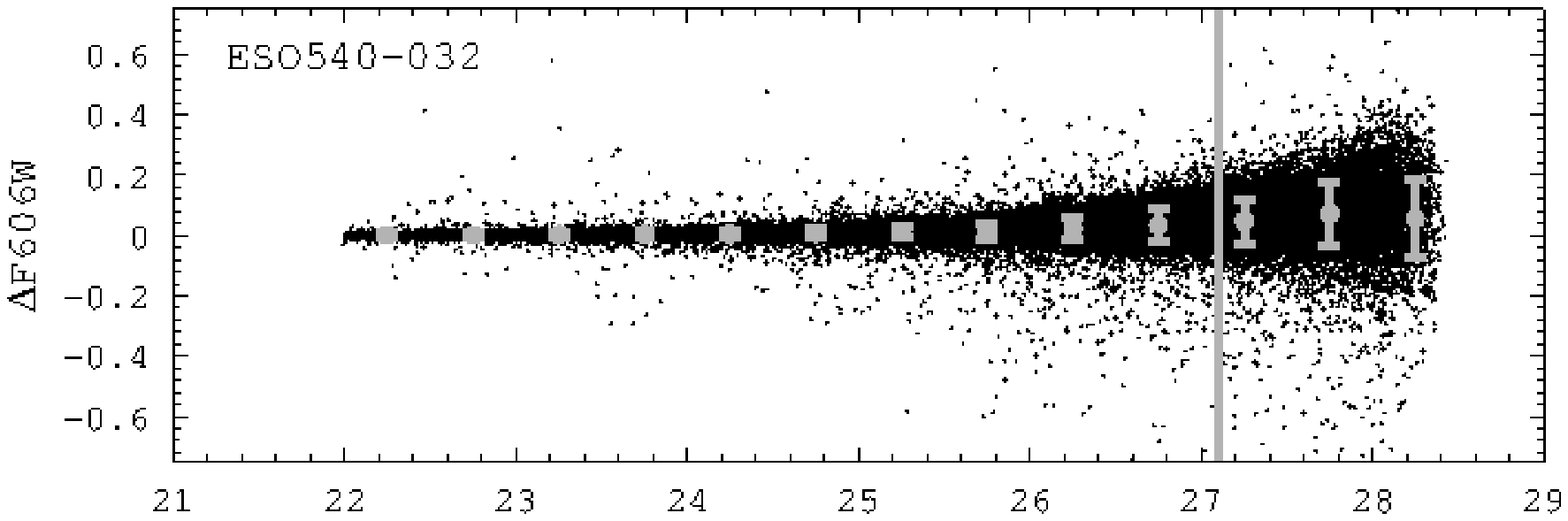}
      \includegraphics[width=8cm,clip]{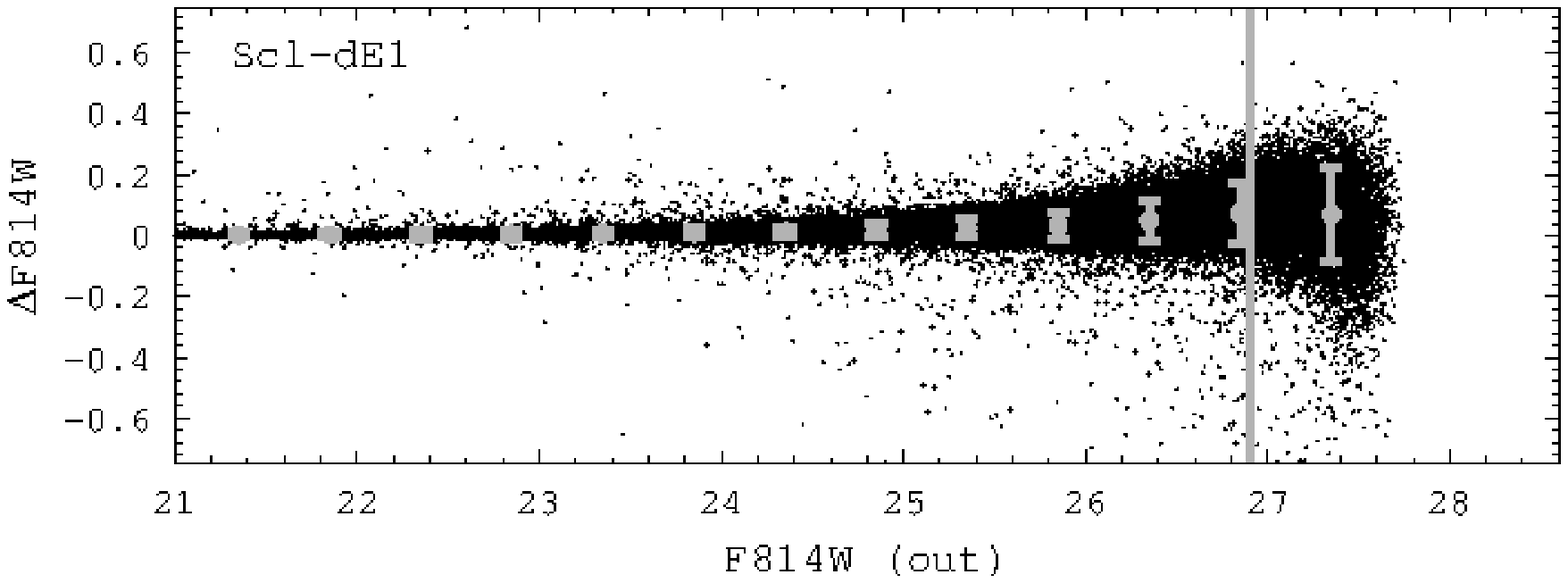}
      \includegraphics[width=8cm,clip]{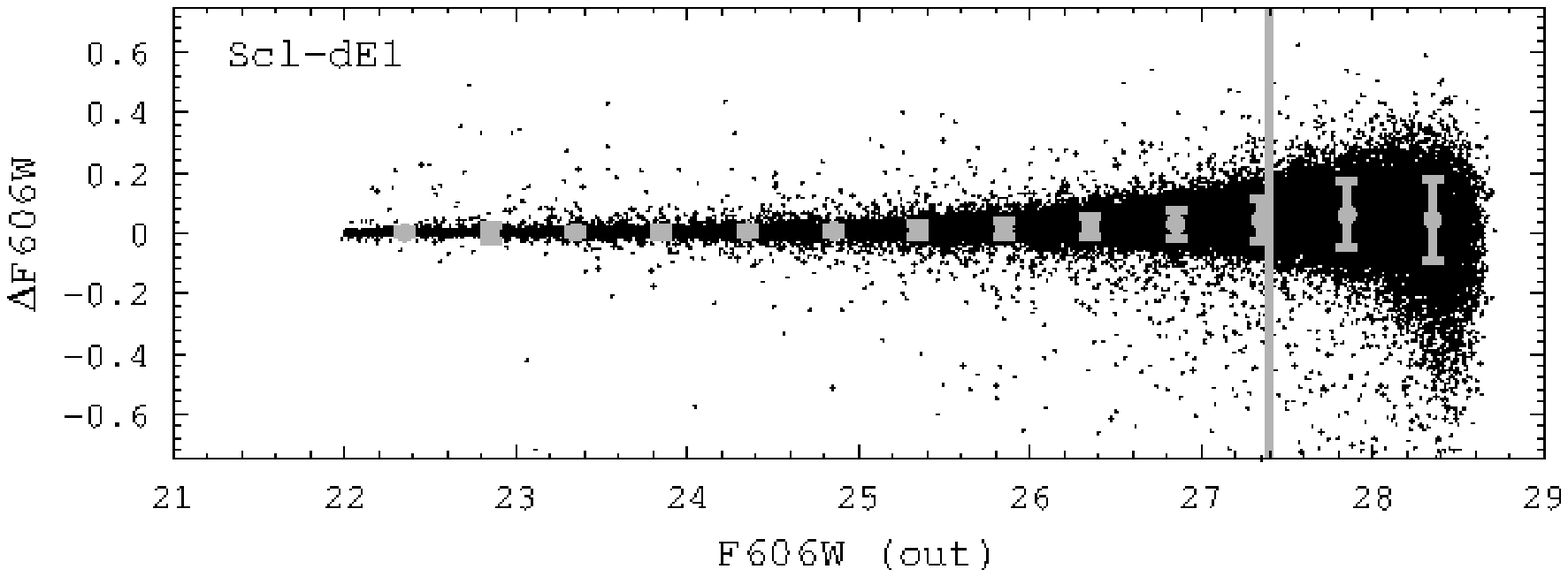}
      \caption{Photometric errors derived from the artificial star tests. The panels show the difference between output and input magnitudes as a function of the output magnitude (left panels: F814W-band; right panels: F606W-band). The vertical gray line indicates the magnitude corresponding to the 50\% completeness factors, while the gray error bars show the mean magnitude difference as a function of the output magnitude.}
      \label{sl_figure02}%
\end{figure*}
%
   We show in Fig.~\ref{sl_figure02} the derived photometric errors in each filter, computed as the output retrieved magnitude minus the input magnitude of each artificial star as a function of its output magnitude. We have applied the same quality cuts as those imposed in the observed stellar catalogues. We use the photometric errors as derived from the artificial tests throughout the whole analysis of this work. 

  \subsection{Crowding effects and foreground contamination}

   We now explore, first, the crowding effects and, second, the foreground contamination that may affect the color-magnitude (CMD) analyses. In order to examine the crowding effects, we use our artificial star tests. We focus on the several stellar evolutionary features we discuss in \S3, such as the main sequence (MS), vertical clump (VC), red giant branch (RGB), luminous asymptotic giant branch (AGB) phases. The definitions of the color and magnitude limits for these stellar evolutionary phases are identical as in $\S$4. We estimate the number of the initially inserted artificial stars in these evolutionary phases, which we compare with the number of stars that were subsequently retrieved in the same evolutionary phases. Their difference will denote the number of stars that have entered in, or migrated from, these evolutionary features, due to the blending of stars. In the case of the MS stars and for all dwarfs where this feature is observed, we find that less than 4\% of the number of stars counted along the MS are due to crowding, while in the case of the VC stars this is less than 3\%. The RGB stars we count are affected by crowding by less than 3.5\%, while the luminous AGB stars are affected by less than 3\%. Therefore, we may conclude that crowding in all dwarfs and in all evolutionary phases examined here is not a major effect on our subsequent analysis and results.

  In order to estimate the foreground contamination, we use the TRILEGAL code (Girardi et al.~\cite{sl_girardi05}; Vanhollebeke, Groenewegen \& Girardi \cite{sl_vanhollebeke09}), and we count the number of foreground stars that fall within the same location in the CMDs of our studied dwarfs. We find that the number of foreground stars is 47 for ESO540-G030, 49 for ESO540-G032, 61 for ESO294-G010, 56 for ESO410-G005, and 58 for Scl-dE1, which translates to less than 0.7\% of the total detected stars (column (3), Table~\ref{table2}). Therefore, foreground contamination is overall negligible in our study. If we focus on the individual stellar evolutionary features, then this translates to foreground contamination of less than 3\% for the MS (for ESO540-G030; 1 foreground star in 33 MS stars), 2\% for the VC (for ESO294-G010; 4 foreground stars in 261 VC stars), less than 1\% for the RGB (for ESO294-G010; 6 foreground stars in 2127 RGB stars), and 10\% for the luminous AGB stars (for Scl-dE1; 5 foreground stars in 50 AGB stars). 

   \section{Color-magnitude diagrams and stellar content}

 \begin{figure*}
   \centering
     \includegraphics[width=6cm,clip]{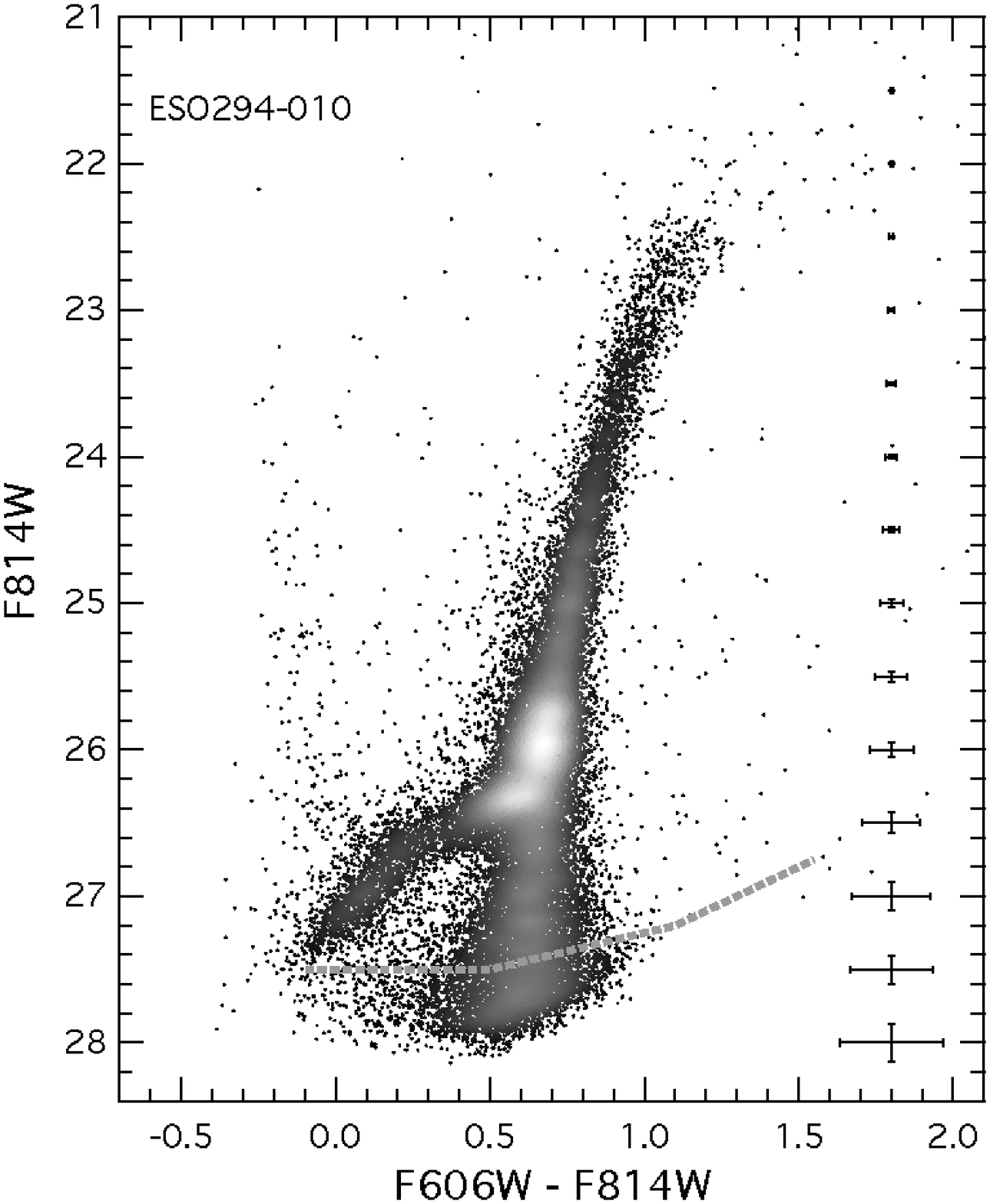}
     \includegraphics[width=6cm,clip]{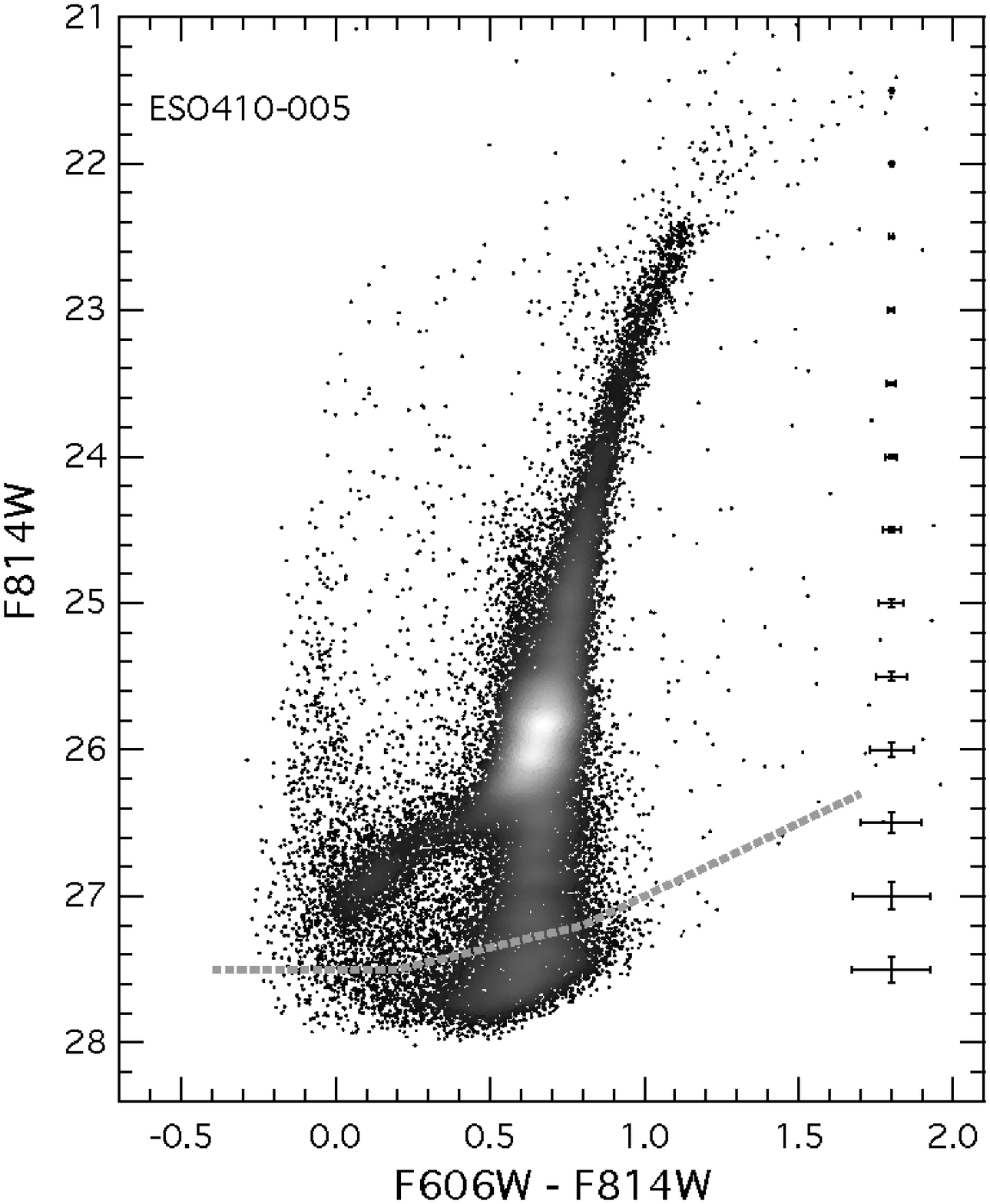}
     \includegraphics[width=6cm,clip]{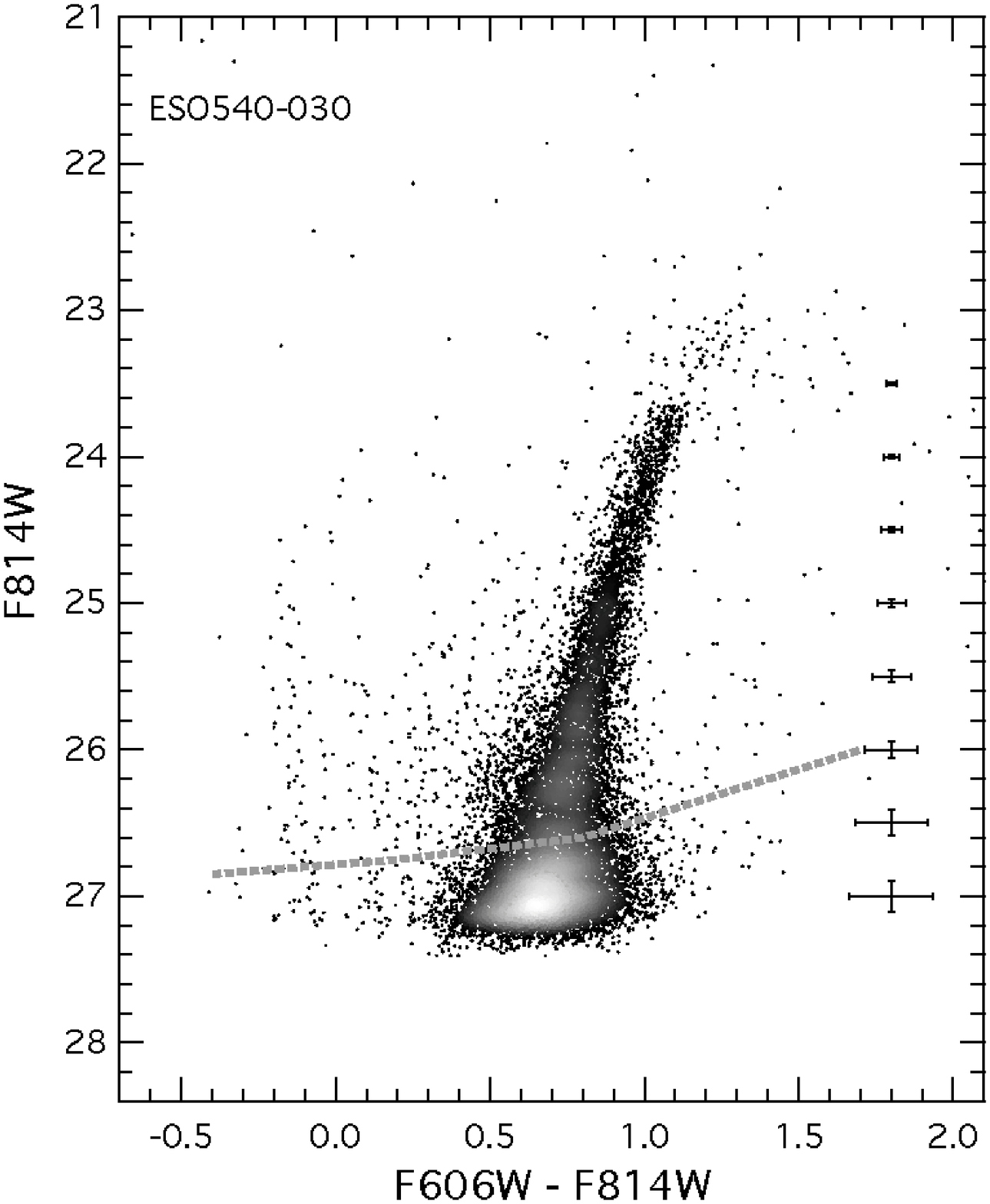}
     \includegraphics[width=6cm,clip]{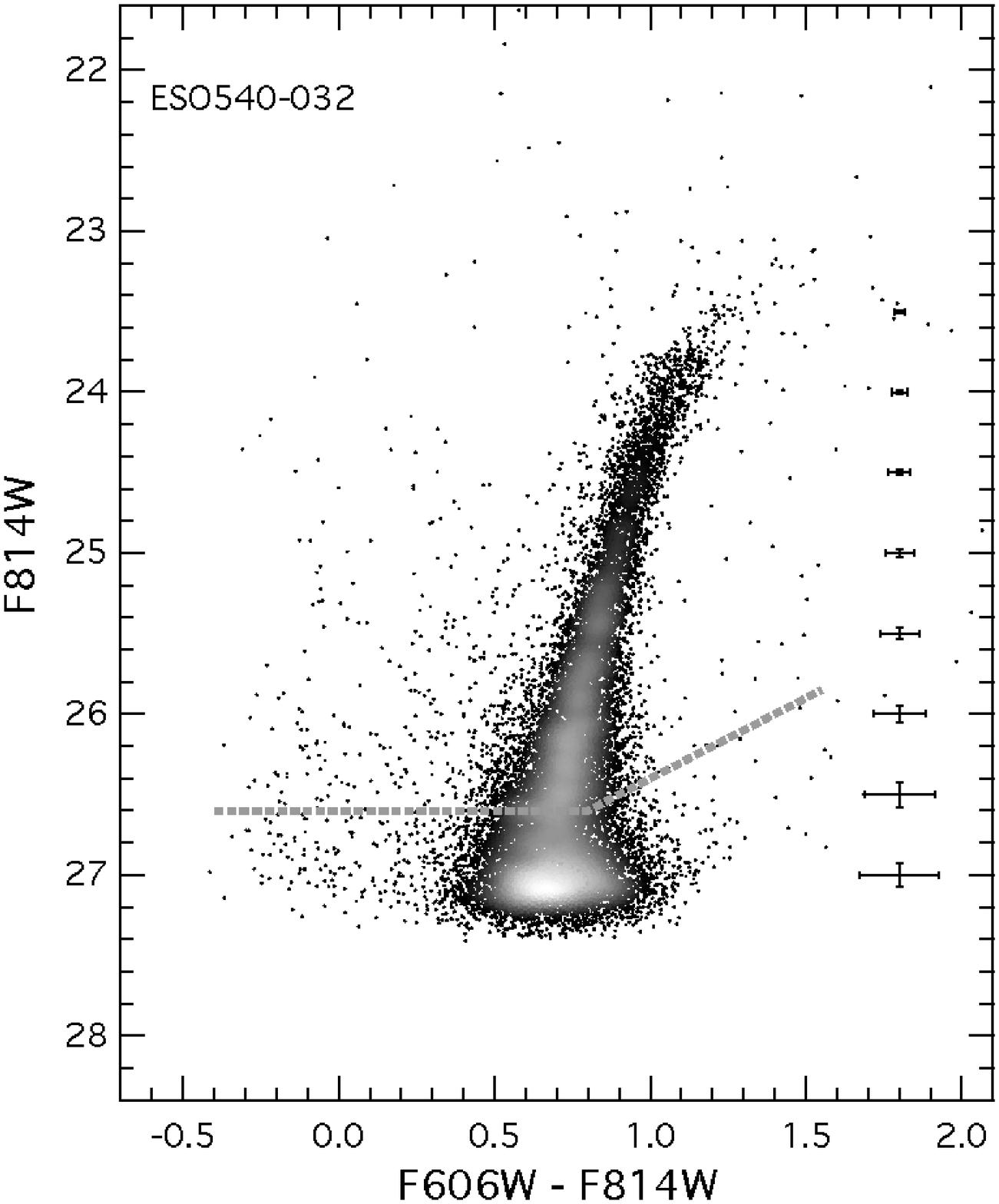}
     \includegraphics[width=6cm,clip]{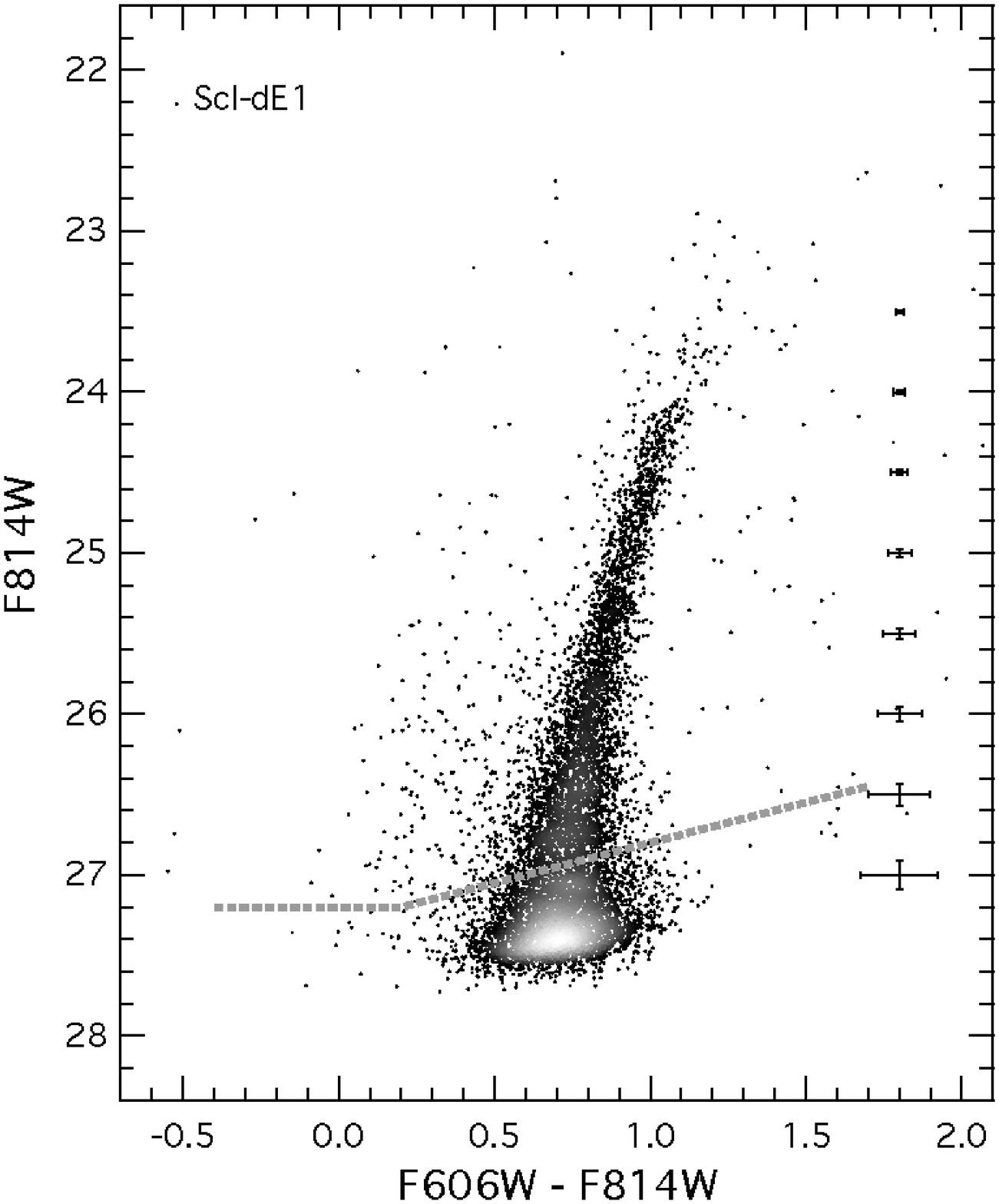}
   \caption{CMDs for the five Sculptor group dwarfs with stellar density maps overlaid to emphasise the high stellar density features. The error bars on the red part of the CMD correspond to the photometric errors as derived from artificial star tests. The grey dotted line indicates the 50\% completeness factors as a function of color and magnitude.}
      \label{sl_figure03}%
\end{figure*}
   We show the CMDs of our studied dwarf galaxy sample in Fig.~\ref{sl_figure03}. A prominent feature in all dwarfs is the RGB. Stars that populate the RGB, depending on the particular SFH of each galaxy, have ages larger than $\sim$1.5~Gyr (e.g., Salaris, Cassisi \& Weiss \cite{sl_salaris02}). Additional features are the young MS, the AGB stars, and the red clump (RC) and VC stars. Another prominent feature in the case of ESO294-G010 and ESO410-G005 is the horizontal branch (HB), which is resolved due to the depth of the observations as well as the proximity of these dwarfs. The fact that ESO294-G010 and E410-G005 contain blue HB stars and RR\,Lyrae variables confirms, for the first time, the existence of ancient stars ($>$10~Gyr) in dwarf galaxies outside the LG (Da Costa et al.~\cite{sl_dacosta10}). 

   In all subsequent analyses, we use the TRGB magnitude and distance modulus we measure for each dwarf using our data in the ACS/WFC filter system, unless explicitly noted otherwise. We use the Sobel-filtering technique on the F814W-band luminosity function (e.g., Lee, Freedman \& Madore \cite{sl_lee93}; Sakai et al.~\cite{sl_sakai96}) in order to measure the F814W-band magnitude of the TRGB. We use this TRGB F814W-band magnitude to compute the distance modulus, using the calibration derived in Rizzi et al.~(\cite{sl_rizzi07}) between the absolute F814W-band magnitude of the TRGB and the color of the TRGB in the ACS filter system, combined with the extinctions listed in Table~\ref{table2}, column (7). The calibration of Rizzi et al.~(\cite{sl_rizzi07}) that we use is the following: $M_{ACS}^{F814W} = -4.06 + 0.20 [(F606W-F814W) - 1.23]$. The mean dereddened color at the TRGB magnitude level is 1.07$\pm$0.02~mag for ESO540-G030, 1.14$\pm$0.02~mag for ESO540-G032, 1.13$\pm$0.02~mag for ESO294-G010, 1.12$\pm$0.01~mag for ESO410-G005, and 1.07$\pm$0.03~mag for Scl-dE1. We list the derived F814W-band magnitudes of the TRGB as well as the distance moduli in Table~\ref{table2}, columns (9) and (10), respectively. The uncertainties indicated for the magnitude of the TRGB take into account the width of the Sobel filter response, and the magnitude difference between the DOLPHOT and DAOPHOT photometry (see Appendix A). The uncertainties indicated for the distance modulus are estimated by taking into account the width of the Sobel filter response for the TRGB magnitude detection, the reddening uncertainty, uncertainties related to the TRGB calibration, and the photometric uncertainties. 

  The F814W-band magnitudes of the TRGB we derive are consistent with those derived in Da Costa et al.~(\cite{sl_dacosta09,sl_dacosta10}) based on the VI photometry from the same images. Da Costa et al. report I-band TRGB values of 24.15$\pm$0.12~mag, 22.45$\pm$0.07~mag and 22.35$\pm$0.07~mag for Scl-dE1, ESO294-G010, and ESO410-G005, respectively. Our TRGB measurements are also consistent with the results from Karachentsev et al.~(\cite{sl_karachentsev00}) for ESO410-G005 (22.4$\pm$0.15~mag) and with the results from Jerjen \& Rejkuba (\cite{sl_jerjen01}) for ESO540-G032 (23.48$\pm$0.09~mag). Finally, these values are in very good agreement with the ones in Dalcanton et al.~(\cite{sl_dalcanton09}), who also use the same observations. 

   In the following, we discuss the ages and metallicities of the stars present in each dwarf using their resolved stellar content, while we leave the RGB
\begin{table}
     \begin{minipage}[t]{\columnwidth}
      \caption{Ages and metallicities of different stellar populations. Upper rows list ages, lower rows list Z metallicities.}
      \label{table3} 
      \centering
      \renewcommand{\footnoterule}{}
      \begin{tabular}{l c c c }
\hline\hline
    Galaxy      	&RC/VC                &luminous AGB             &MS                \\ 
    (1)                 &(2)                  &(3)                      &(4)               \\ 
\hline 
    ESO540-G030         &...                  &1-2~Gyr            &30~Myr            \\
                        &...                  &...                      &0.001             \\
\hline
    ESO540-G032         &...                  &1-2~Gyr            &70~Myr            \\
                        &...                  &...                      &0.001             \\
\hline
    ESO294-G010   	&400~Myr-8~Gyr        &1-2~Gyr            &50~Myr            \\
                        &0.001-0.003          &...                      &0.001             \\
\hline
    ESO410-G005  	&400~Myr-8~Gyr        &1-2~Gyr            &100~Myr           \\ 
                        &0.001-0.002          &...                      &0.001             \\ 
\hline
    Scl-dE1	   	&...                  &1-2~Gyr            &...               \\
                        &...                  &...                      &...               \\
\hline
\end{tabular} 
\end{minipage}
\end{table}
%
as the focus of a subsequent section. The results on the ages and Z metallicity estimates of the following paragraphs are summarised in Table~\ref{table3}, upper rows for the ages and lower rows for the Z metallicity, in each galaxy respectively.

   \subsection{Red clump and vertical clump stars}

   The CMDs of ESO294-G010 and ESO410-G005 (Fig.~\ref{sl_figure03}) reveal the presence of RC and VC stars. The RC stars are core helium-burning stars with intermediate ages larger than 1~Gyr (e.g., Girardi \& Salaris \cite{sl_girardi01}). The VC stars are also core helium-burning stars with younger ages between 400~Myr to less than 1~Gyr (e.g., Beaulieu \& Sackett \cite{sl_beaulieu98}). For the RC stars, the mean F814W magnitude is 25.86$\pm$0.17~mag for ESO294-G010, and 25.89$\pm$0.15~mag for ESO410-G005, while their mean color is 0.68$\pm$0.04~mag, and 0.67$\pm$0.04~mag, respectively. The VC stars have a mean color of 0.61$\pm$0.03~mag for ESO294-G010, and 0.60$\pm$0.03~mag for ESO410-G005, while they extend in magnitude from 24.4~mag to fainter magnitudes reaching the RC. We use these two features in order to constrain the ages and metallicities of the stars in these phases, following the method outlined in Beaulieu \& Sackett (\cite{sl_beaulieu98}). These authors compute the mean magnitudes of the RC and VC phases using isochrones, with the aim to demonstrate that the VC stars observed in several Large Magellanic Cloud fields represent an age sequence of younger and more massive stars than those found at the primary RC locus. In our study, we compute the mean F606W magnitude of the RC and VC phases using Padova isochrones (Marigo et al.~\cite{sl_marigo08}; Girardi et al.~\cite{sl_girardi08}) for a range of metallicities and ages. We use the F606W magnitude since it is more sensitive to metallicity (e.g., Sarajedini \cite{sl_sarajedini99}). Keeping the metallicity of the isochrones constant, we compute the path on the CMD that the RC and VC phases can take, varying the ages of the isochrones from $\sim$400~Myr to $\sim$8~Gyr in steps of 0.1 in log(t). We then compute several RC and 
 \begin{figure*}
   \centering
      \includegraphics[scale=0.25,clip]{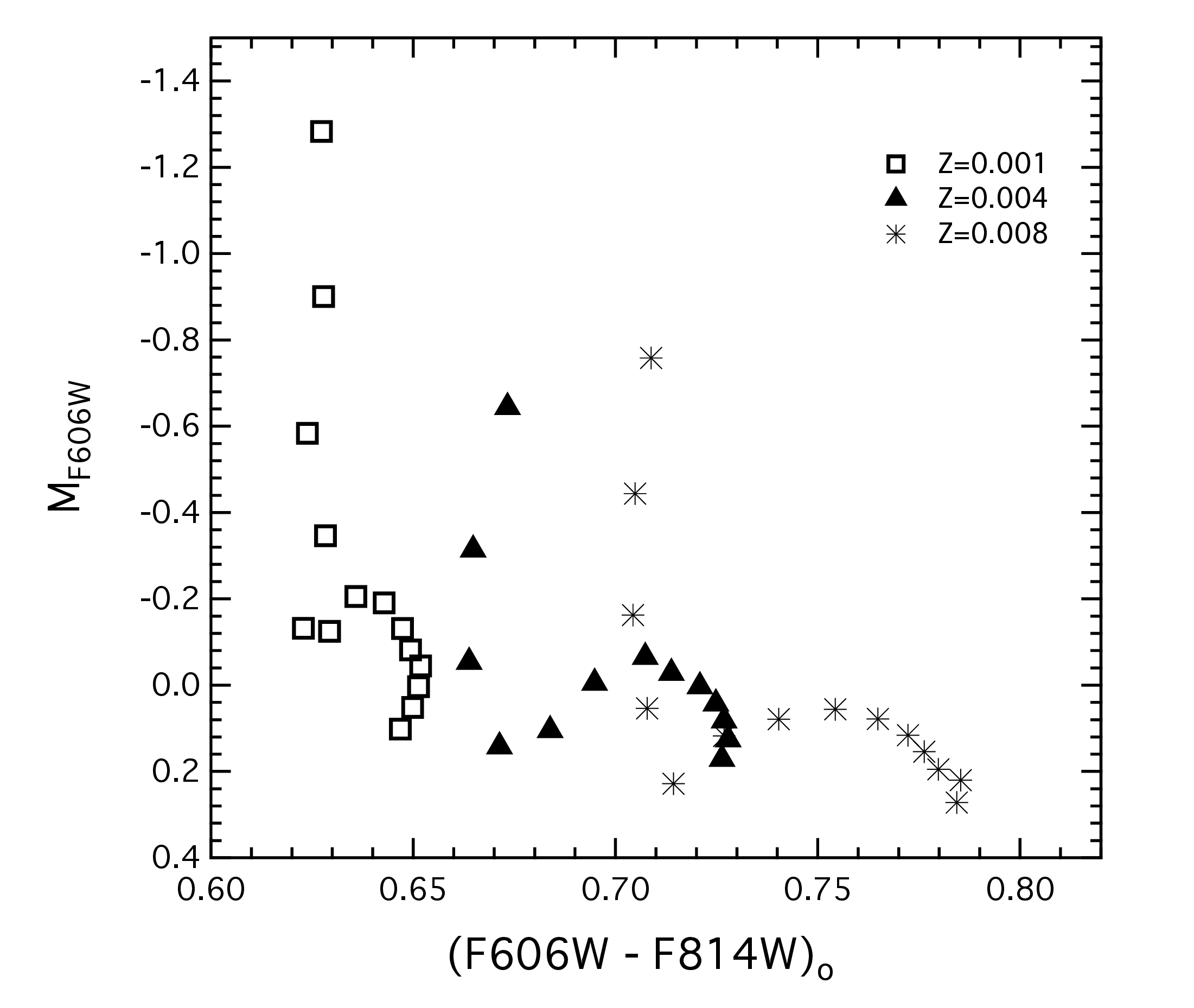}
      \includegraphics[scale=0.26,clip]{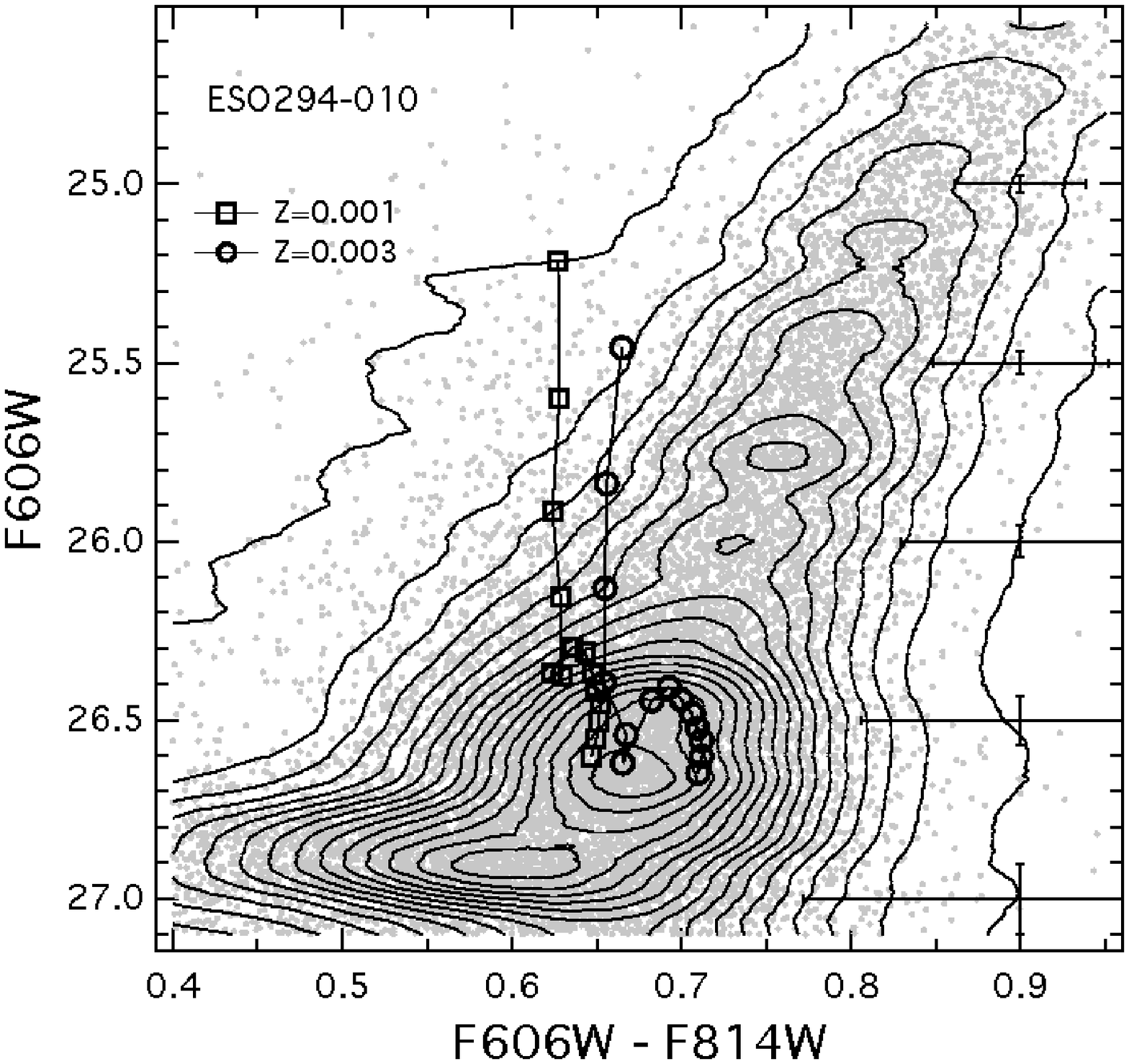}
      \includegraphics[scale=0.26,clip]{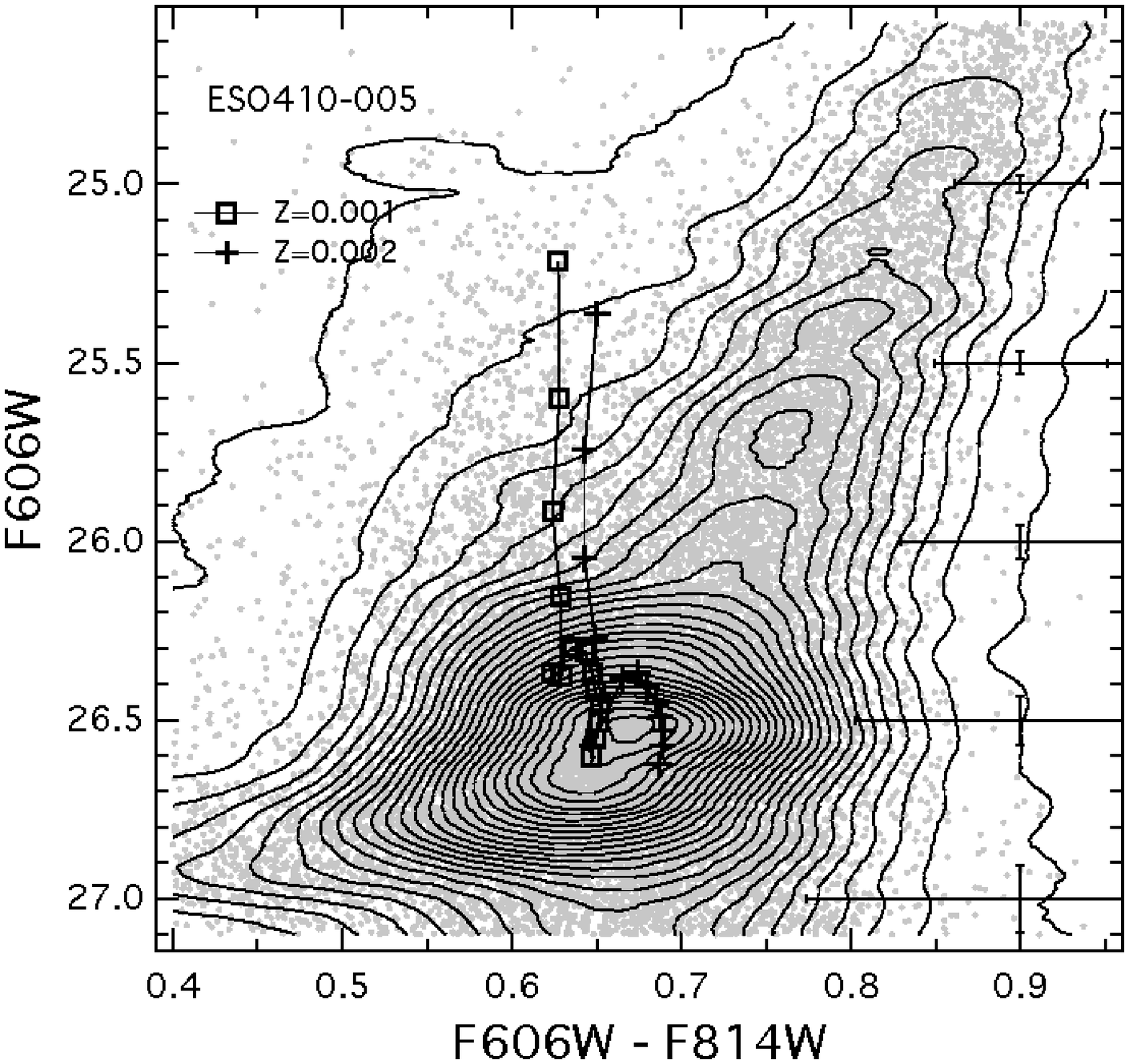}
      \caption{\textit{Left panel}: Mean F606W-band magnitudes of the RC and VC phases for stars with ages ranging from $\sim$400~Myr (bright magnitude symbols) to $\sim$8~Gyr (faint magnitude symbols) with a step of 0.1 in log(t), and with Z metallicities of 0.001 (squares), 0.004 (triangles), and 0.008 (asterisks), computed using Padova isochrones. 
\textit{Middle panel}: Zoom in of the CMD on the RC and VC phases, with isodensity contours overplotted, for ESO294-G010. The RC and VC paths correspond to Z metallicities of 0.001 (line connecting the open square symbols) and 0.003 (line connecting the open circle symbols), and to ages as in the paths of the left panel. 
\textit{Right panel}: The same as the middle panel, but for the case of ESO410-G005. The RC and VC paths correspond to Z metallicities of 0.001 (line connecting the open square symbols) and 0.002 (line connecting the plus symbols), and to ages as in the paths of the left panel.}
\label{sl_figure04}%
\end{figure*}
%
VC paths on the CMD in the same way but now for a range of metallicities. We show an example of the VC and RC paths in the left panel of Fig.~\ref{sl_figure04}. 

   In the middle and right panels of Fig.~\ref{sl_figure04} we zoom in the CMDs of ESO294-G010 and ESO410-G005, respectively, to focus on the RC and VC phases, which are highlighted with the overlaid contours. These features in both dwarfs lie above the magnitude limit set by the 50\% completeness factors. We investigate which of the computed VC and RC paths can adequately encompass the bulk of the RC and VC features observed and shown in the middle and right panels of the same figure. The best VC and RC paths for both galaxies are shown in Fig.~\ref{sl_figure04}, for ESO294-G010 by a solid line connecting the open circle symbols and corresponding to a Z metallicity of 0.003 (middle panel), while for ESO410-G005 with a line connecting the plus symbols corresponding to a Z metallicity of 0.002 (right panel). In both galaxies a VC path of a lower metallicity Z equal to 0.001, shown as a line connecting the square symbols can also adequately reproduce the observed feature. We note though that due to the diffuseness and the size of the color errors of the VC feature, the range of the metallicities can be only poorly constrained. Therefore, with the current analysis on the VC and RC features we can place constraints on the ages and metallicities of the bulk of the stars populating these two stellar evolutionary phases. 

   Our analysis suggests the presence of core He burning stars in the RC and VC phases with ages between 400~Myr to 8~Gyr. The bulk of the stars appears to have Z metallicities ranging from 0.001 to 0.003 for ESO294-G010, and from 0.001 to 0.002 for ESO410-G005. For scaled solar abundances, Z = 0.001 corresponds to [Fe/H] = $-$1.28~dex, using the approximation [Fe/H] = $\rm log ($Z$/$Z$_{\sun} )$ and Z$_{\sun}$ = 0.019 (Girardi et al.~\cite{sl_girardi00}). The VC in ESO294-G010 seems less populated than in ESO410-G005, and we note that the presence of any significant VC population needs to be confirmed after a detailed characterisation of the stellar populations in comparison with simulated CMDs. More detailed analyses of the SFHs can reveal the full range of stellar ages, metallicities and fractions present within each dwarf. Our studied dwarf galaxy sample is included within the Local Volume dwarf galaxy sample analysed in Weisz et al.~(\cite{sl_weisz11a}). Using CMD modelling of ESO294-G010 and ESO410-G005, these authors find intermediate-age stars ($1<$~age~$<10$~Gyr) with fractions of 19\% and 36\%, respectively. These are values in between the ones derived for transition-type dwarfs in and around the LG, as for instance in the case of DDO210 and Antlia that have intermediate-age fractions of 12\% and 43\%, respectively (Orban et al.~\cite{sl_orban08}).

   \subsection{Luminous asymptotic giant branch stars}

   All dwarfs in our sample contain luminous AGB stars. We estimate their ages using the bolometric magnitude of the brightest AGB stars, which is a function of age (e.g., Mould \& Da Costa \cite{sl_mould88}). This method is not free from age-metallicity degeneracy when using V, I magnitudes (e.g., Bertelli et al.~\cite{sl_bertelli94}), therefore it only provides an estimate of the ages of the luminous AGB stars. For each galaxy, we first estimate the bolometric luminosity of the luminous AGB stars, using the relation between the bolometric correction of the I-band magnitude and the intrinsic color, provided in Da Costa \& Armandroff (\cite{sl_dacosta90}; eq. 2). We note that for estimating the bolometric magnitudes, we use values in the UBVRI photometric system whenever needed, for instance, the V, I magnitudes as given by DOLPHOT, the TRGB adopted from Da Costa et al.~(\cite{sl_dacosta09,sl_dacosta10}), Karachentsev et al.~(\cite{sl_karachentsev03}). 

  Then, we use the estimated bolometric luminosity of the brightest AGB stars in conjunction with the relation given in Fig.~19 of Rejkuba et al.~(\cite{sl_rejkuba06}), in order to estimate their age. Our results give an age for the brightest luminous AGB stars between 1~Gyr to 2~Gyr in all five dwarfs, with bolometric magnitudes of $-$5.12~mag for ESO540-G030, $-$5.32~mag for ESO540-G032, $-$5.31~mag for ESO294-G010, $-$5.32~mag for ESO410-G005, and $-$5.50~mag for Scl-dE1.

   \subsection{Young main sequence stars}

   The four transition-type dwarfs have more luminous blue stars that belong to a younger MS.  
 \begin{figure*}
   \centering
     \includegraphics[scale=0.45,clip]{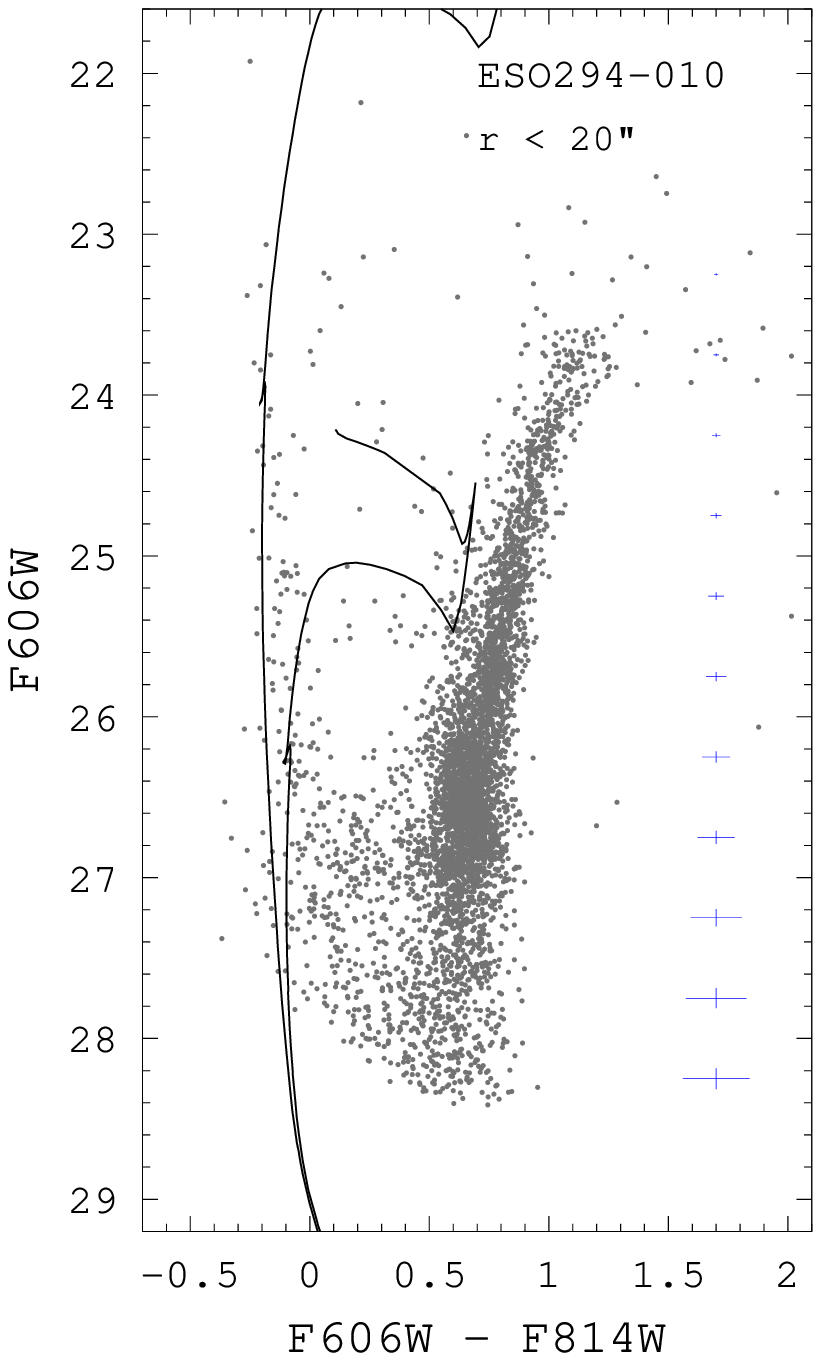}
     \includegraphics[scale=0.45,clip]{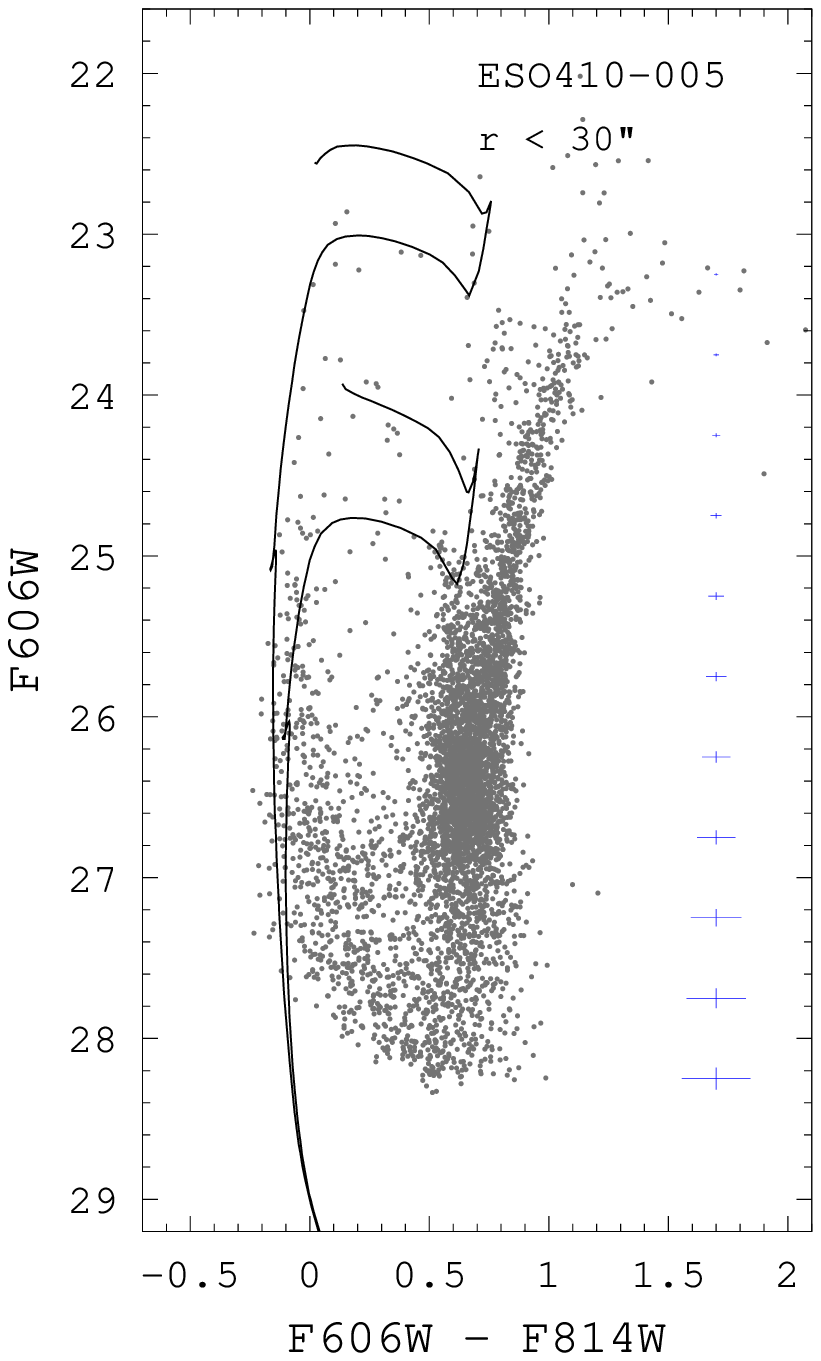}
     \includegraphics[scale=0.45,clip]{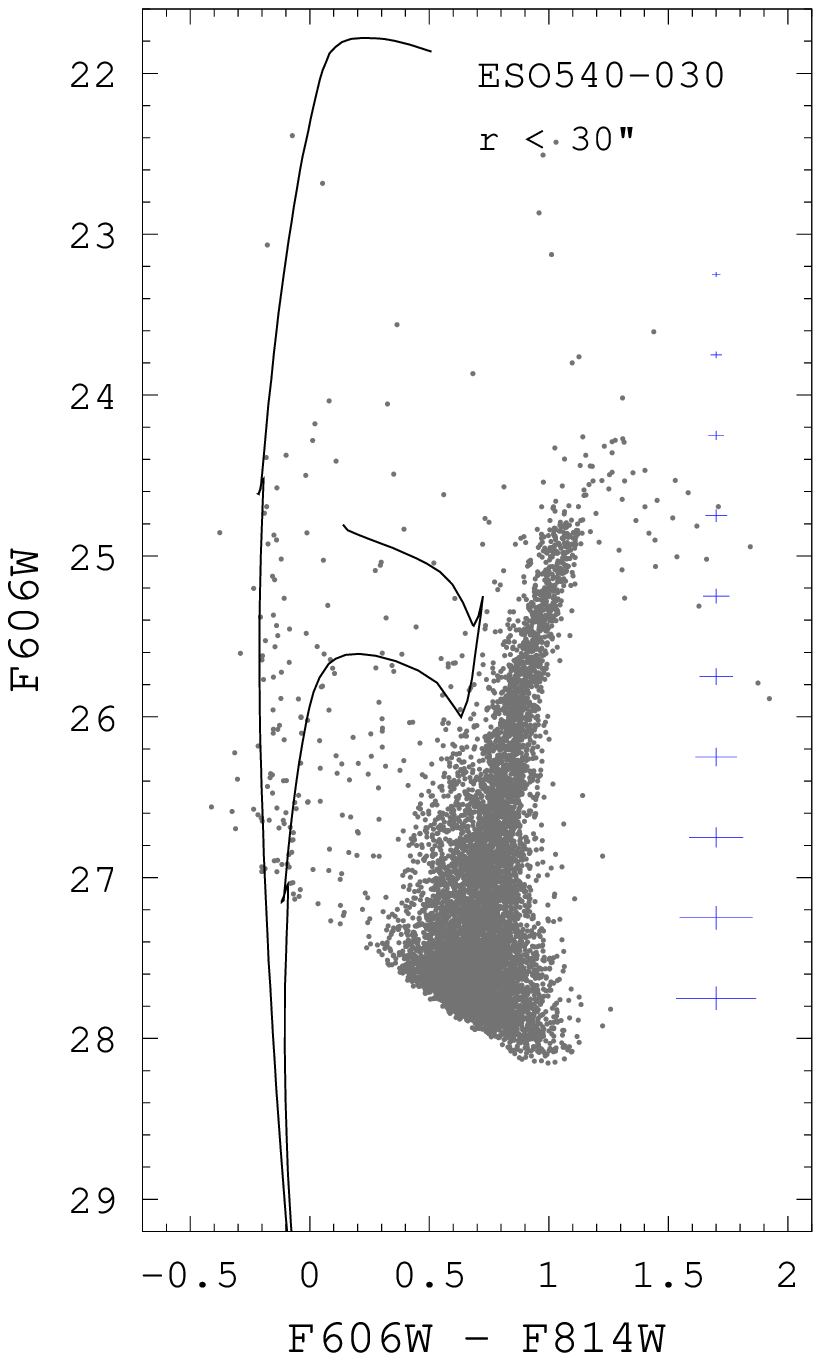}
     \includegraphics[scale=0.45,clip]{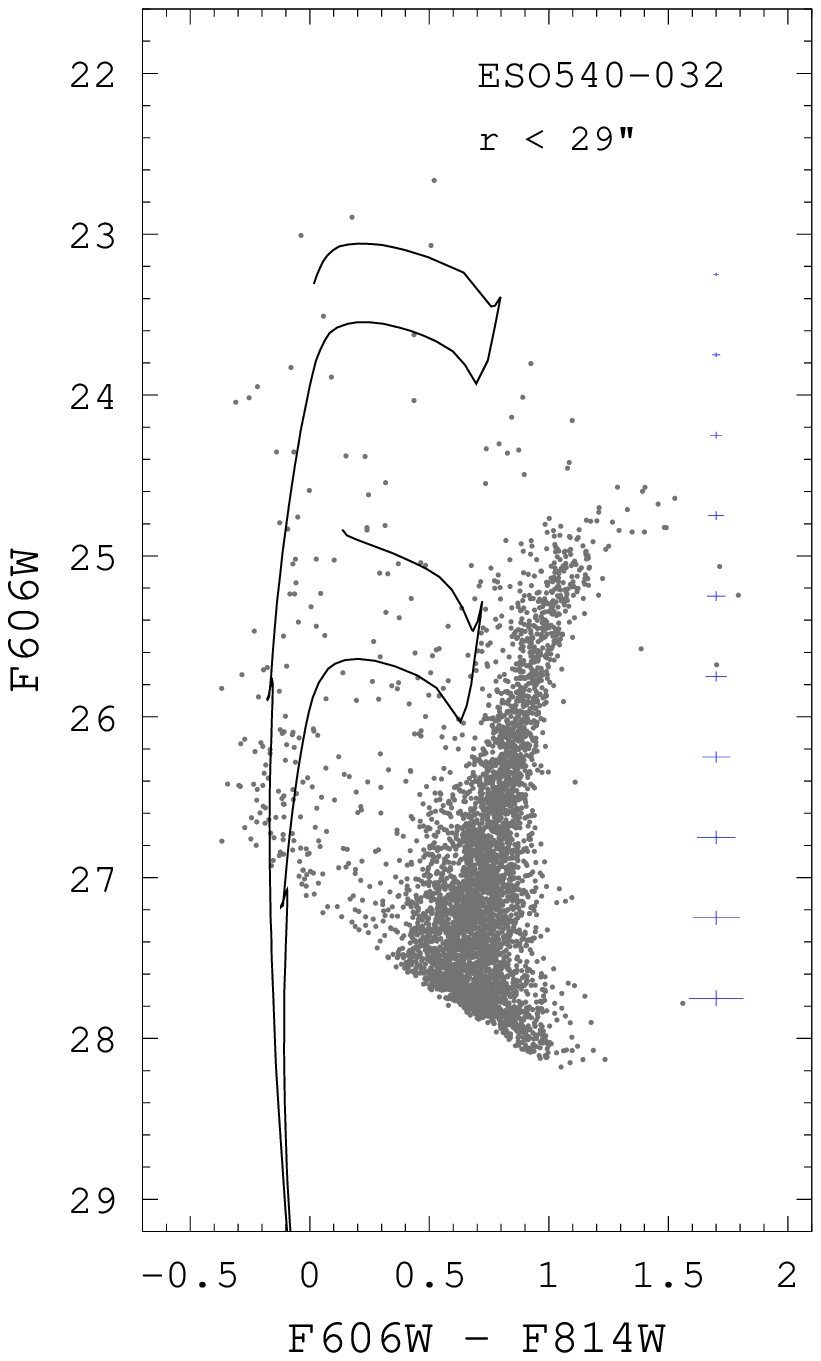}
     \includegraphics[scale=0.45,clip]{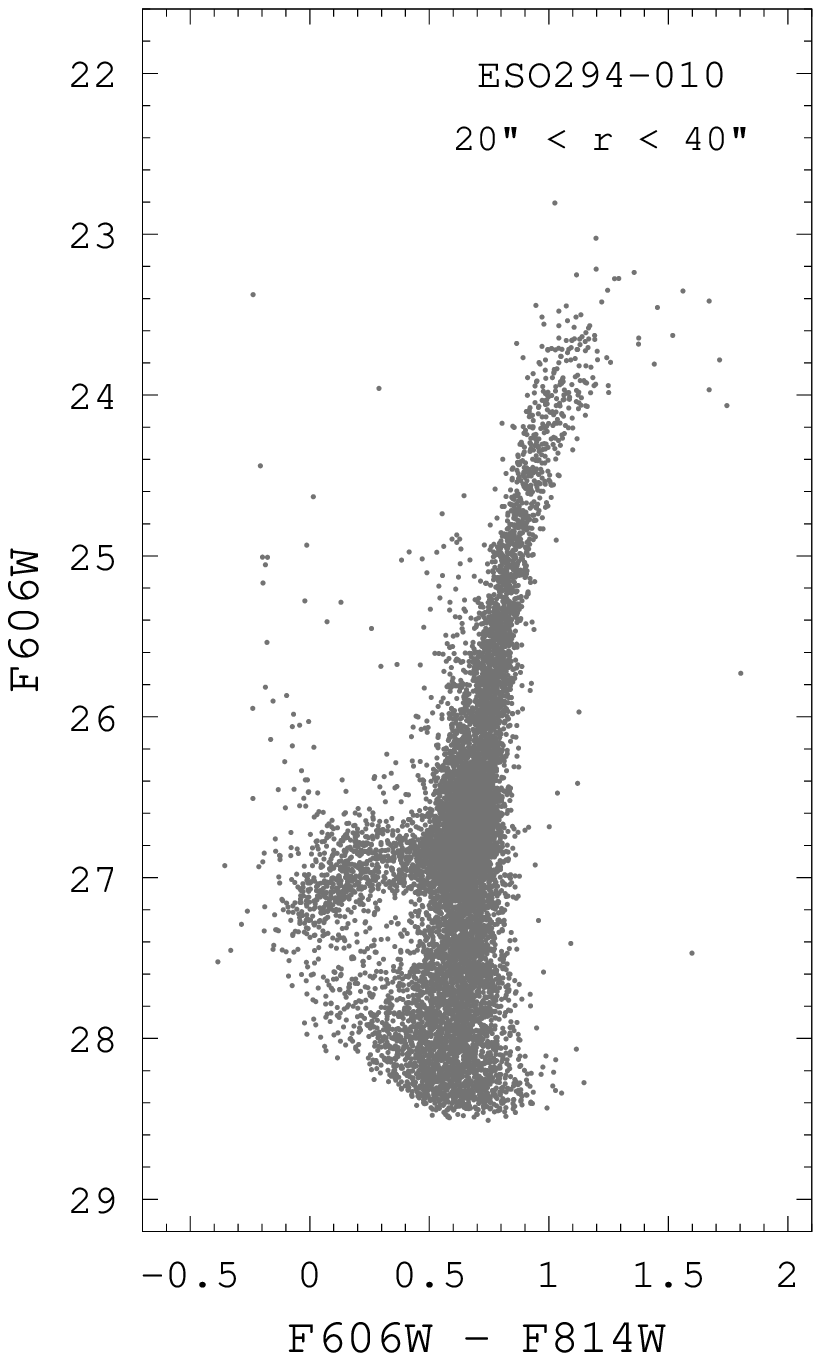}
     \includegraphics[scale=0.45,clip]{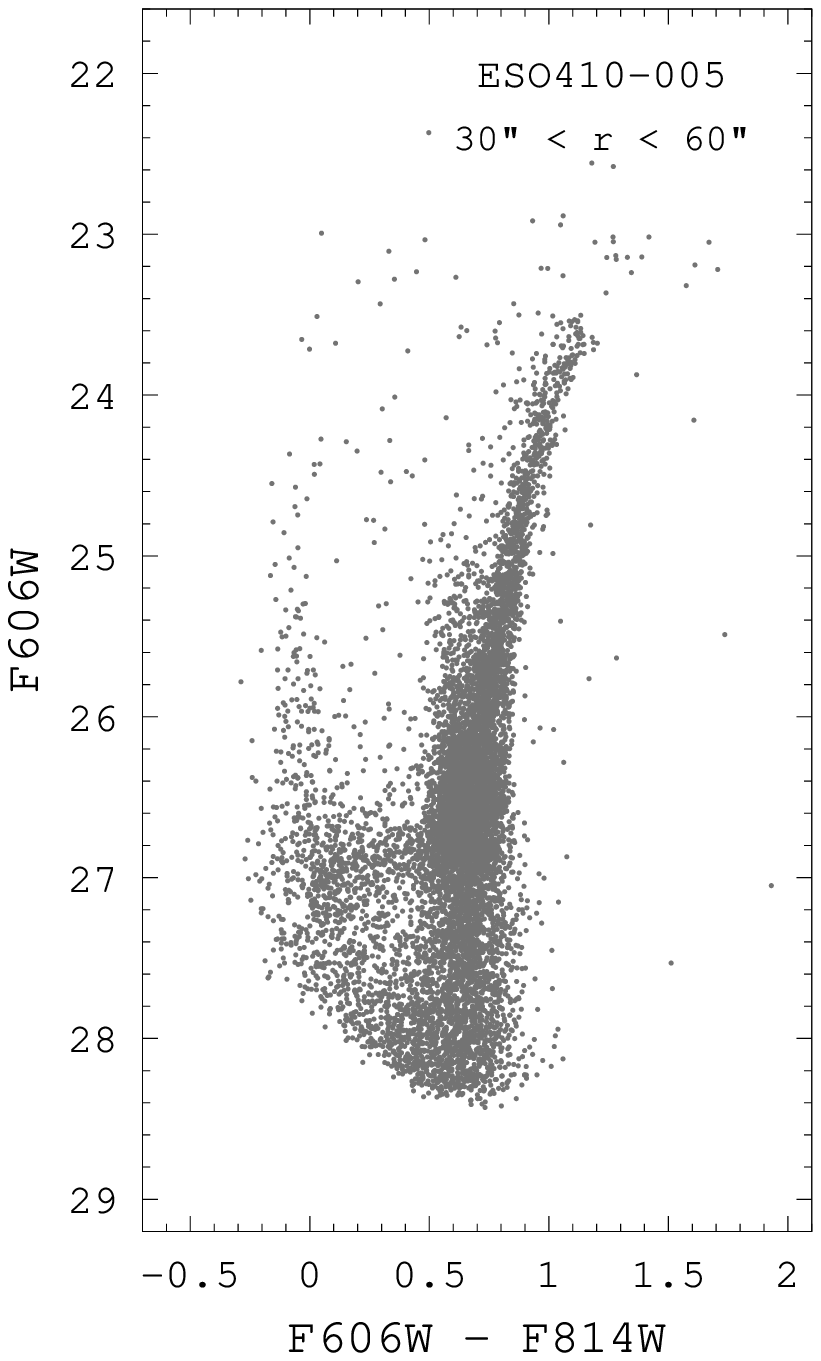}
     \includegraphics[scale=0.45,clip]{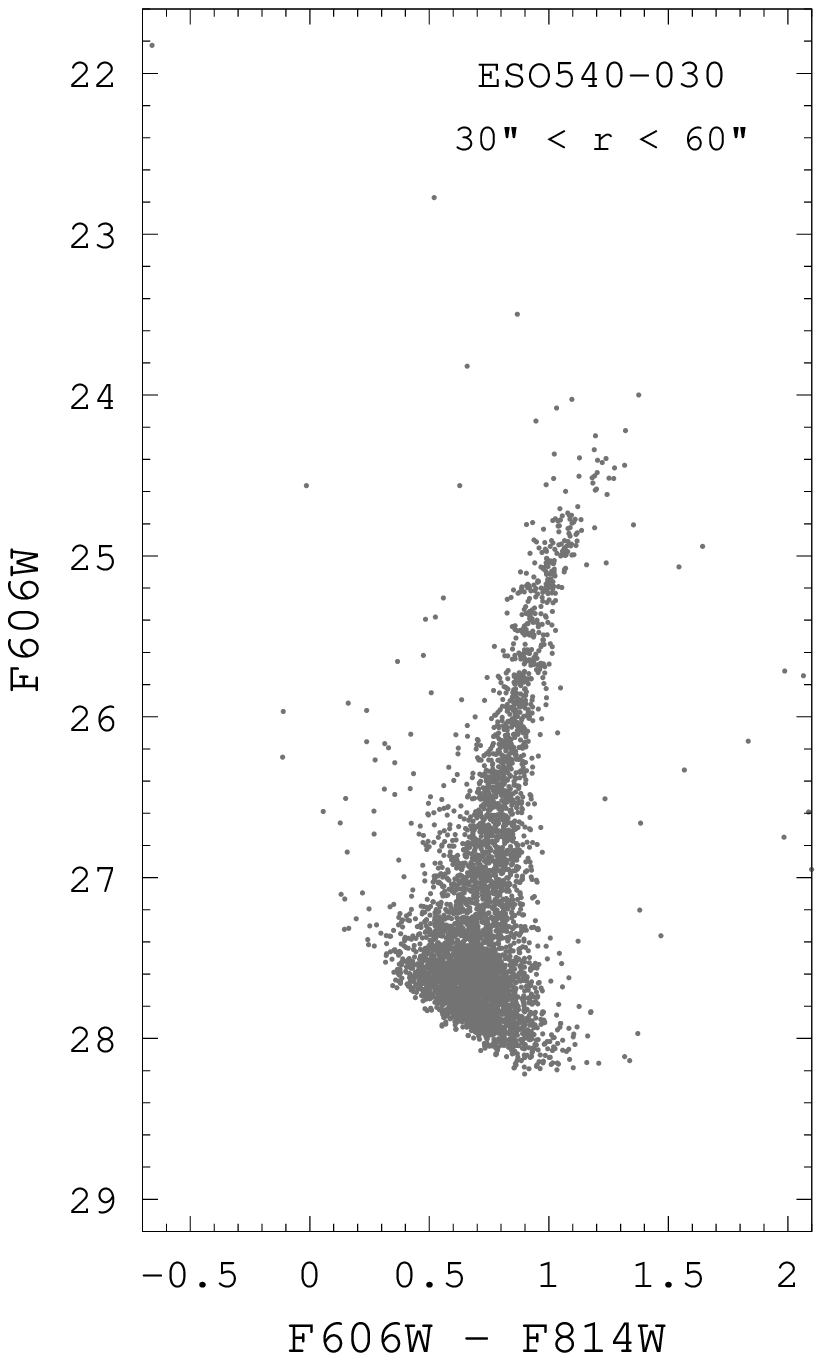}
     \includegraphics[scale=0.45,clip]{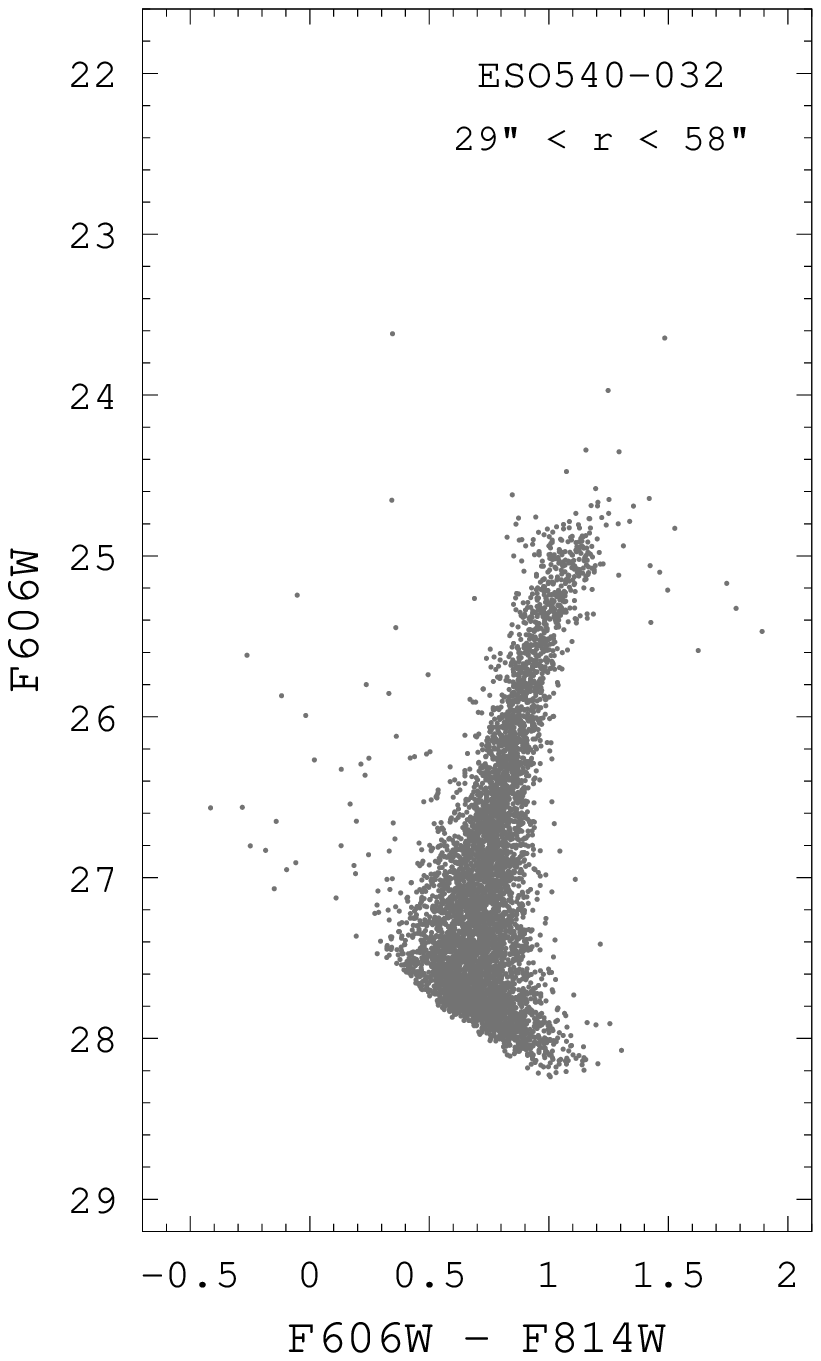}
     \includegraphics[scale=0.45,clip]{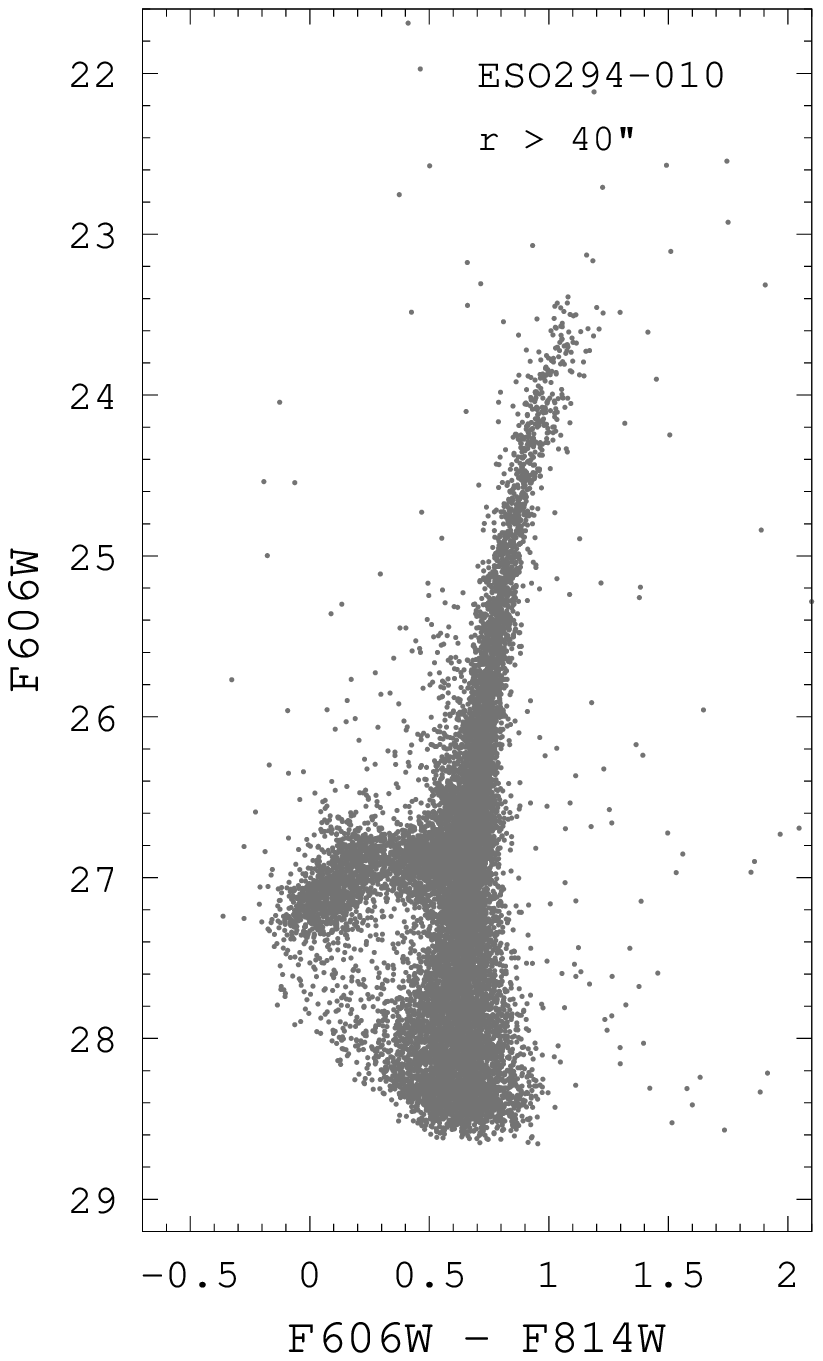}
     \includegraphics[scale=0.45,clip]{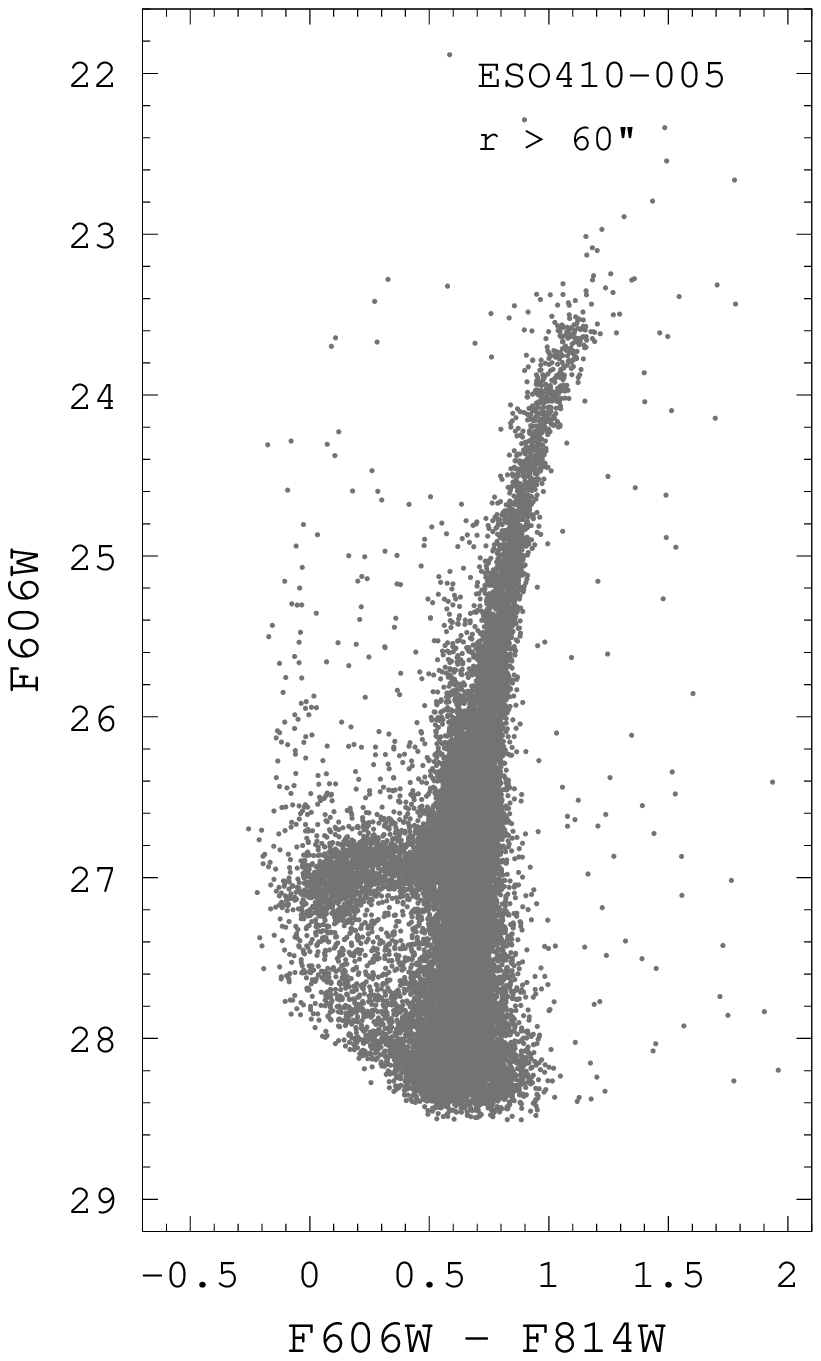}
     \includegraphics[scale=0.45,clip]{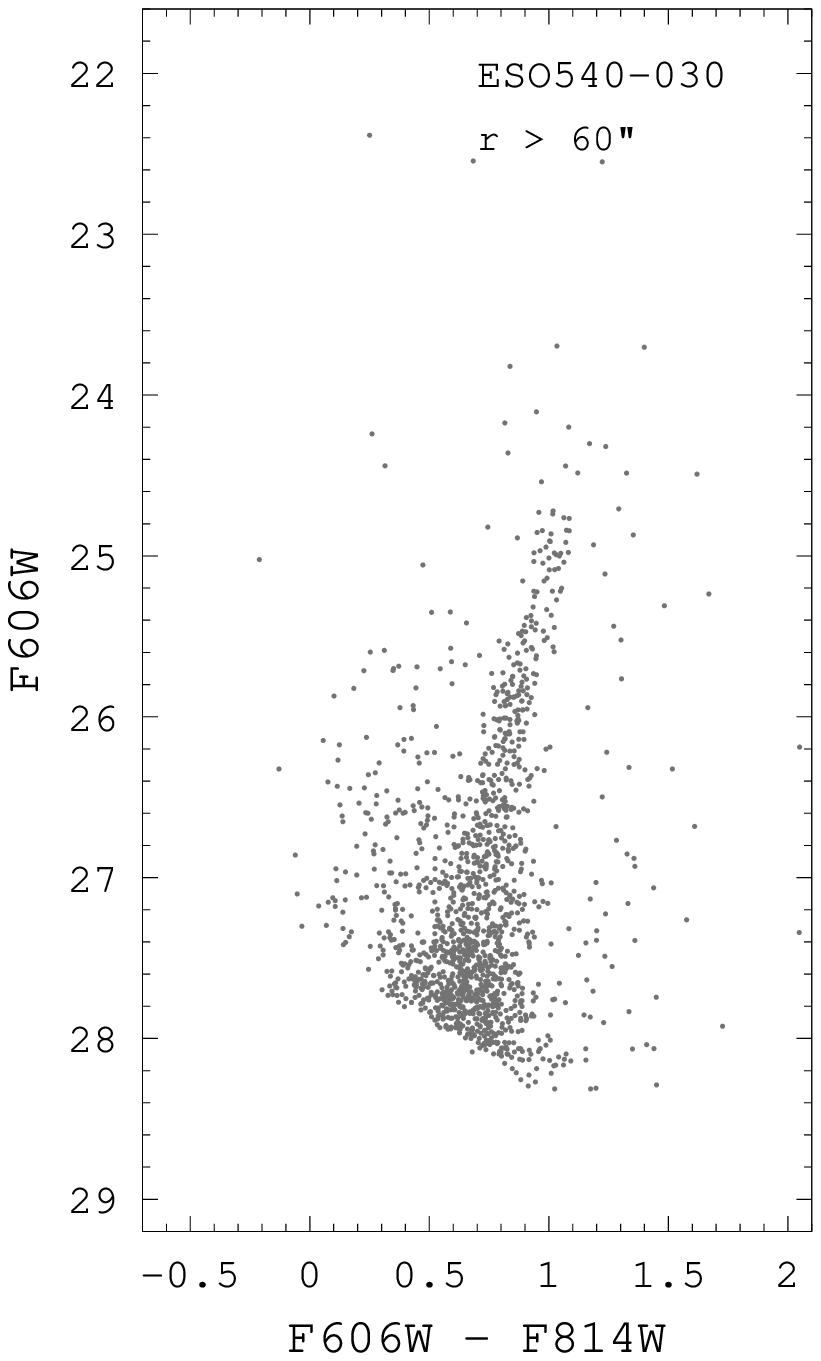}
     \includegraphics[scale=0.45,clip]{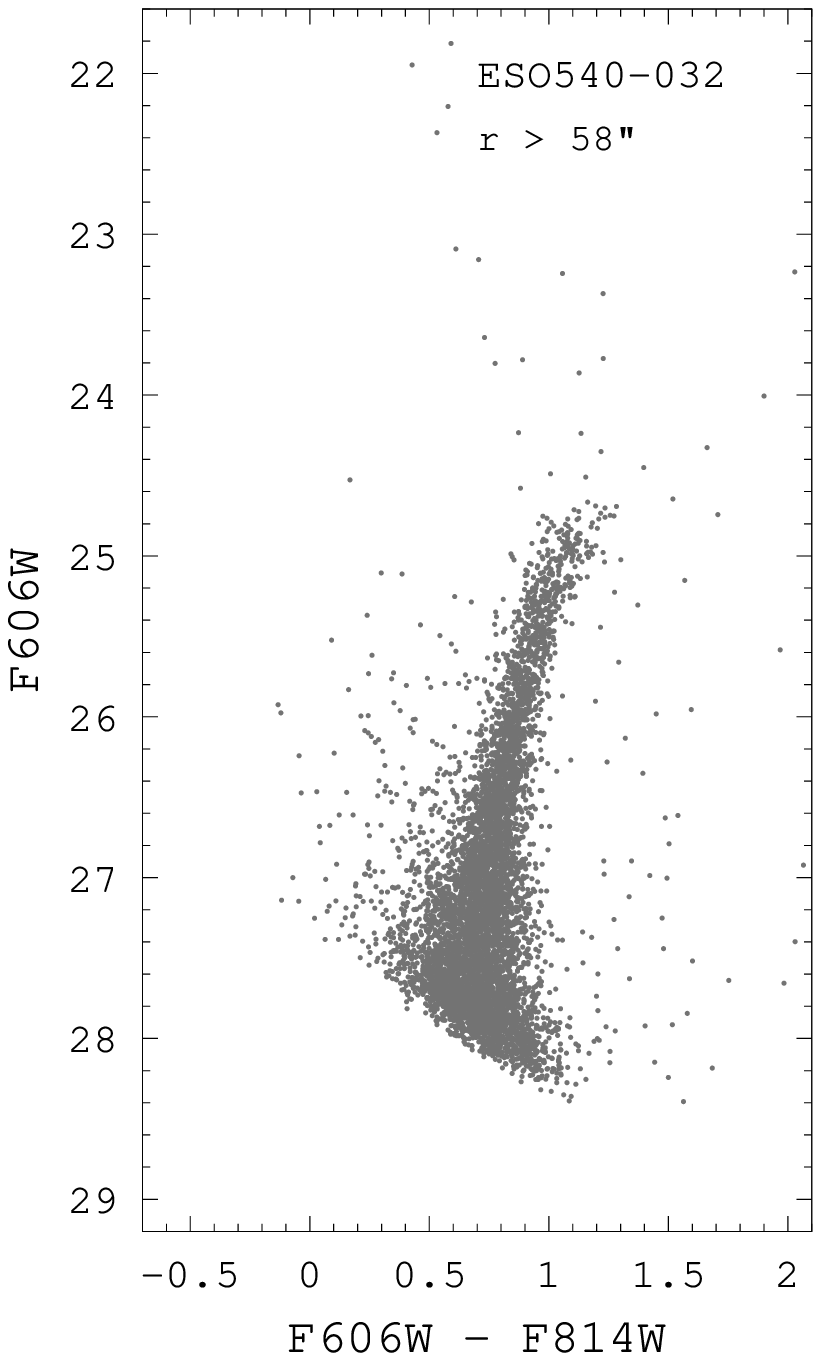}
   \caption{CMDs of stars within three selected elliptical radius annuli. The limits of the annuli are indicated within each panel. In the first annulus (upper panels) we overplot Padova isochrones, shown in black, of a range in ages, and with Z metallicity of 0.001, as detailed in \S~3.3.}
      \label{sl_figure05}%
\end{figure*}
In Fig.~\ref{sl_figure05} we show the CMDs of our studied transition-type dwarfs with their stars selected within three annuli. While constructing the annuli, we account for the elliptical shape of each dwarf using the elliptical radius $r$, defined as: $r = [x^{2}+y^{2}/(1 - \epsilon)^{2}]^{1/2}$, where $x$ and $y$ are the distance along the major and minor axis, and $\epsilon$ is the ellipticity (Kirby et al.~\cite{sl_kirby08}; Lee et al.~\cite{sl_lee11} for ESO410-G005). The first annulus is within one effective radius r$_{\rm eff}$ (upper panels), the second one from 1~r$_{\rm eff}$ to 2~r$_{\rm eff}$ (middle panels), and the last one outside 2~r$_{\rm eff}$ (lower panels). The r$_{\rm eff}$ values are taken from Kirby et al.~(\cite{sl_kirby08}), apart from ESO410-G005 for which we adopt the value from Sharina et al.~(\cite{sl_sharina08}). Their centers are computed using the equation: $\rm{x_{C}}=\Sigma(\rm{mag}_{i}\times \rm{x}_{i})/\Sigma \rm{mag}_{i}$, where $\rm{mag}_{i}$ corresponds to the F814W-band magnitude of each star, and $\rm{x}_{i}$ to their x (or y) coordinates, in pixels. 
\begin{table}
     \begin{minipage}[t]{\columnwidth}
      \caption{Adopted galaxy centers.}
      \label{table4}
      \centering
      \renewcommand{\footnoterule}{}
      \begin{tabular}{l c c}
\hline\hline
Galaxy        &x$_{C}$               &y$_{C}$             \\ 
              &(pixels)              &(pixels)           \\
(1)           &(2)                   &(3)                \\
\hline
ESO540-G030   &2059.8$\pm$24.3      &2391.2$\pm$28.1     \\
ESO540-G032   &2204.3$\pm$27.5      &2365.3$\pm$29.7     \\
ESO294-G010   &2173.7$\pm$40.6      &2409.3$\pm$52.2     \\
ESO410-G005   &2248.3$\pm$37.9      &2421.7$\pm$54.1     \\
Scl-dE1       &2014.7$\pm$37.0      &2241.5$\pm$32.1     \\

\hline
\end{tabular} 
\end{minipage}
\end{table}
%
  Table~\ref{table4} lists in columns (2) and (3) the x and y coordinates of the adopted centers, respectively, along with their corresponding standard deviation in which the photometric errors are taken into account. We use stars with magnitudes brighter than the magnitude corresponding to the 50\% completeness limit in order to compute the centers. From Fig.~\ref{sl_figure05}, it is evident that in most dwarfs the bulk of the young MS stars is concentrated in the inner annuli. In the case of ESO540-G032, our result is consistent with the earlier analysis of Jerjen \& Rejkuba (\cite{sl_jerjen01}) based on ground-based observations, and of Da Costa et al.~(\cite{sl_dacosta07}) based on a preliminary analysis of the same data, who also find the young stars to be distributed towards the central part of the dwarf. 

   In the upper panels of Fig.~\ref{sl_figure05}, we overplot Padova isochrones for ages less than 1~Gyr in the MS locus and with a Z metallicity of 0.001 for all isochrones. The corresponding ages of the young MS isochrones are 30~Myr and 200~Myr for ESO540-G030, 70~Myr and 200~Myr for ESO540-G032, 50~Myr and 300~Myr for ESO294-G010, and 100~Myr and 250~Myr for ESO410-G005. All transition-type dwarfs have MS stars as young as $\sim$100~Myr, and even younger in some cases. 

 \begin{figure*}
   \centering
     \includegraphics[scale=0.5,clip]{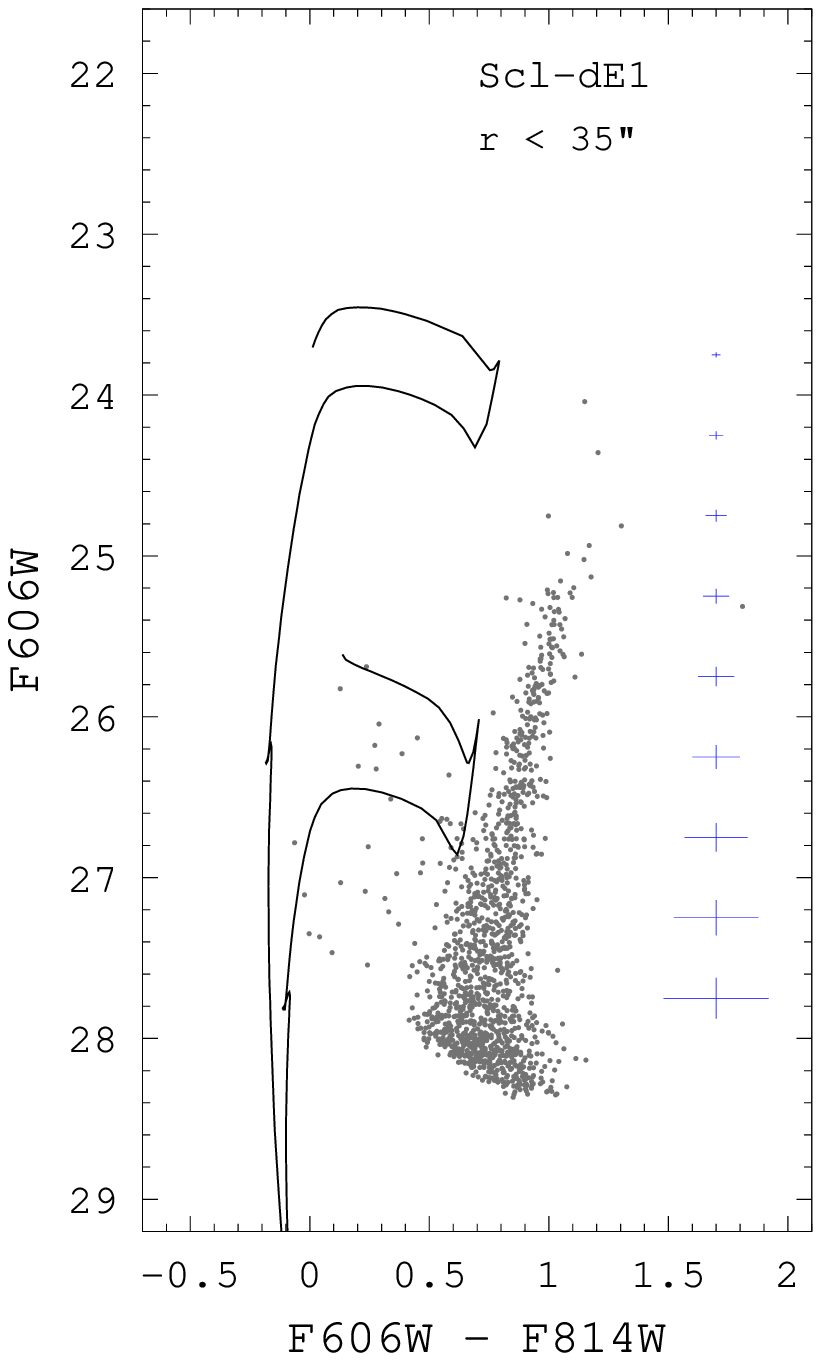}
     \includegraphics[scale=0.5,clip]{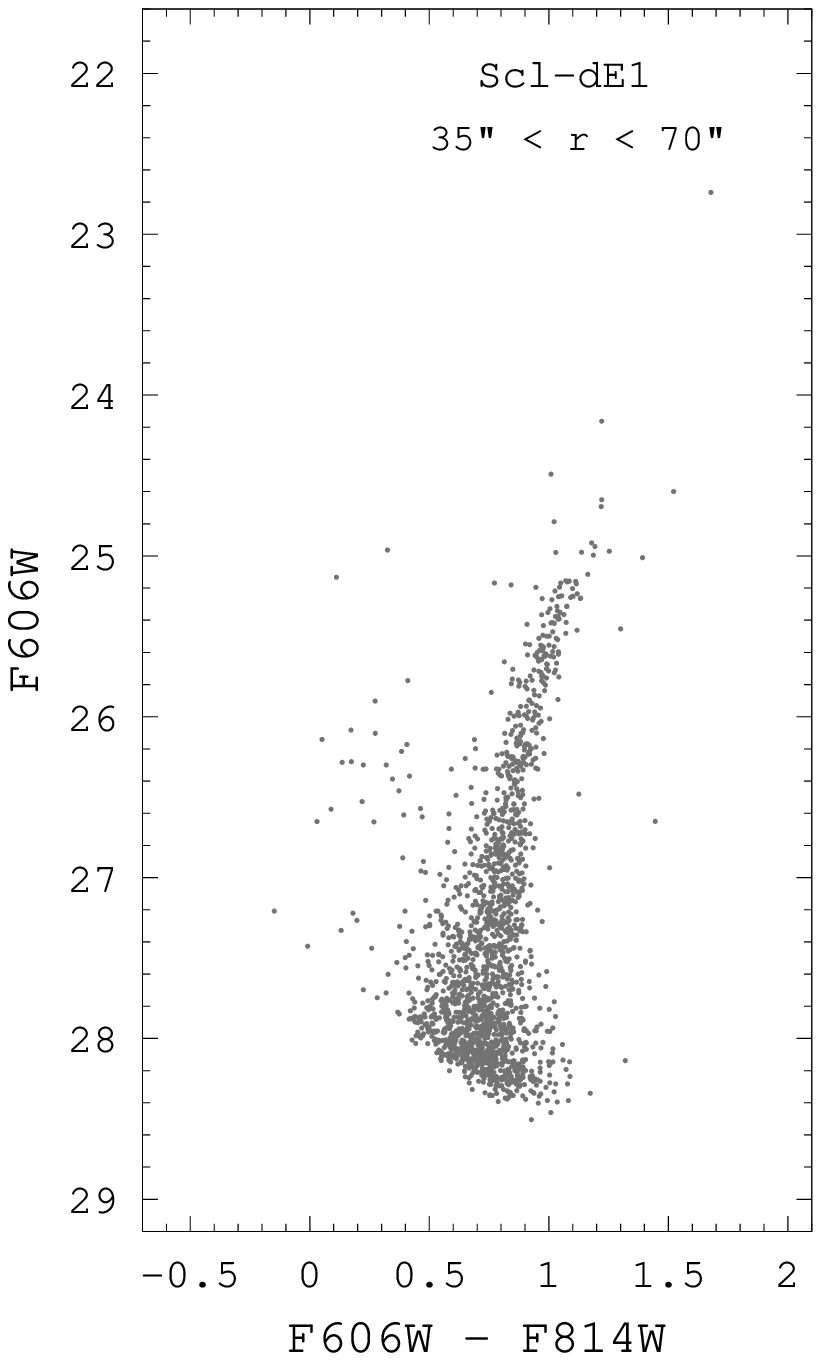}
     \includegraphics[scale=0.5,clip]{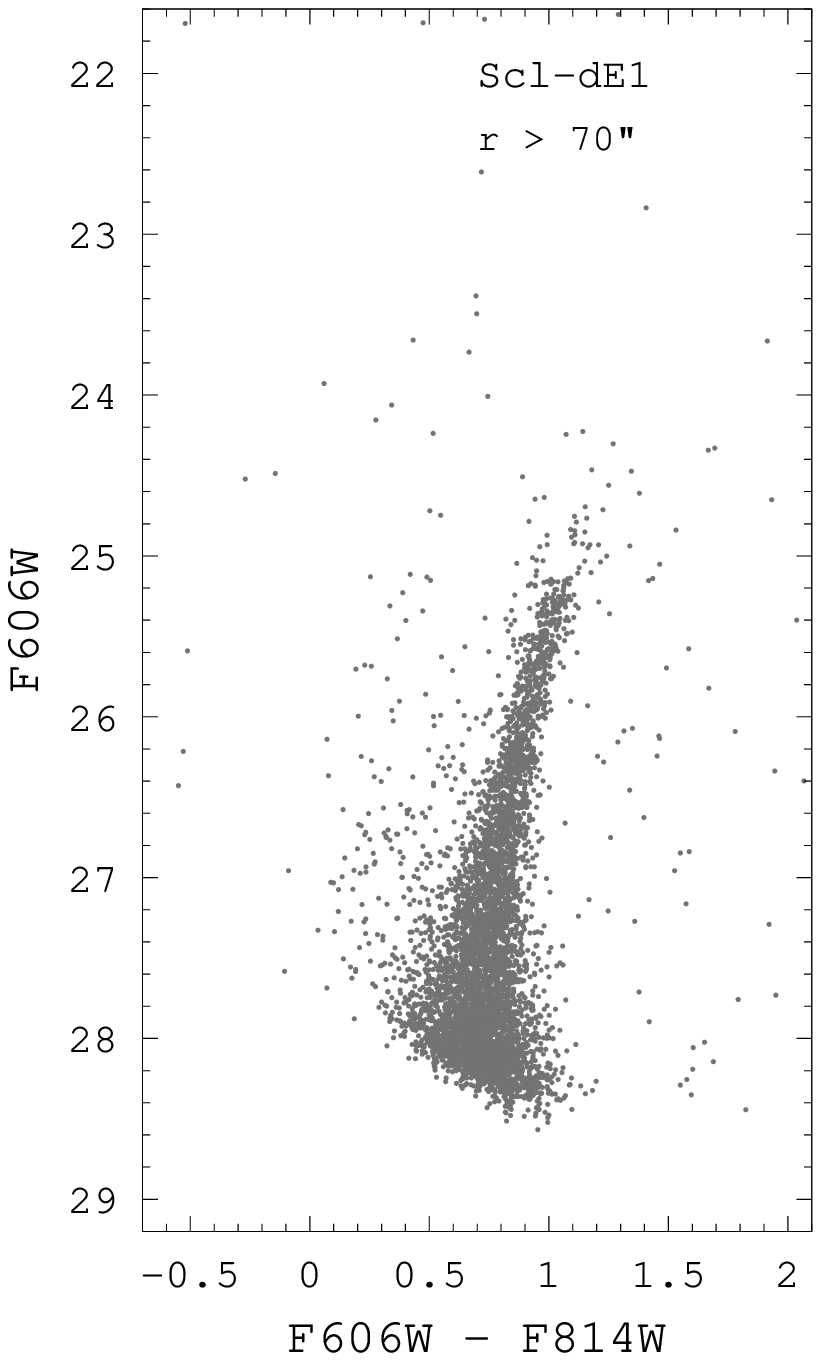}
   \caption{The same as in Fig.~\ref{sl_figure05} but here for Scl-dE1.}
      \label{sl_figure06}
\end{figure*}
%
   Fig.~\ref{sl_figure06} shows the same as Fig.~\ref{sl_figure05} but for Scl-dE1. Scl-dE1 does not show indications of young MS stars, at least not for such stars with ages younger than $\sim$70~Myr as indicated by the overplotted 70~Myr isochrone, shown in the left panel. The stars in the CMD located between the RGB and MS, as well as the stars brighter and bluer than the TRGB, are consistent with evolved post-MS stars, as indicated with the isochrone of 70~Myr and 250~Myr, and these stars seem to be present at all radii. The contamination from foreground stars is negligible in this color-magnitude space (1 foreground star out of $\sim$350~stars), while $\sim$5~stars (or $\sim$1.4\%) are due to crowding in this part of the CMD. Such post-MS stars seem to be present in all transition-type dwarfs (see also Da Costa et al.~\cite{sl_dacosta07} for ESO540-G032). 

  \section{Stellar density maps and population gradients}
  
   \subsection{Transition-type dwarfs}
    
  We examine the spatial distribution as well as the cumulative distribution functions of stars that belong to different evolutionary phases in the case of the four transition-type dwarfs. While constructing the cumulative distribution functions, we use the elliptical radius $r$, defined in $\S$3.~3, to account for the elliptical shape of each dwarf. We adopt for each dwarf the centers listed in Table~\ref{table4}. 

  We select stars in the blue HB, RGB, RC, luminous AGB, VC, and MS phases, wherever resolved. Stars in the HB, RC and VC phase are resolved only in the cases of ESO294-G010 and ESO410-G005. The HB stars are selected within the color limits from -0.1~mag to 0.3~mag, and F814W-band magnitude limits of 27.6~mag to 26.2~mag. The RC stars are selected within $\pm$0.5$\sigma$ from the mean color and magnitude given in $\S$3.1, corresponding to ages within the range from $\sim$1.3~Gyr to less than 8~Gyr. The VC stars are selected within the color limits of 0.53~mag to 0.67~mag, and F814W-band magnitude limits of 24.4~mag to 25.4~mag, and correspond to ages from $\sim$400~Myr to less than $\sim$900~Myr. 

  The MS stars were selected such that they are fainter than the youngest isochrone we use in Sec.~3.3 in order to place constraints on their ages, and brighter than the magnitude of 26.4~mag in order to avoid contamination of the young MS stars by HB stars in the case of ESO294-G010 and ESO410-G005. The color limits which we use to select the MS stars range from $-$0.35~mag to $-$0.05~mag for ESO540-G030 and ESO540-G032, and from $-$0.25~mag to 0.1~mag for ESO294-G010 and ESO410-G005. The luminous AGB stars were selected to be brighter by 0.15~mag than the TRGB (Armandroff et al.~\cite{sl_armandroff93}) and to lie within 1~mag above ($I_{TRGB}-0.15$)~mag, and within the color range of $\rm{c}<(V-I)_0<\rm{c}+$~2.5~(mag), where $c$ is equal to the color of the TRGB of the most metal-poor Dartmouth isochrone, reddened for each dwarf using the extinction values listed in columns (7) and (8) of Table~\ref{table2}. We note here that we use Dartmouth isochrones (Dotter et al.~\cite{sl_dotter07,sl_dotter08}) to define the luminous AGB stars in order to remain consistent with our previous investigations (Lianou et al.~\cite{sl_lianou10,sl_lianou11}).

  The RGB stars are selected such that the brighter magnitude limit is equal to the TRGB magnitude, while the fainter magnitude limit is equal to 1~mag fainter than the TRGB magnitude. The blue and red color limits are chosen to encompass the RGB stars, following the RGB slope. We use the RGB populations as generic populations to which we compare all other population distributions. Considering that the RGB potentially contains a mixture of stars with ages larger than $\sim$2~Gyr, we expect that the older ones of these RGB stars will indeed trace the most extended spatial distribution of each galaxy. For the two dwarfs with deep enough data to resolve the HB, the blue HB stars provide a better tracer of the most extended spatial distribution of their genuine old population, as other studies in dwarf galaxies have revealed (e.g., Harbeck et al.~\cite{sl_harbeck01}; Lee et al.~\cite{sl_lee03}; Bellazzini, Gennari \& Ferraro \cite{sl_bellazzini05}). We note that the blue HB populations of ESO294-G010 and ESO410-G005 are mixed with their young MS stars. Given that these young MS stars are confined to the central parts of the dwarfs (Sec.~3.3; see also the following paragraph), the potentially extended spatial distribution of the blue HB stars is not affected by the more central location of the contaminating young MS stars.

 \begin{figure}
    \centering
       \includegraphics[width=3.3cm,clip]{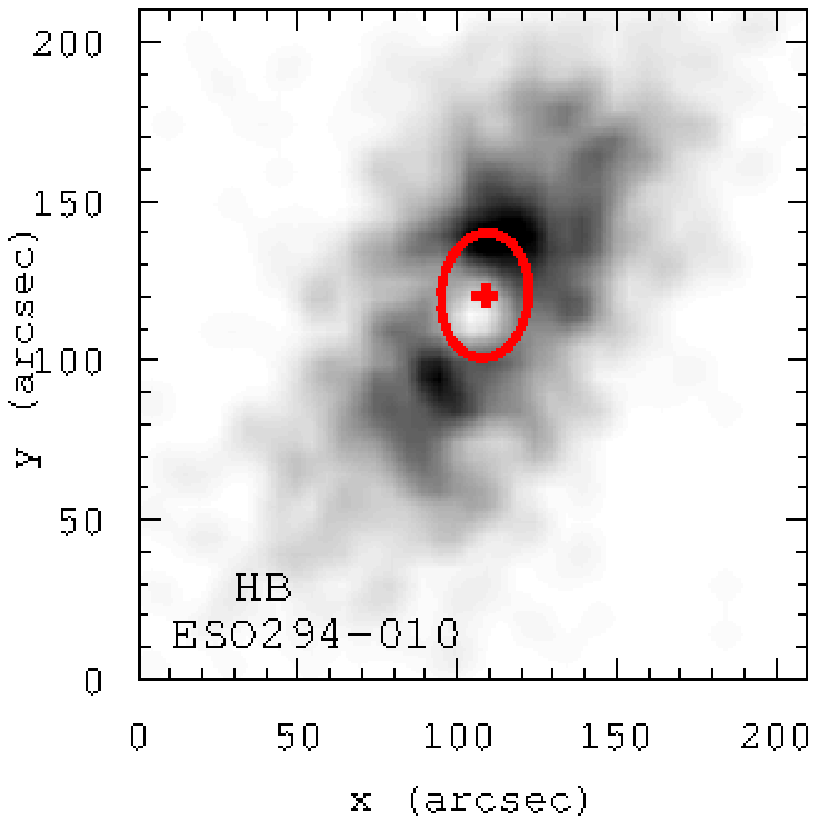}
       \includegraphics[width=3.3cm,clip]{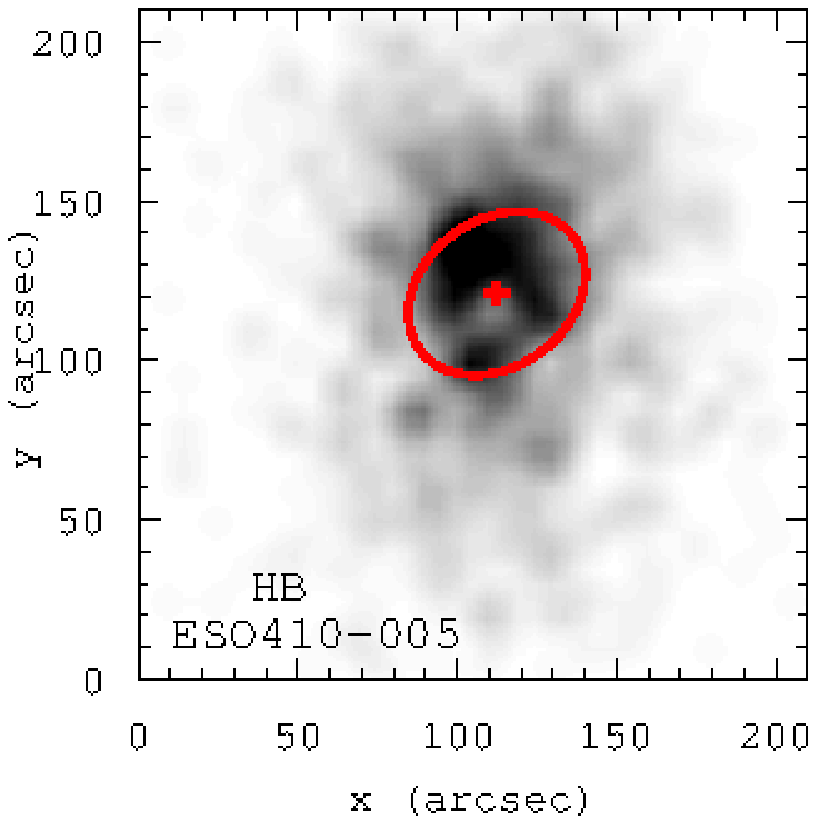}
       \includegraphics[width=3.3cm,clip]{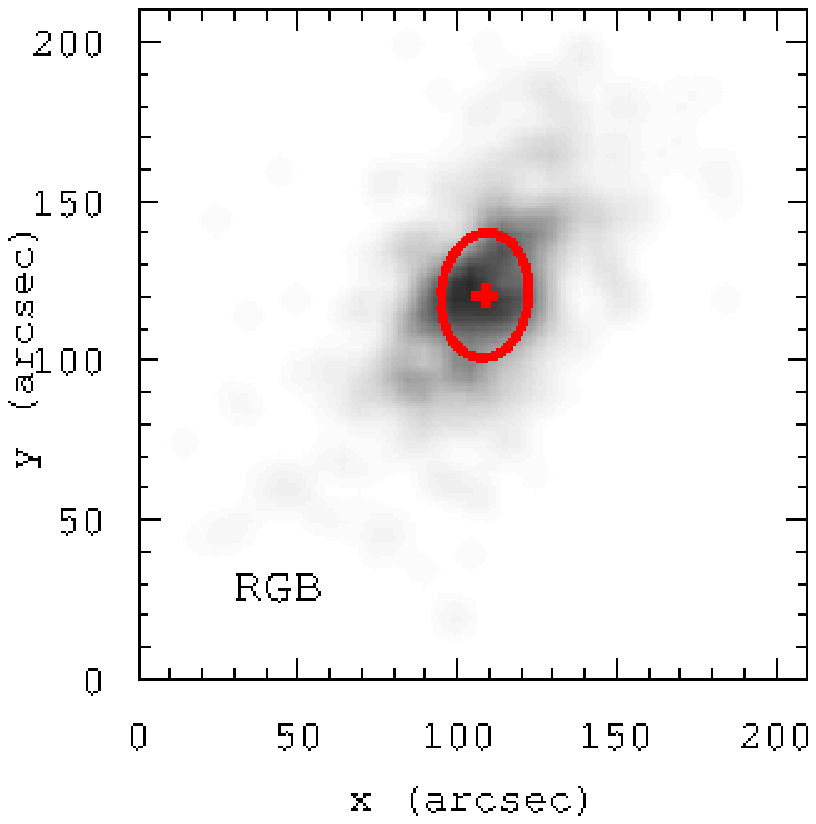}
       \includegraphics[width=3.3cm,clip]{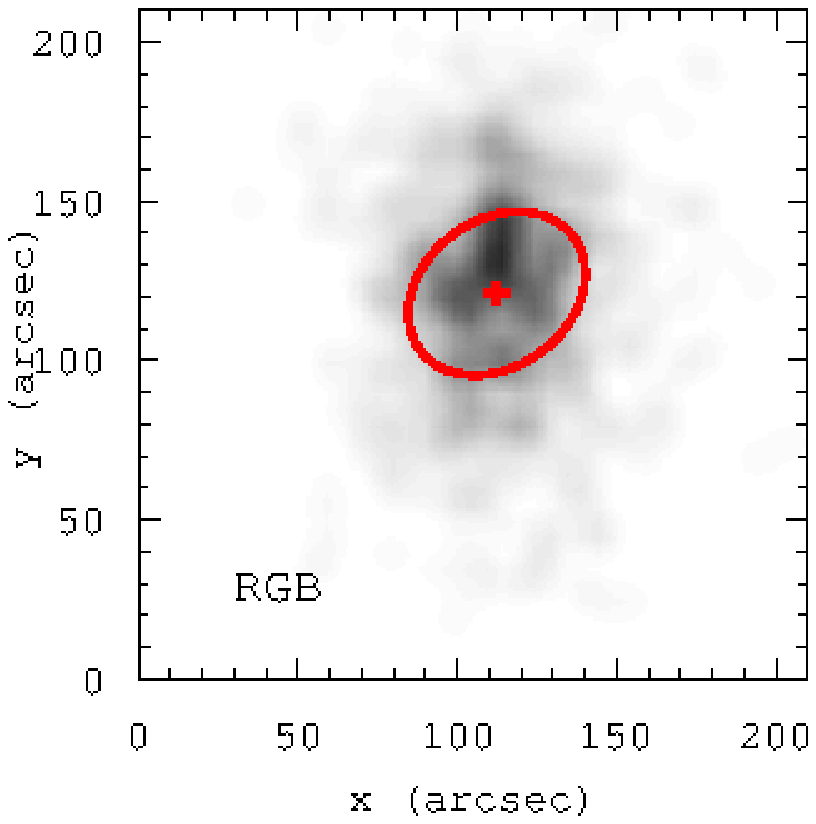}
       \includegraphics[width=3.3cm,clip]{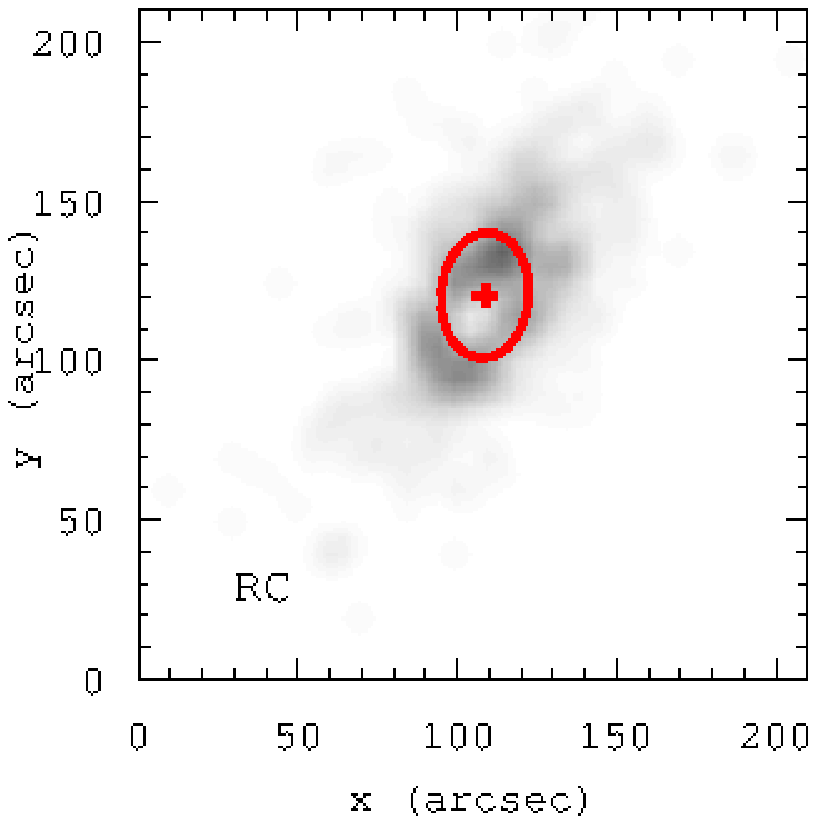}
       \includegraphics[width=3.3cm,clip]{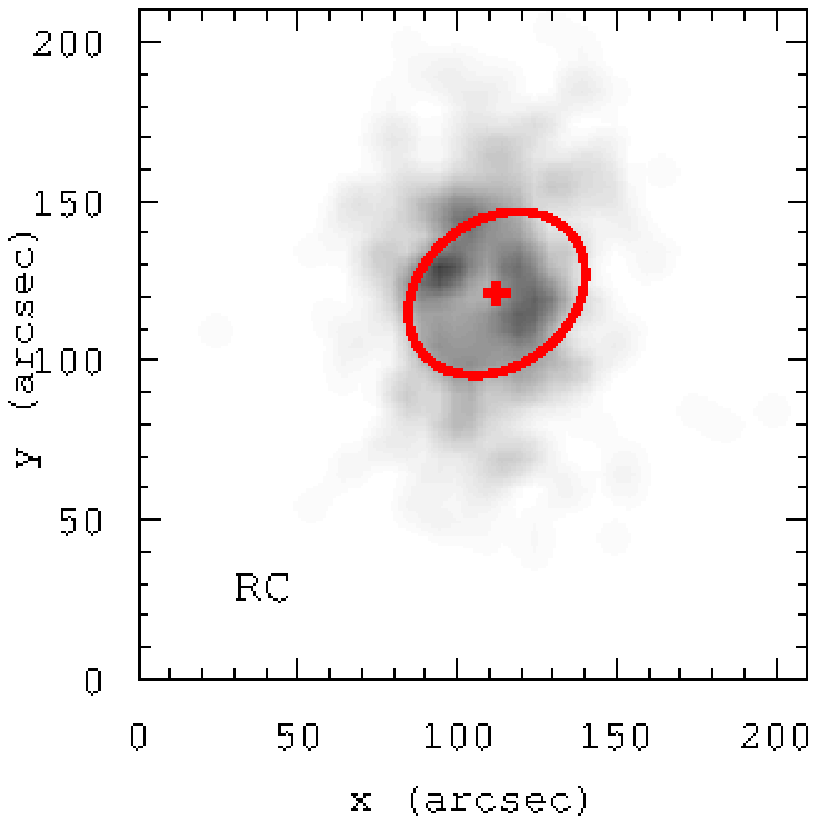}
       \includegraphics[width=3.3cm,clip]{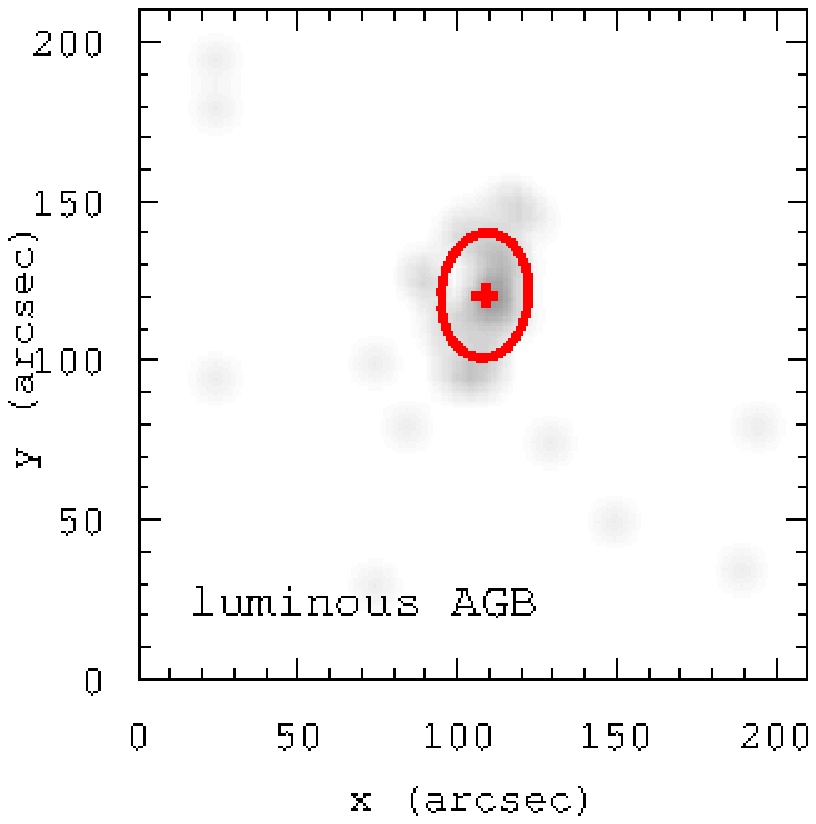}
       \includegraphics[width=3.3cm,clip]{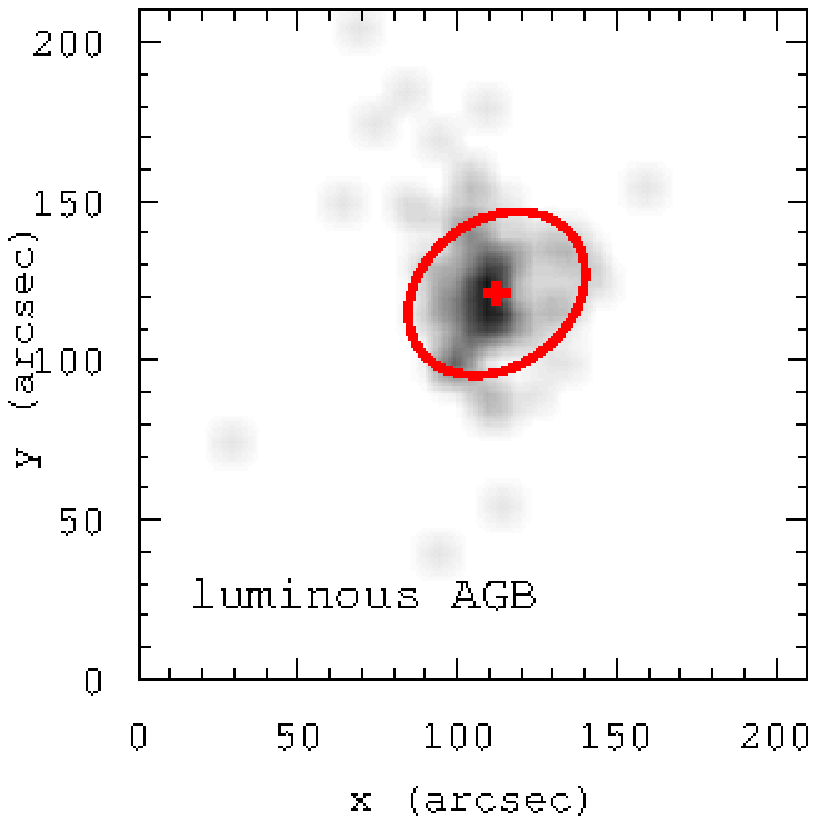}
       \includegraphics[width=3.3cm,clip]{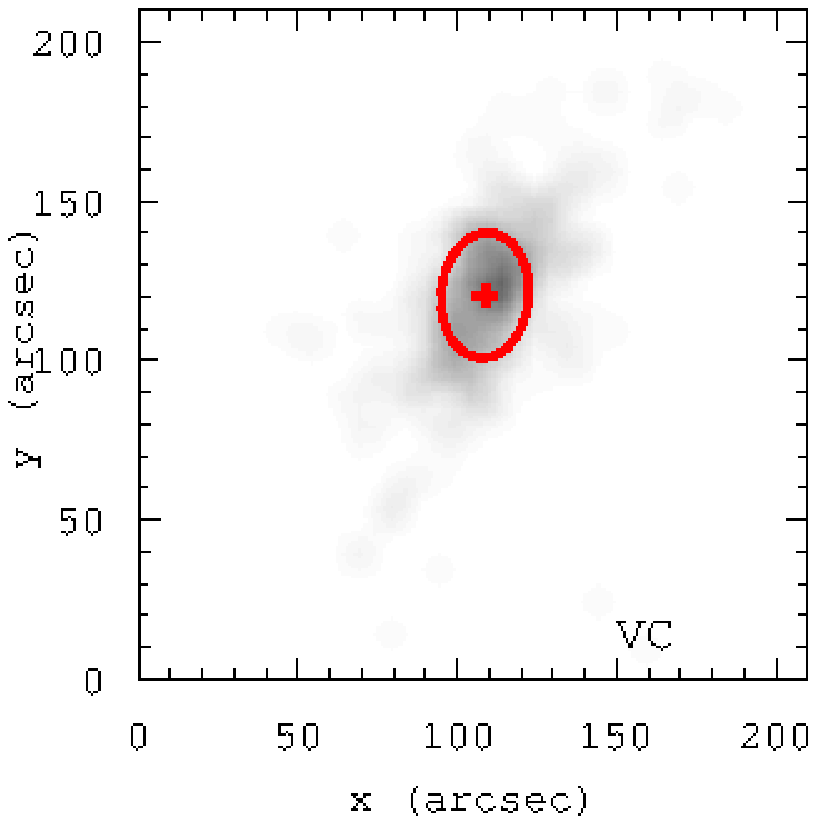}
       \includegraphics[width=3.3cm,clip]{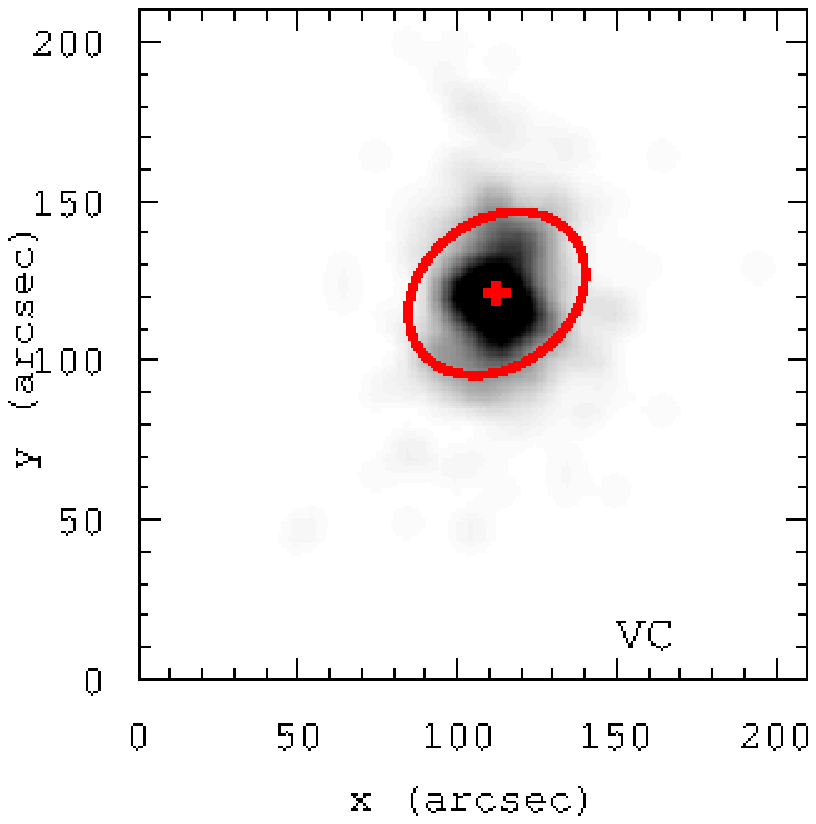}
       \includegraphics[width=3.3cm,clip]{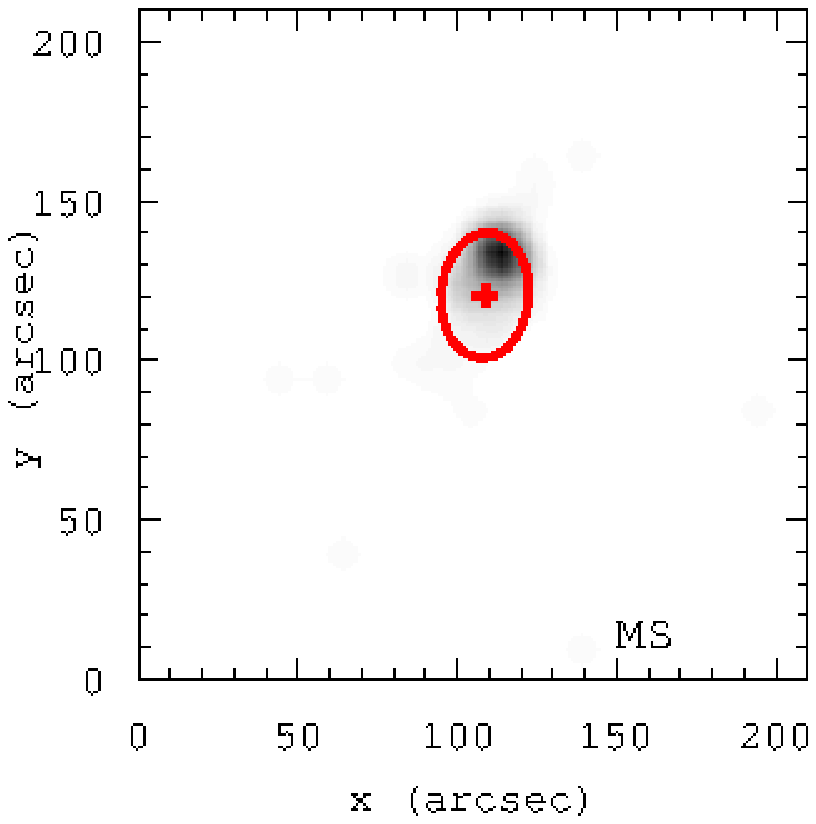}
       \includegraphics[width=3.3cm,clip]{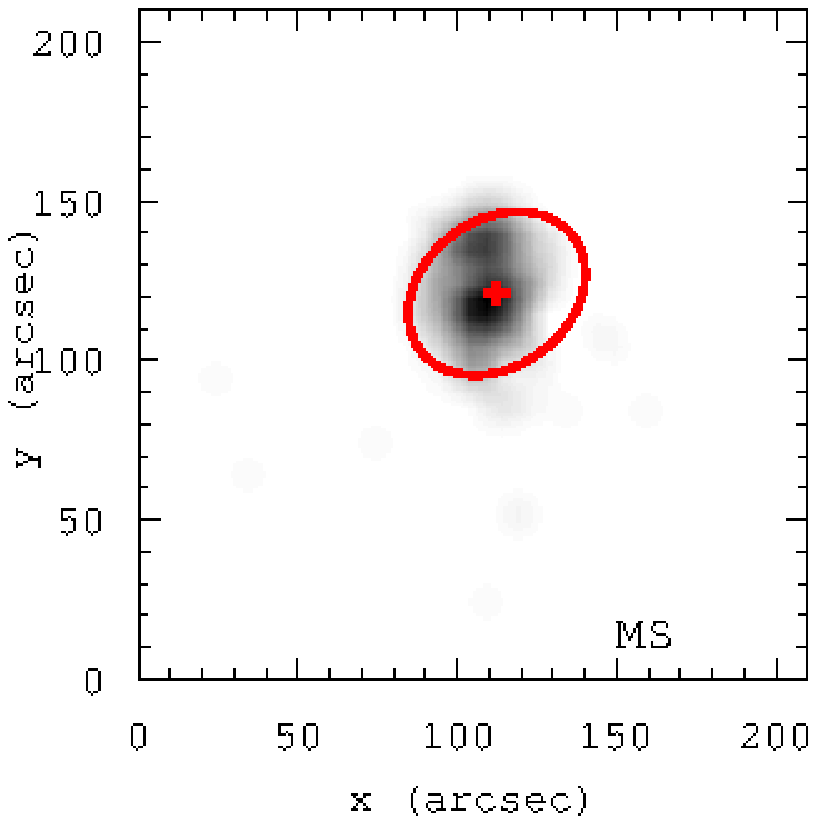}
       \caption{From top to bottom, Gaussian-smoothed density maps of blue HB, RGB, RC, luminous AGB, VC, and MS stars, for ESO294-G010 (left panels) and ESO410-G005 (right panels). The density maps are color coded such that the black color corresponds to a density of 0.30 stars per sq.~arcsec for ESO294-G010, and 0.27 stars per sq.~arcsec for ESO410-G005, except for the luminous AGB maps with values of 0.1 stars per sq.~arcsec and 0.07 stars per sq.~arcsec, respectively. In both galaxies, a white color corresponds to zero density. The ellipses indicate a major axis equal to the r$_{eff}$, while the plus sign indicates their center. The plus sign is wide enough to include the uncertainty, with the latter being less than 3~arcsec.}
    \label{sl_figure07}
\end{figure}
%
 \begin{figure}
    \centering
       \includegraphics[width=3.5cm,clip]{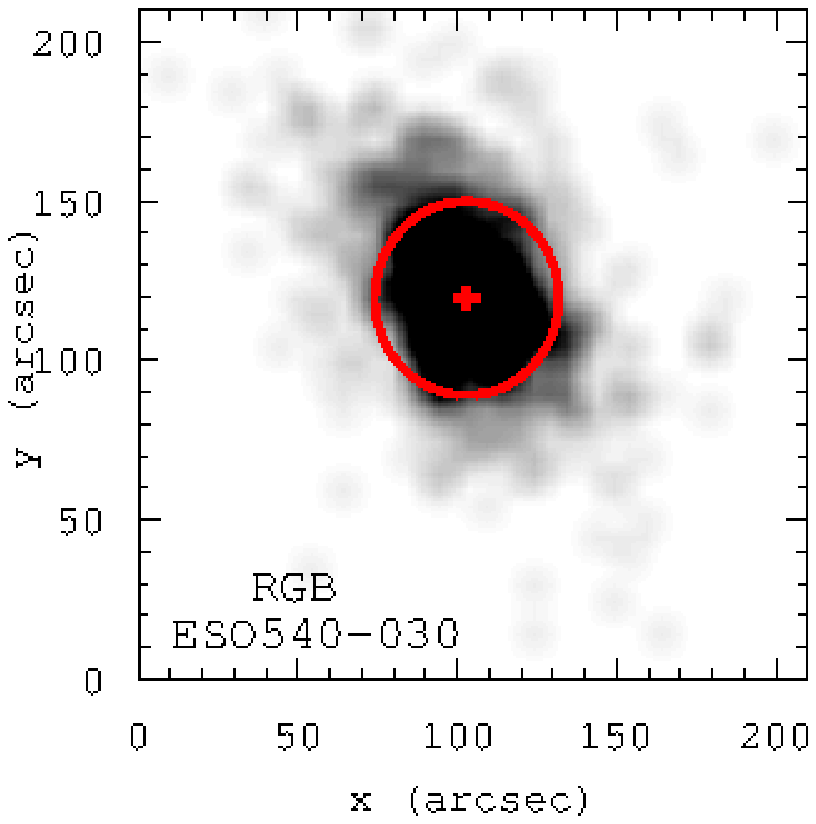}
       \includegraphics[width=3.5cm,clip]{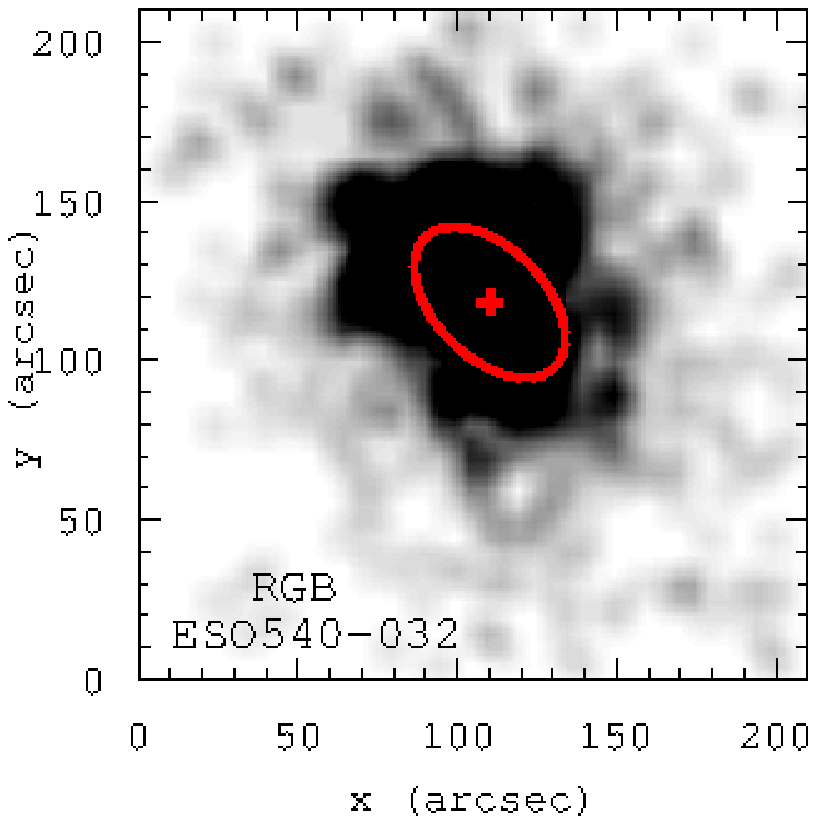}
       \includegraphics[width=3.5cm,clip]{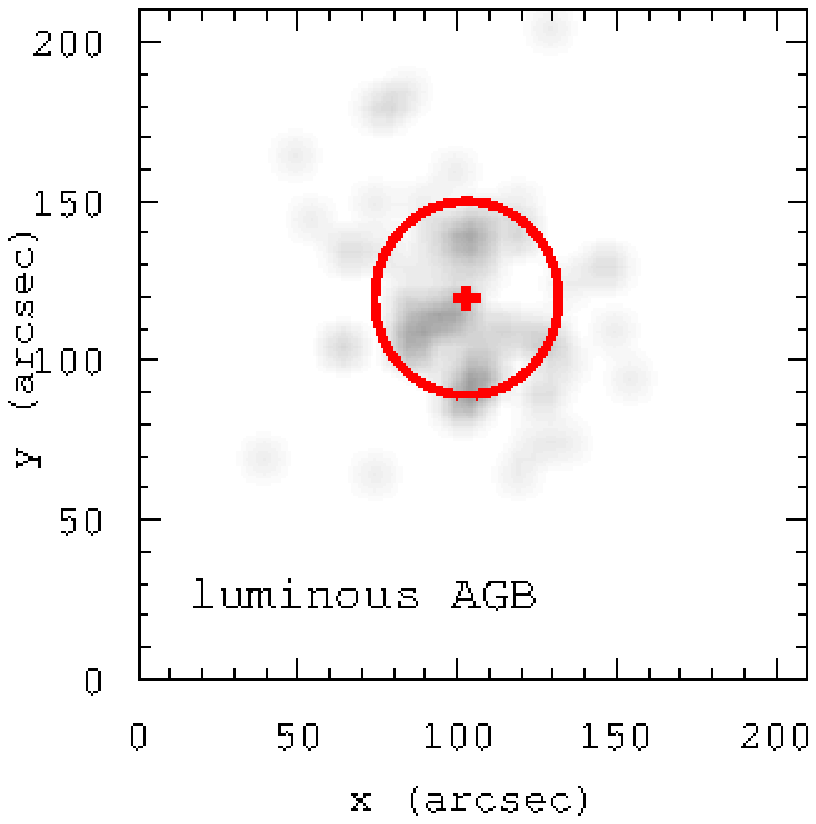}
       \includegraphics[width=3.5cm,clip]{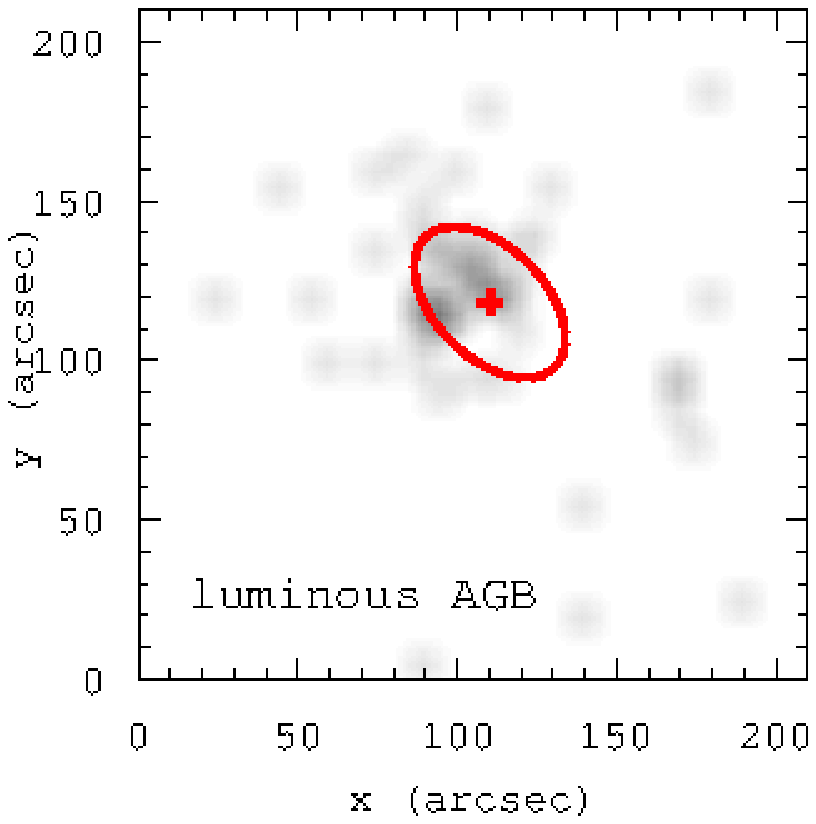}
       \includegraphics[width=3.5cm,clip]{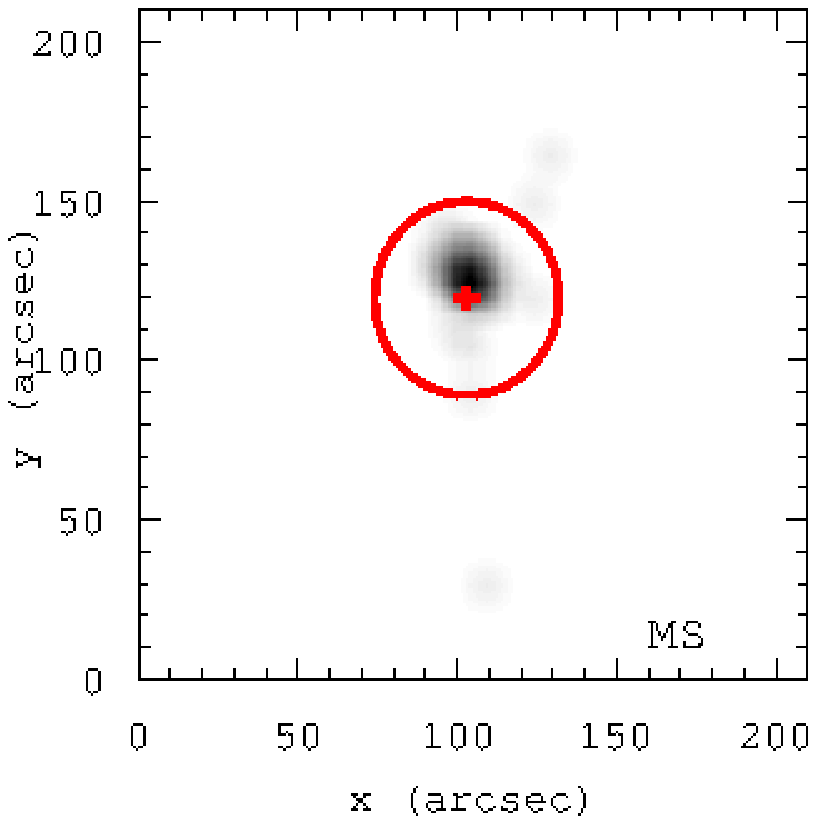}
       \includegraphics[width=3.5cm,clip]{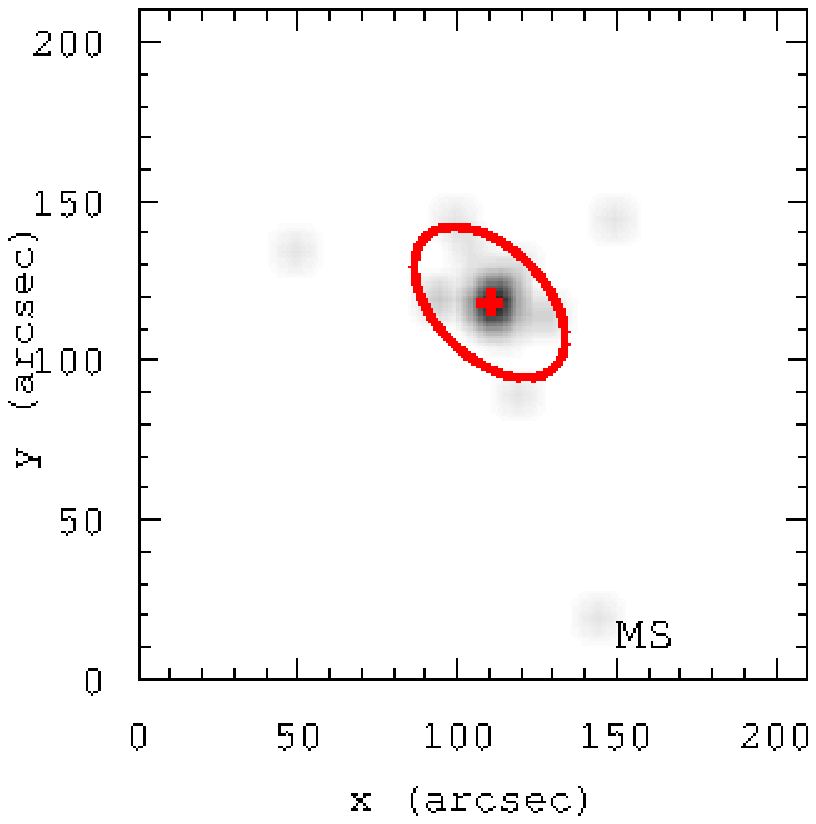}
       \caption{Similar as in Fig.~\ref{sl_figure07} but for the two remaining transition-type dwarfs. The density maps are color coded such that black color corresponds to a density of 0.10 stars per sq. arcsec for ESO540-G030, and 0.07 stars per sq. arcsec for ESO540-G032, with the density decreasing from black to white colors. The ellipses indicate a major axis equal to the r$_{eff}$, while the plus sign indicates their center.}
       \label{sl_figure08}
\end{figure}
%
  In Fig.~\ref{sl_figure07} and Fig.~\ref{sl_figure08} we show the spatial distribution of stars for the selected evolutionary phases. All density maps are Gaussian-smoothed with a 3$\times$3 kernel size and a 5$\arcsec \times$5$\arcsec$ pixelsize. The spatial distribution analysis of the four transition-type dwarfs indicates that in all cases the MS stars are confined to the central parts of the galaxies, consistent with previous analyses of the MS stars within elliptical radius or circular annuli (this study; Karachentsev et al.~\cite{sl_karachentsev00}; Jerjen \& Rejkuba \cite{sl_jerjen01}; Da Costa et al.~\cite{sl_dacosta07}). In order to facilitate the comparison among the various spatial distributions of stars belonging to different stellar evolutionary phases, we overplot the center of each dwarf marked with a plus sign in Fig.~\ref{sl_figure07} and Fig.~\ref{sl_figure08}, as well as an ellipse with the major axis equal to the r$_{eff}$, and with position angle and ellipticity taken from Kirby et al.~(\cite{sl_kirby08}), apart from ESO410-G005, which are taken from Lee et al.~(\cite{sl_lee11}). 

   Looking at the spatial distribution of the VC stars, in the case of ESO294-G010, this is more extended than the MS stellar distribution, while for ESO410-G005 the VC and MS stars are similarly distributed. It is interesting to note that the highest density of the MS stars seems to be distributed off from each dwarf's center, stronger in the case of ESO294-G010 and less stronger for ESO410-G005 and ESO540-G030. This is suggestive of local sites of recent star formation as in LG dIrrs (e.g., Dohm-Palmer et al.~\cite{sl_dohm-palmer97}; Cole et al.~\cite{sl_cole99}) and transition-type dwarfs (e.g., Gallagher et al.~\cite{sl_gallagher98}) rather than pure central location. Note that such localised star formation is also seen in the Fornax dSph (Stetson et al.~\cite{sl_stetson98}), a massive dSph galaxy that also shows signs of a tidal interaction event (Coleman et al.~\cite{sl_coleman05}; Amorisco \& Evans \cite{sl_amorisco12}).

   The spatial distribution of the luminous AGB stars is coincident with the distribution of the MS stars in both ESO294-G010 and ESO410-G005. In the case of ESO540-G030 and ESO540-G032, the luminous AGB stars are more extended than the MS stars, and in both cases the luminous AGB stars are extended following the distribution of the RGB stars. In all cases, the number of the luminous AGB stars is low, with values of 42 stars in ESO294-G010, 82 stars in ESO410-G005, 73 stars in ESO540-G030, and 51 stars in ESO540-G032. We compute the fraction of the luminous AGB stars, f$_{AGB}$, defined as the number of the luminous AGB stars over the number of the RGB stars within one magnitude below the TRGB. The results give 9\% for ESO294-G010, 14\% for ESO410-G005, 13\% for ESO540-G030, and 6\% for ESO540-G032. We note that luminous AGB stars are best studied through infrared observations, since these are dust-enshrouded objects with red colors that lead to an underestimate of their true numbers based on optical observations (e.g., Boyer et al.~\cite{sl_boyer09}). Nevertheless, luminous AGB stars estimated in the optical regime may provide useful information in conjunction with stellar evolution models (e.g., Girardi et al.~\cite{sl_girardi10}). Studies of luminous AGB stars in the infrared regime exist for LG galaxies (e.g., Battinelli \& Demers \cite{sl_battinelli04}; Groenewegen \cite{sl_groenewegen07}; Gullieuszik et al.~\cite{sl_gullieuszik07}; Whitelock et al.~\cite{sl_whitelock09}) as well as dwarfs outside the LG (e.g., Da Costa \cite{sl_dacosta04}; Rejkuba et al.~\cite{sl_rejkuba06}, Crnojevic et al.~\cite{sl_crnojevic11}; Dalcanton et al.~\cite{sl_dalcanton12}). 

 \begin{figure}
  \centering
       \includegraphics[width=4cm,clip]{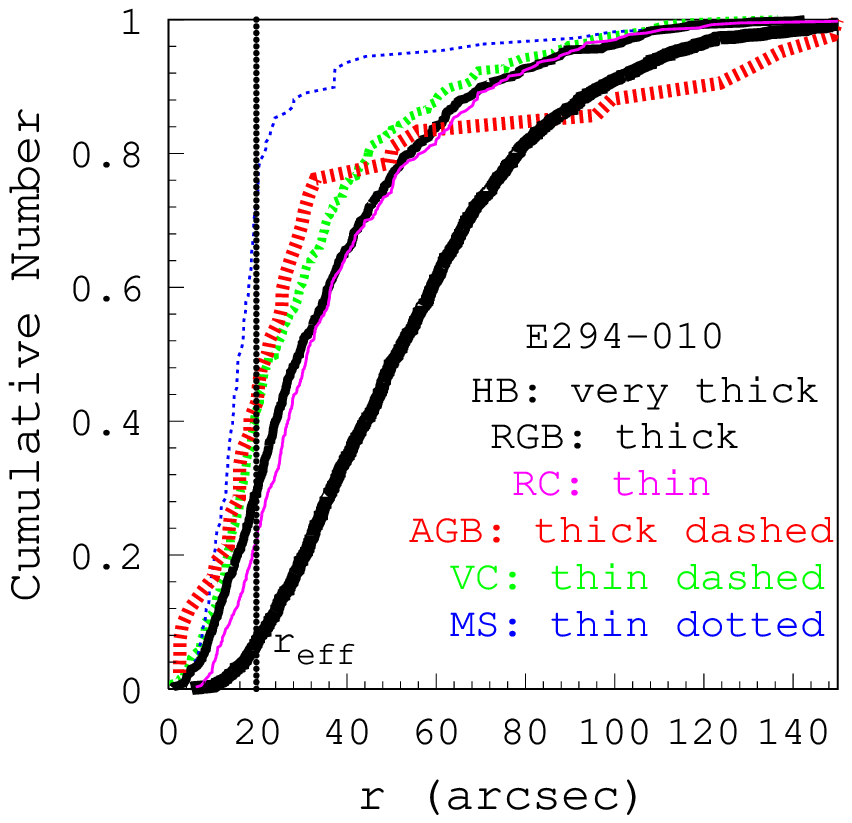}
       \includegraphics[width=4cm,clip]{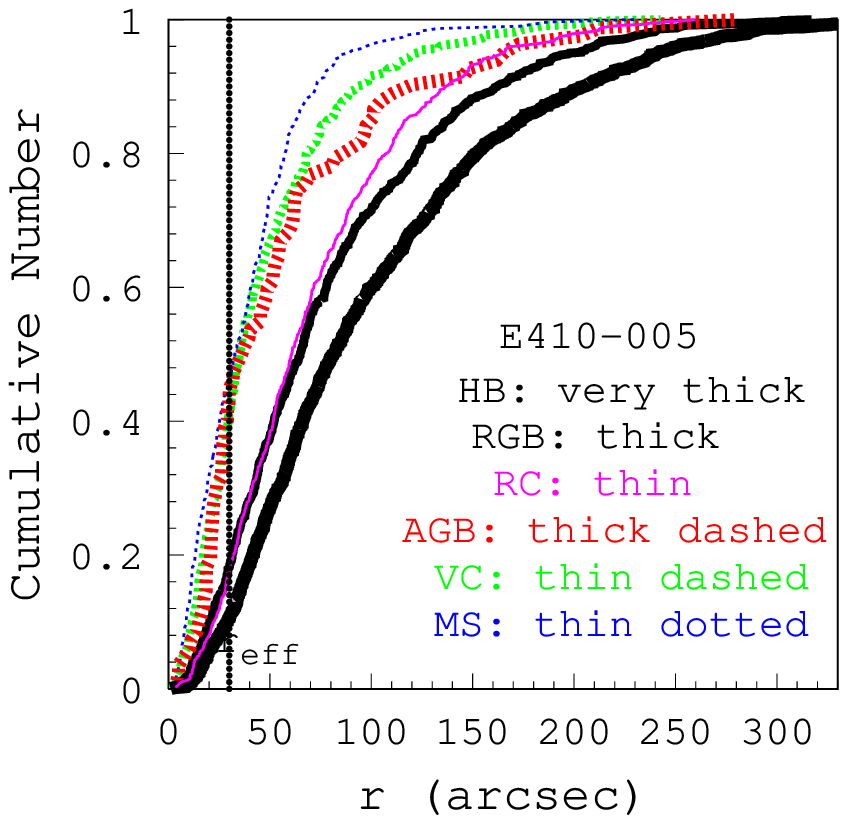}
       \includegraphics[width=4cm,clip]{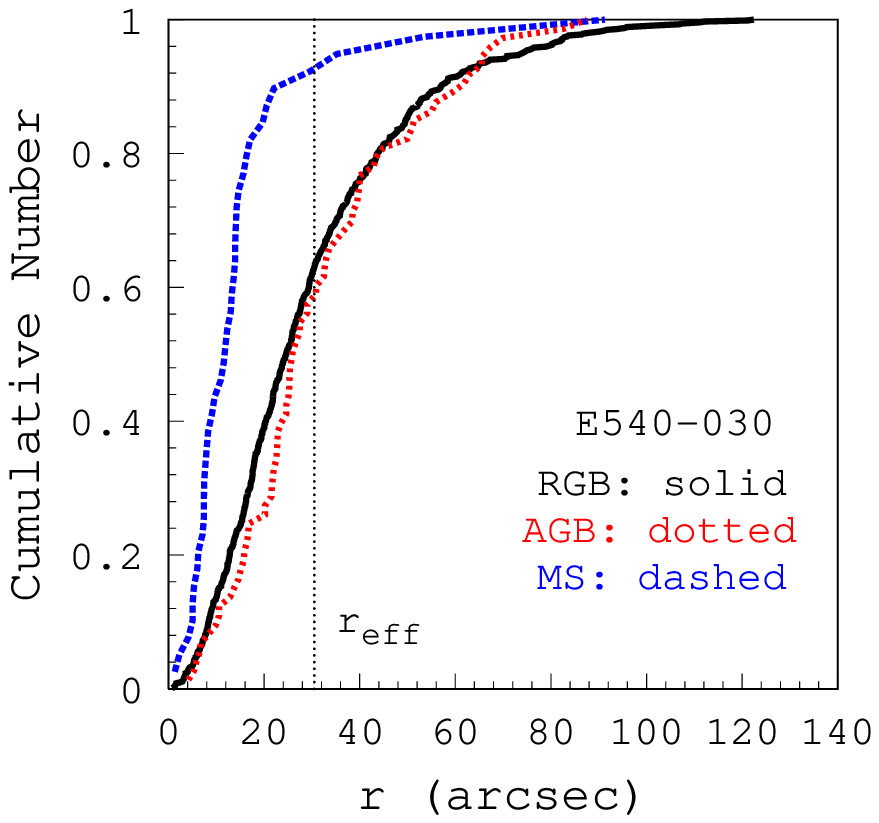}
       \includegraphics[width=4cm,clip]{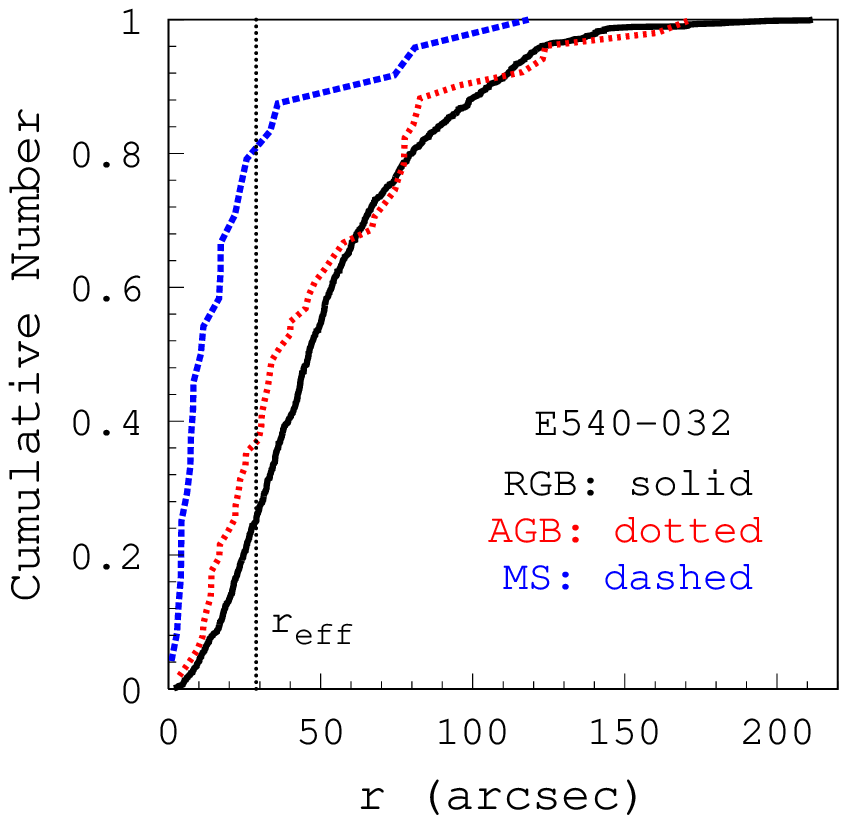}
       \includegraphics[width=4cm,clip]{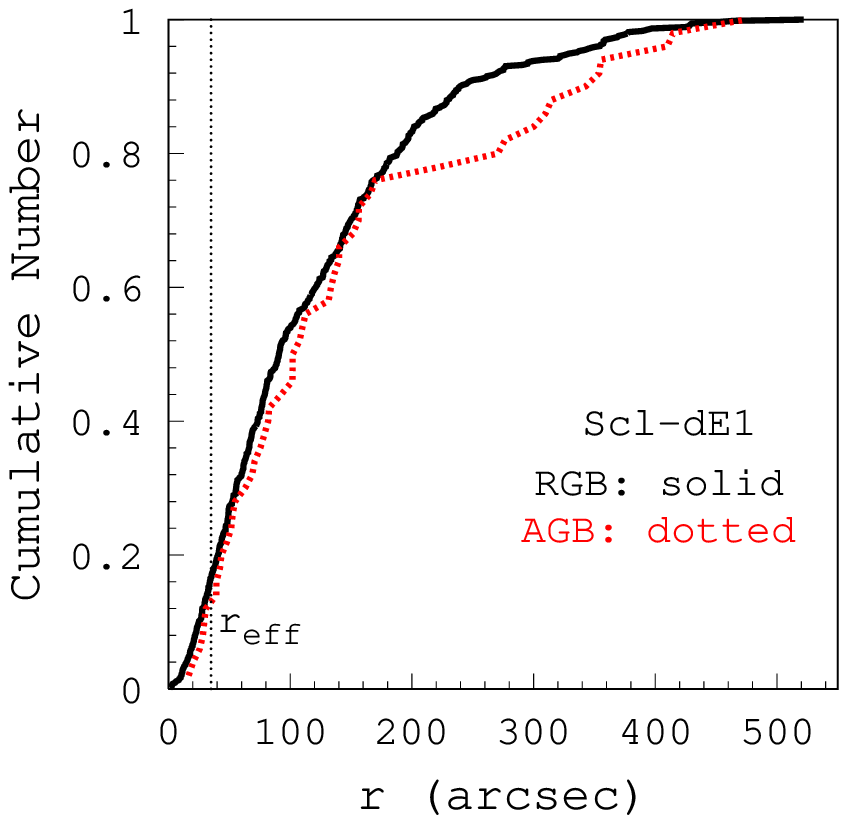}
       \caption{Cumulative distribution functions of stars belonging to different stellar evolutionary phases for our studied dwarfs. The r$_{eff}$ radius is shown with the vertical dotted line. } 
  \label{sl_figure09}
  \end{figure}
%
   The RC stars are spatially more extended than the VC stars, or the luminous AGB stars. The cumulative distribution functions of the several stellar populations in each dwarf are shown in Fig.~\ref{sl_figure09}. In the case of ESO294-G010 and ESO410-G005 the spatial distribution of the HB stars, tracing the most ancient star formation, appears more extended than that of the RGB and RC stars. We compare the cumulative distributions of each of the stellar populations with the cumulative distribution of the RGB population. The two-sided Kolmogorov-Smirnov (K-S) tests show results consistent with spatially separated populations in all cases at the 95\% confidence level and higher, or at 90\% and 87\% confidence level for the cumulative distributions of the AGB populations of ESO540-G030 and ESO540-G032, respectively.

  \subsection{Scl-dE1}
   
 \begin{figure}
    \centering
      \includegraphics[width=4cm,clip]{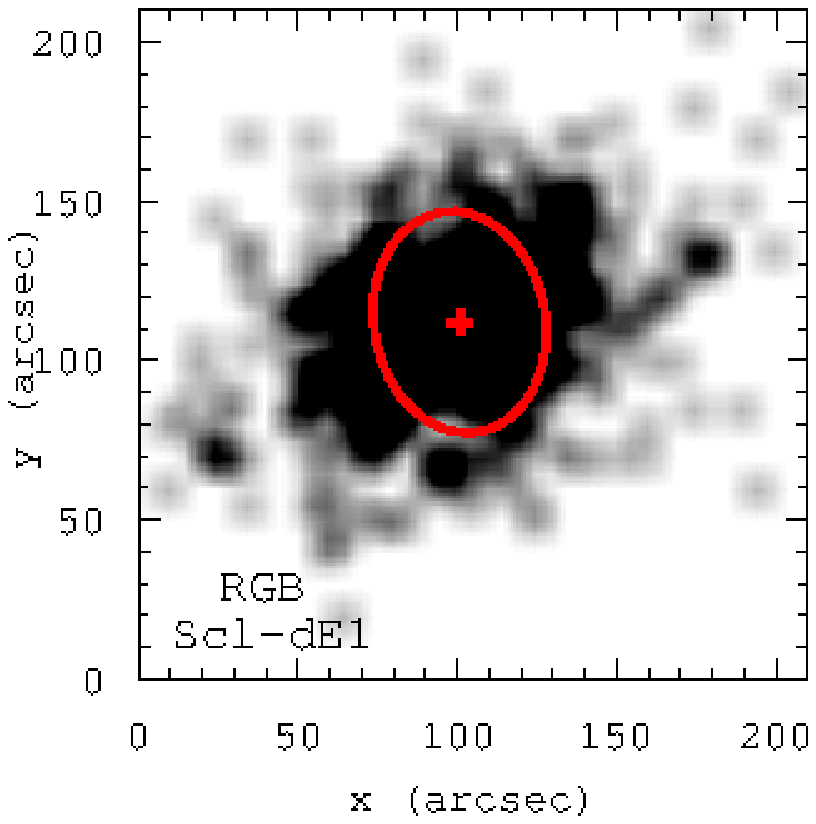}
      \includegraphics[width=4cm,clip]{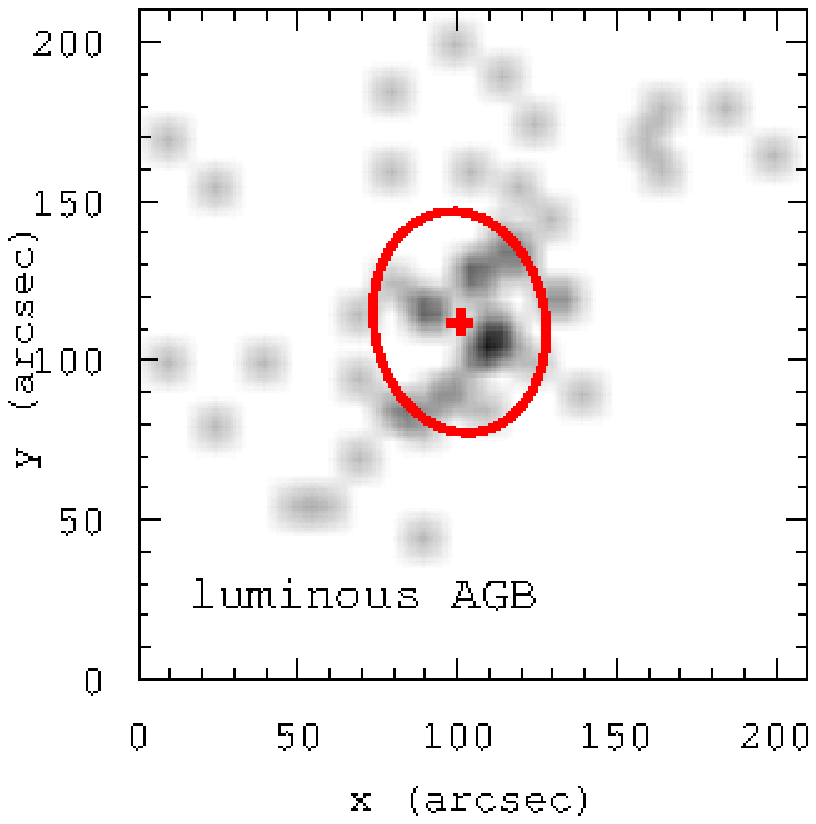}
        \caption{Gaussian-smoothed density maps of RGB stars, and luminous AGB stars for Scl-dE1. The density maps are color coded such that the black color corresponds to a density of 0.1 stars per sq. arcsec, with the density decreasing towards white colors. Also overplotted are the center of the dwarf, shown with a plus sign, and an ellipse. The major axis of the ellipse is equal to the r$_{eff}$. }
       \label{sl_figure10}
\end{figure}
%
   Scl-dE1 is the only dwarf in the Sculptor group that is classified as a dSph and has no HI or H$\alpha$ detection (Bouchard et al.~\cite{sl_bouchard05, sl_bouchard09}). In Fig.~\ref{sl_figure10} we show the density maps of the stars in the RGB (left panel) and luminous AGB phase (right panel). The selection of the stars in these phases is the same as described previously for the transition-type dwarfs. In the same figure, we overplot the center of Scl-dE1 and the ellipse corresponding to a major axis equal to r$_{eff}$, with position angle and ellipticity adopted from Lee et al.~(\cite{sl_lee11}). The 50 luminous AGB stars do not show the same spatial distribution as the RGB stars. This is also evident from the cumulative distributions of the RGB and AGB population, shown in Fig.~\ref{sl_figure09}. Again, the K-S results indicate that the AGB and RGB populations are spatially separated at the 70 \% confidence level. The luminous AGB fraction, i.e., the number of luminous AGB stars over the number of RGB stars within 1~mag below the TRGB, is 12\%, similar to the f$_{AGB}$ of the transition-type dwarfs in the Sculptor group. 

   \section{Photometric Metallicity Distribution Functions}

  The importance of the metallicity distribution functions (MDFs) lies in their ability to constrain the global, integrated chemical evolution history and galaxy formation models (e.g., Mouhcine et al.~\cite{sl_mouhcine05}; Prantzos~\cite{sl_prantzos08}; Mould \& Spitler \cite{sl_mould10}). We derive the photometric MDFs for the five dwarf galaxies using linear interpolation between isochrones. In practice, we interpolate between the two closest isochrones that bracket the color of a star, in order to find the metallicity of that star. The isochrones have a metallicity resolution of 0.1~dex. We choose a fixed age of 12.5~Gyr for the whole set of isochrones, assuming that the RGB consists of mainly old stars in an age range of about 10~Gyr to 13~Gyr. We note at this point that the analysis of the MDFs relies significantly on the assumption of a single, old age, a fact that may be valid only for the stellar populations of Scl-dE1. We resume regarding this point later on in our analysis, and we use for the moment this assumption to understand the early chemical enrichment of our dwarf galaxy sample as a first approach.  

   For each galaxy, we derive its MDF using RGB stars within a selection box such that the brighter magnitude limit is equal to the TRGB magnitude, while the fainter magnitude limit is defined such that up to this limit and brighter, the intrinsic metallicity spread is larger than the metallicity errors. The faint magnitude limit for each dwarf reaches to 25~mag for ESO540-G030 and ESO540-G032, to 24.5~mag for ESO294-G010, to 24.25~mag for ESO410-G005, and to 25.1~mag for Scl-dE1, different for each dwarf due to the photometric depth of the data. These limits are between roughly 1~mag to 1.5~mag fainter than the F814W-band TRGB magnitude, and also by more than 1~mag brighter than the 50\% magnitude limit. Typical photometric errors at the magnitude of 25~mag have values less than 0.03~mag in F814W, and 0.04~mag in (F606W-F814W). The blue and red color limits of the selection box are such to encompass the RGB stars, following the RGB slope. 

  We use for our analysis Dartmouth isochrones (Dotter et al.~\cite{sl_dotter07,sl_dotter08}), in order to remain consistent with our previous investigations on photometric metallicities (Lianou et al.~\cite{sl_lianou10,sl_lianou11}). Using Padova isochrones instead to derive the RGB photometric metallicities introduces an offset in metallicity of the order of $\sim$0.3~dex, making the stars appear more metal-rich than when using Dartmouth isochrones (see Appendix B). 

  The Dartmouth isochrone set we use ranges in [Fe/H] from $-$2.5~dex to $-$0.5~dex, assuming a scaled-solar [$\alpha$/Fe]. We note that these isochrone metallicity ranges are to limit the metallicities of the stars that we include in the MDFs. We redden the isochrones and shift them to the distance of each dwarf according to the values of the Galactic foreground extinction and distance moduli listed in Table~\ref{table2}. We do not correct for internal extinction since this can be assumed to be negligible (Da Costa et al.~\cite{sl_dacosta10}). 

 \begin{figure}
  \centering
       \includegraphics[width=4cm,clip]{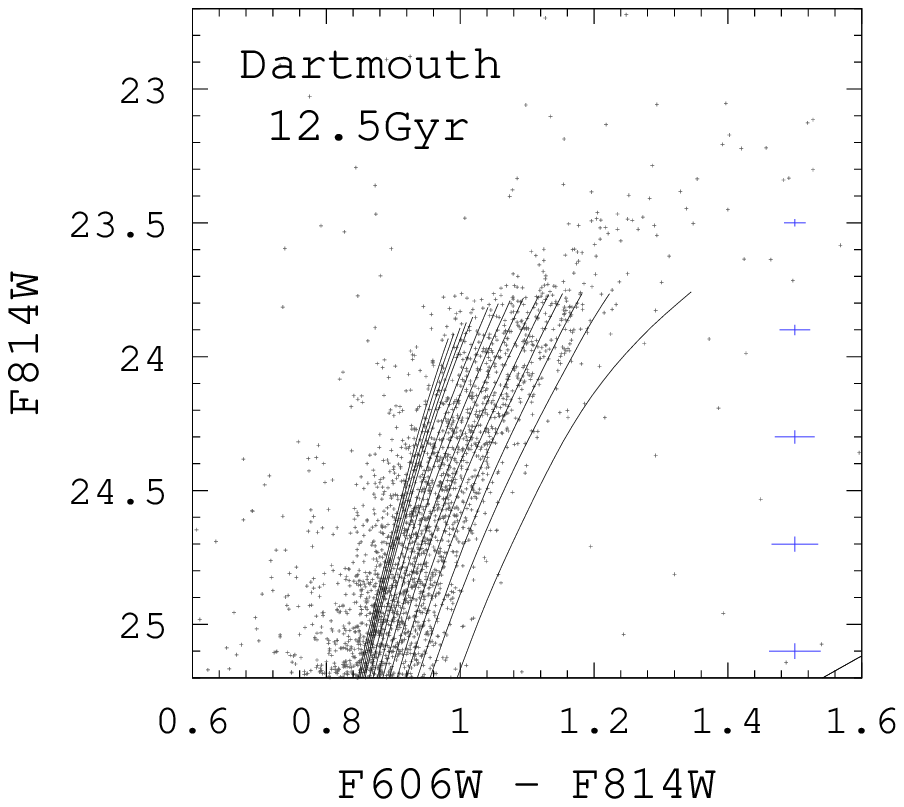}
       \includegraphics[width=4cm,clip]{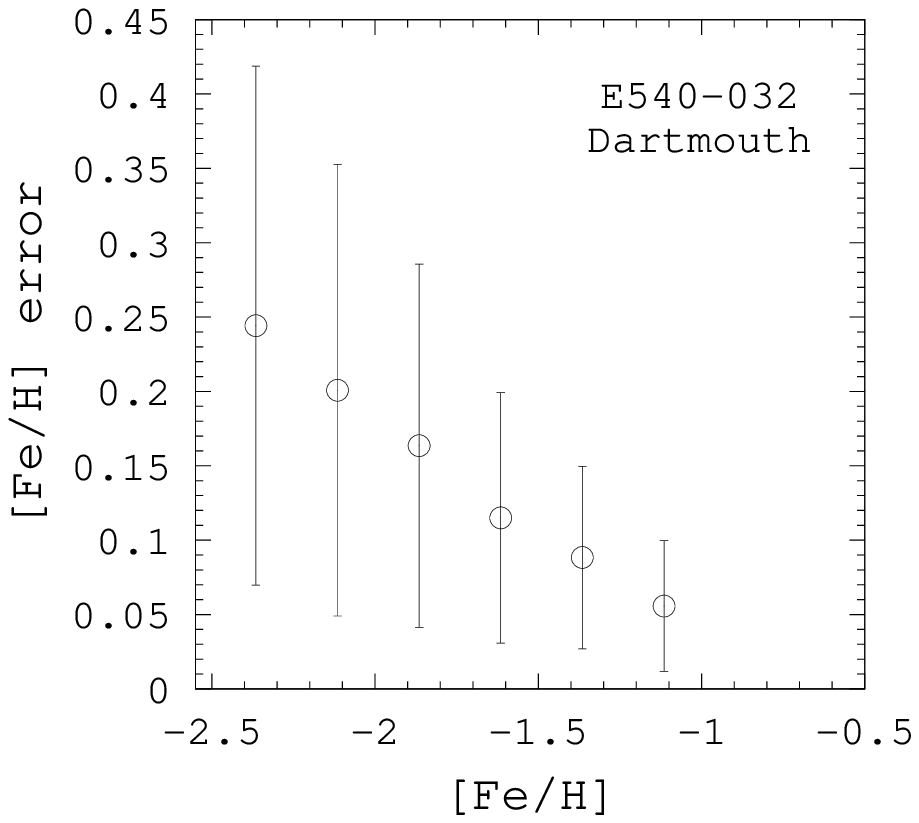}
       \caption{Left panel: Zoom in on the RGB of ESO540-G032 with a subset of Dartmouth isochrones overplotted. The isochrones have a step in metallicity of 0.1~dex, and they range from $-$2.5~dex to $-$0.9~dex in [Fe/H], where the metallicities increase towards redder colors. 
Right panel: The open circles indicate the metallicity errors as a function of metallicities based on Dartmouth isochrones, while the error bars indicate the standard deviation of the metallicity errors. } 
  \label{sl_figure11}
  \end{figure}
   In the left panel of Fig.~\ref{sl_figure11} we show a subset of the Dartmouth isochrones overplotted on the RGB, on which we focus in the case of ESO540-G032, used here as an example. The isochrones range from $-$0.9~dex at the metal-rich end to $-$2.5~dex at the metal-poor end. We note that at the metal-rich end we restrict the isochrones to $-$0.9~dex for demonstration purposes, while in our MDF analysis we use the full range from $-$2.5~dex to $-$0.5~dex. The metallicity step of the isochrones in the figure is 0.1~dex, which is the same resolution we use in the interpolation. As is evident, the spacing between the Dartmouth isochrones decreases rapidly towards the metal-poor part. We note that the stars bluer than the most metal-poor isochrone may be either more metal-poor RGB stars, or younger RGB stars, or old AGB stars with somewhat higher metallicity and with ages typically greater or equal to 10~Gyr. Such old AGB stars that have the same luminosity as RGB stars were also noted, for example, by Harris et al.~(\cite{sl_harris99}). As mentioned earlier, these bluer stars are excluded from the MDFs, since their metallicities are extrapolated values with respect to the most metal-poor isochrone used. The fraction of these excluded stars is 43\% for ESO540-G030, 33\% for ESO540-G032, 30\% for ESO294-G010, 33\% for ESO410-G005, and 40\% for Scl-dE1, where the fraction here represents the number ratio of the stars bluer than the bluest isochrone versus all stars within the RGB selection box. These fractions are significantly lower if the Padova isochrones are used (see Appendix B for details). These fractions depend on the details of the isochrones used, such as the choice of the isochrone age or the [$\alpha$/Fe]. Using a younger (6.5~Gyr) single isochrone age gives smaller fractions, of the order of 15\% to 22\%, i.e. half those when using the 12.5~Gyr isochrones, while using a higher (lower) [$\alpha$/Fe] for the isochrones leads to an increase (decrease) of the fractions, of the order of an additional 3-5\%. The excluded fraction of stars leads to an overestimate of the {\em mean} metallicities by up to a maximum of 0.2~dex in [Fe/H] (see Appendix~C). Further effects of the choice of the isochrone age, as well as systematic biases on the MDFs are discussed in the next section. 

   The errors in metallicity are computed from a set of Monte Carlo simulations, in which each star is varied by its photometric uncertainties (both in color and magnitude, as given by the artificial star tests) and re-fit using the identical isochrone interpolation as described above. The 1$\sigma$ scatter of the output random realisations was then adopted as the formal metallicity error for each star. In the right panel of Fig.~\ref{sl_figure11} we show the distribution of the errors in metallicity as a function of the metallicity, using ESO540-G032 as an example.

 \begin{figure*}
  \centering
       \includegraphics[width=5cm,clip]{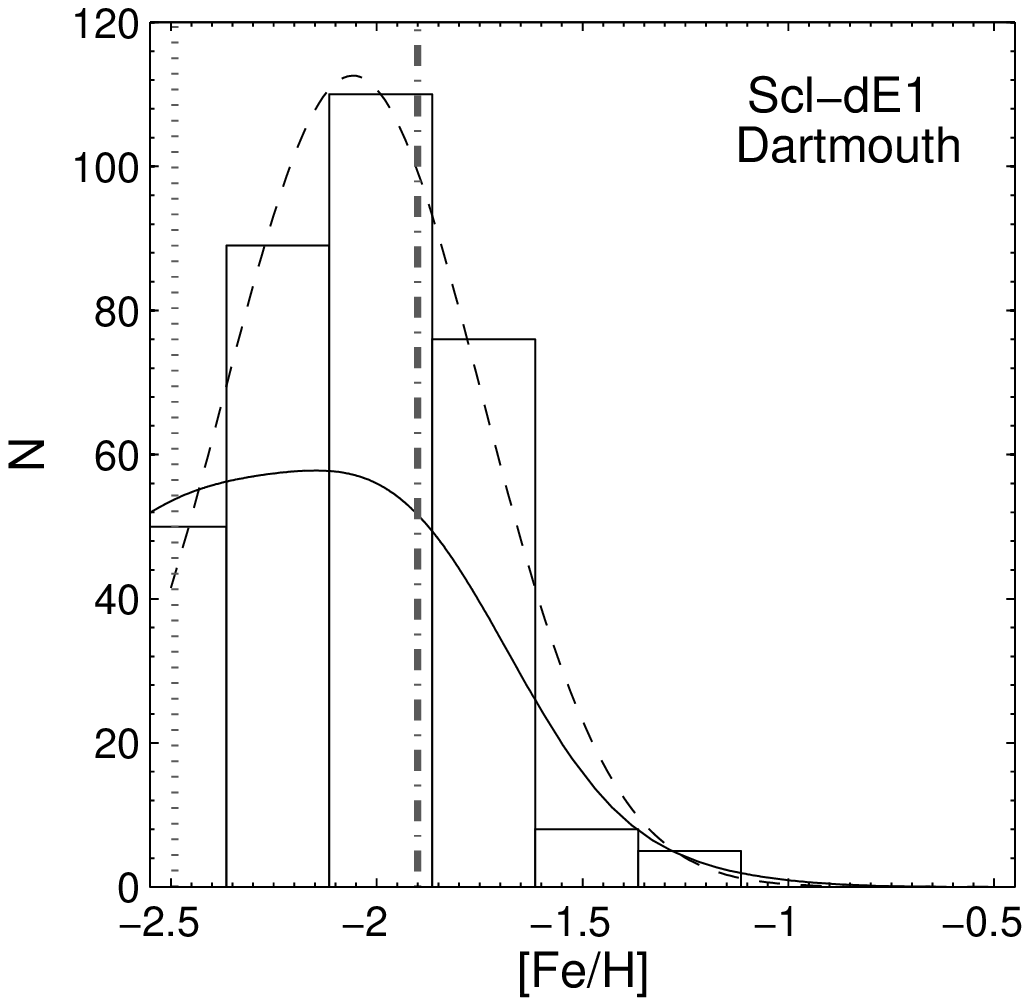}
       \includegraphics[width=5cm,clip]{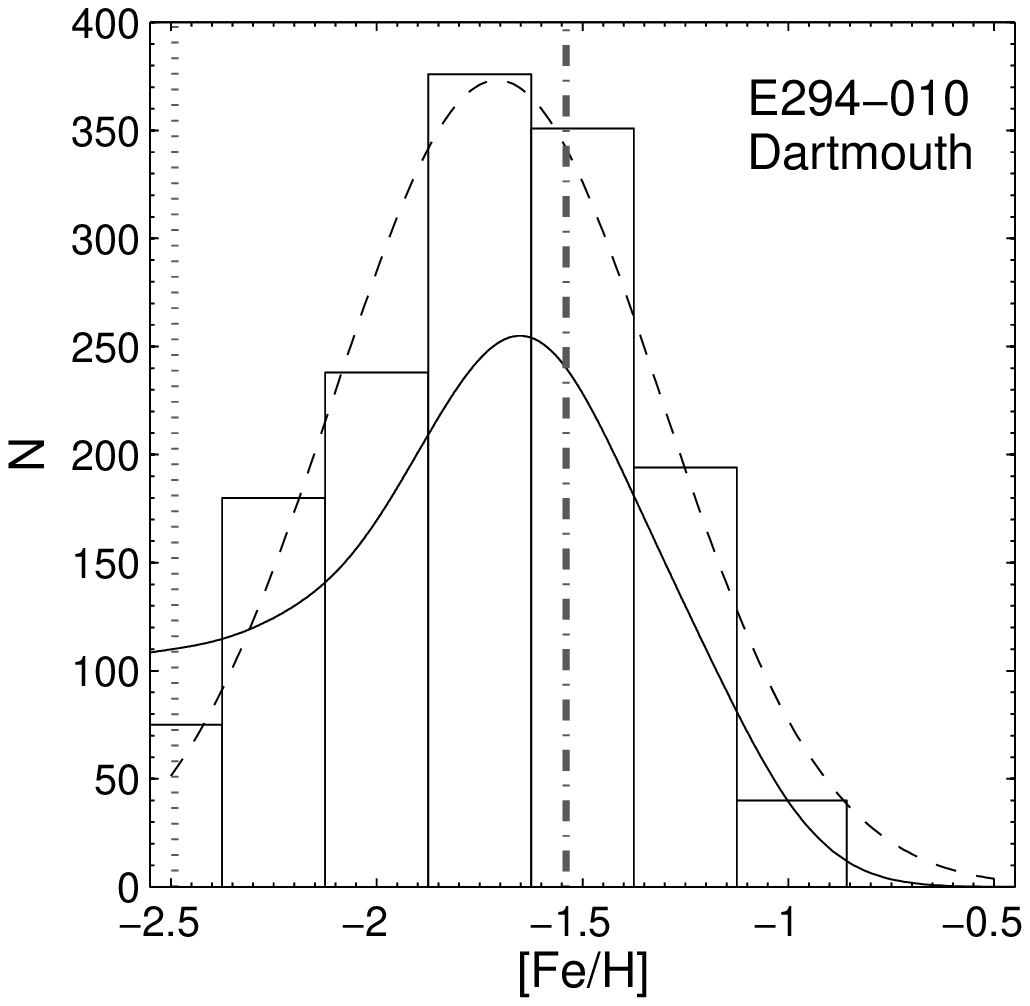}
       \includegraphics[width=5cm,clip]{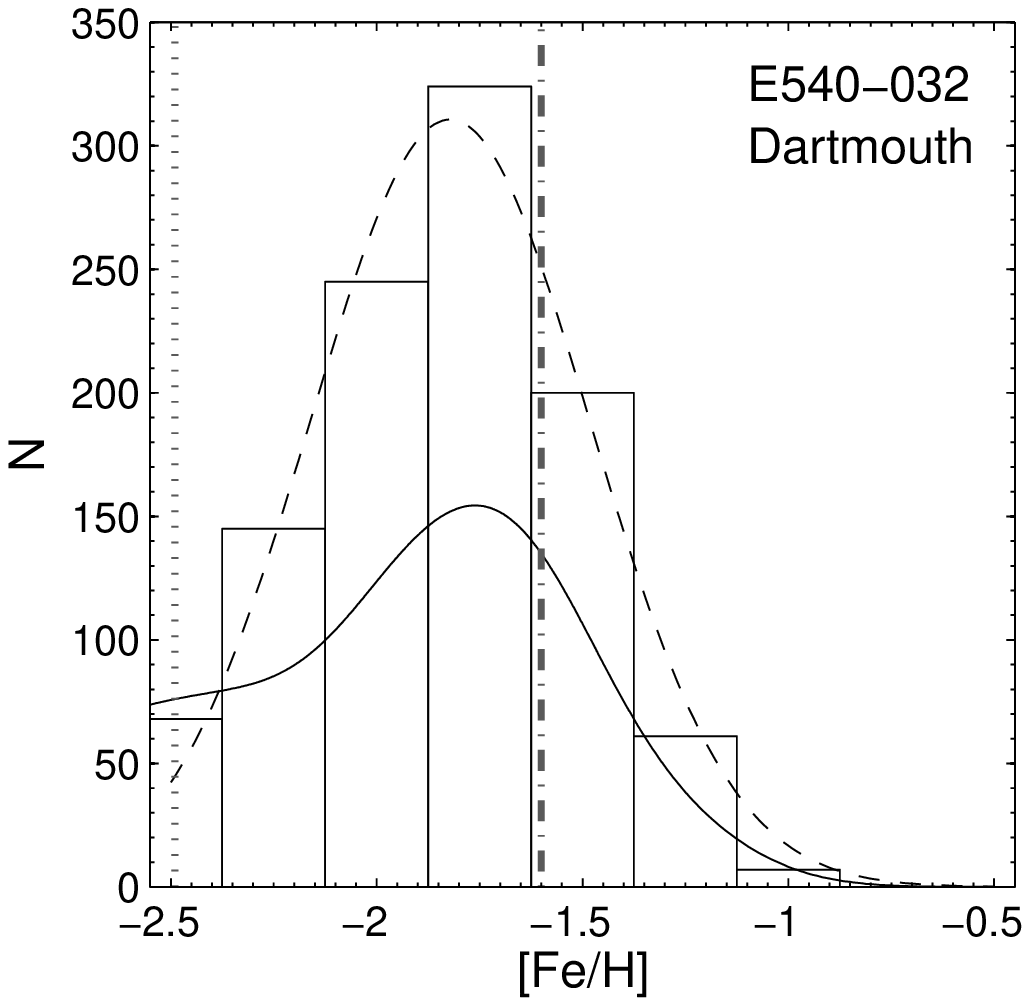}
       \includegraphics[width=5cm,clip]{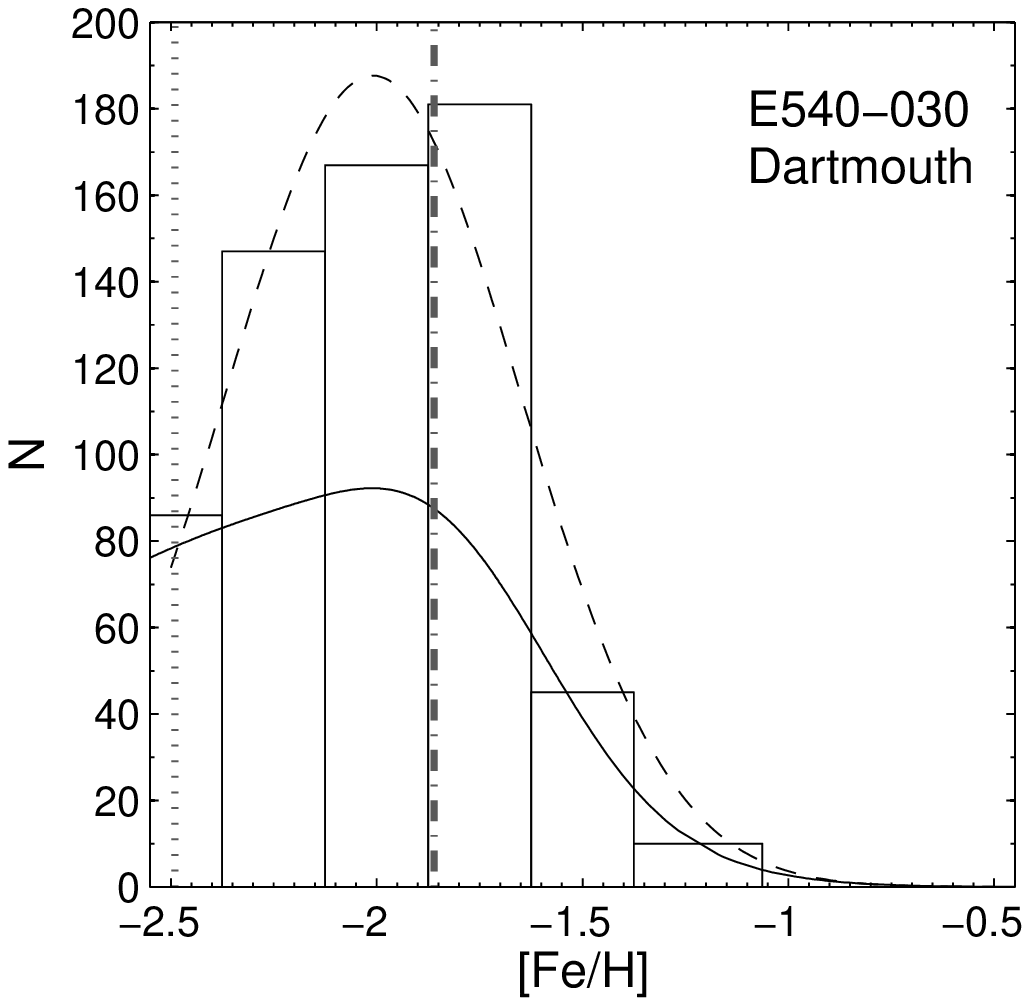}
       \includegraphics[width=5cm,clip]{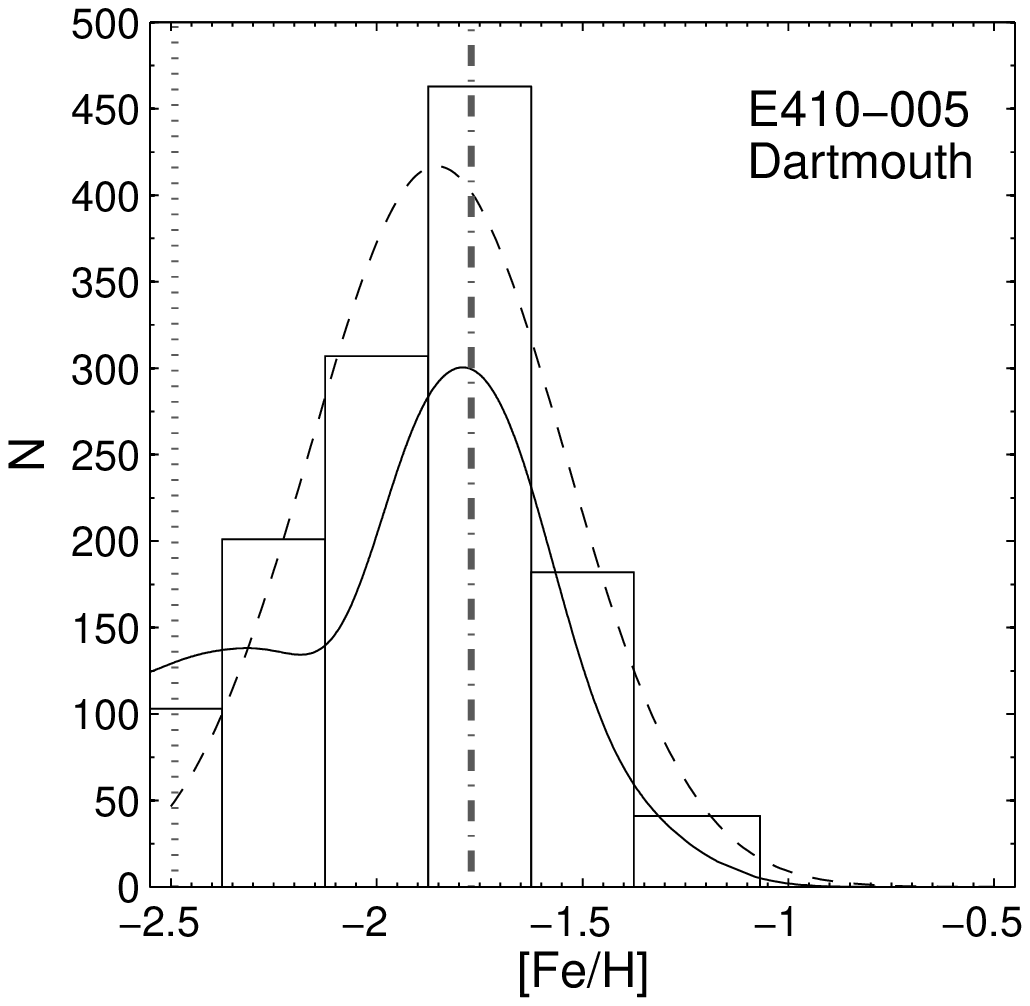}
       \caption{MDFs for our dwarf galaxy sample. The solid lines result from convolving the histograms with the errors of the individual measurements, and the dashed lines show the fitted Gaussian distributions. The vertical dotted-dashed lines indicate the value of the error-weighted mean metallicity, while the dotted vertical line show the most metal-poor value of the isochrone used in the interpolation. The MDFs were derived using Dartmouth isochrones and the assumption of a single old age for the upper RGB stars. Each dwarf's MDF is constructed using the stars that lie redder than the bluest, most metal-poor isochrone and that are within the selection box described in $\S$5. } 
  \label{sl_figure12}
  \end{figure*}
%
   The resulting MDFs are shown in Fig.~\ref{sl_figure12}. In all cases, we restrict the metallicities to within the ranges defined by the metallicities of the isochrones we use in the interpolation, i.e., from $-$2.5~dex to $-$0.5~dex. We also show in Fig.~\ref{sl_figure12} as a solid curve the error-convolved histograms, i.e., the sum of unit Gaussians with a mean equal to the metallicity determination for each star and the $\sigma$ its metallicity error, as well as with the dotted lines the best-fit Gaussian distribution. 
%
\begin{table}
\begin{minipage}[t]{\columnwidth}
\caption[]{Derived properties based on Dartmouth isochrones. }
\label{table5}
\centering
\renewcommand{\footnoterule}{}
\begin{tabular}{l c c c}
\hline\hline
  Galaxy      &$\langle$[Fe/H]$\rangle\pm\sigma$  &$\langle$[Fe/H]$\rangle_{w}\pm\sigma$     &[Fe/H]$_{med}$  \\
              &(dex)                              &(dex)                                     &(dex)          \\
   (1)        &(2)                                &(3)                                       &(4)            \\
\hline                                                                 
 Scl-dE1      &$-2.1\pm0.3$                       &$-1.9\pm0.2$                              &$-2.1$        \\
 ESO294-G010  &$-1.7\pm0.4$                       &$-1.5\pm0.1$                              &$-1.7$        \\
 ESO540-G032  &$-1.8\pm0.3$                       &$-1.6\pm0.2$                              &$-1.8$        \\
 ESO540-G030  &$-2\pm0.3$                         &$-1.8\pm0.2$                              &$-2$          \\
 ESO410-G005  &$-1.9\pm0.3$                       &$-1.8\pm0.2$                              &$-1.9$        \\
\hline
\end{tabular}
\end{minipage}
\end{table}
%
We list the derived \emph{mean} metallicity properties in Table~\ref{table5}. For each dwarf we compute the mean metallicity, $\langle$[Fe/H]$\rangle$, the weighted mean metallicity, $\langle$[Fe/H]$\rangle_{w}$, along with their corresponding 1$\sigma$ dispersion, and the median metallicity, [Fe/H]$_{med}$. We note that, given that we use solar [$\alpha$/Fe] values for the Dartmouth isochrones, the [Fe/H] is equivalent to the global metallicity [M/H] (Salaris et al.~\cite{sl_salaris93}). 

   Comparing our derived metallicity values with those from the literature, the error-weighted mean metallicities for ESO294-G010 and ESO410-G005 based on the Dartmouth isochrones are in very good agreement with the ones derived in Da Costa et al.~(\cite{sl_dacosta10}), who use the mean color of the RGB at the luminosity of M$_{I}=-$3.5\,mag to derive the mean metallicity (Da Costa \& Armandroff \cite{sl_dacosta90}; Armandroff et al.~\cite{sl_armandroff93}; Caldwell et al.~\cite{sl_caldwell98}). These authors find a mean metallicity of $-$1.7$\pm$0.1~dex and $-$1.8$\pm$0.1~dex for ESO294-G010 and ESO410-G005, respectively. Similarly good agreement is found between our error-weighted mean metallicity for Scl-dE1 and the mean metallicity of $-$1.73$\pm$0.17~dex from Da Costa et al.~(\cite{sl_dacosta09}). This is not surprising, since the authors in that work use the same method as also used by Da Costa et al.~(\cite{sl_dacosta10}). We obtain a similarly good agreement when we compare our results for ESO540-G032 with those of Jerjen \& Rejkuba (\cite{sl_jerjen01}), who derive a value of $-$1.7$\pm$0.3~dex based on the mean color as in the above cases, and for ESO410-G005 with those of Karachentsev et al.~(\cite{sl_karachentsev00}), who derive a value of $-$1.8$\pm$0.4~dex using the mean color of the RGB and the calibration of Lee et al.~(\cite{sl_lee93}). Finally, when we compare with the metallicities derived in Sharina et al.~(\cite{sl_sharina08}), we again have consistent results, with metallicities of $-$1.48$\pm$0.13~dex for ESO294-G010, $-$1.93$\pm$0.15~dex for ESO410-G005, $-$1.75$\pm$0.15~dex for ESO540-G030, and $-$1.45$\pm$0.13~dex for ESO540-G032, derived using the mean color of the RGB and the calibration of Lee et al.~(\cite{sl_lee93}). 
\begin{table*}
\begin{minipage}[t]{\textwidth}
\caption[]{Comparison between the metallicity values from this work and from the literature. }
\label{table6}
\centering
\begin{tabular}{l c c c c c}
\hline\hline
  Galaxy     &This work      &Sharina et al.~(2008)   &Da Costa et al.~(2009/2010)    &Jerjen \& Rejkuba (2001)        &Karachentsev et al.~(2000) \\
             &(dex)          &(dex)                   &(dex)                          &(dex)                           &(dex)          \\
   (1)       &(2)            &(3)                     &(4)                            &(5)                             &(6)            \\
\hline 
 Scl-dE1      &$-1.9\pm0.2$   &...                     &$-1.73\pm0.17$                 &...                             &...            \\
 ESO294-G010  &$-1.5\pm0.1$   &$-1.48\pm0.13$          &$-1.7\pm0.1$                   &...                             &...            \\
 ESO540-G032  &$-1.6\pm0.2$   &$-1.45\pm0.13$          &...                            &$-1.7\pm0.3$                    &...            \\
 ESO540-G030  &$-1.8\pm0.2$   &$-1.75\pm0.15$          &...                            &...                             &...            \\
 ESO410-G005  &$-1.8\pm0.2$   &$-1.93\pm0.15$          &$-1.8\pm0.1$                   &...                             &$-1.8\pm0.4$  \\
\hline
\end{tabular}
\end{minipage}
\end{table*}
%
These comparisons are summarised in Table~\ref{table6}. The consistency between the metallicities derived using the Dartmouth isochrones and those using the mean color of the RGB at the luminosity of M$_{I}=-$3.5\,mag on the Zinn \& West (\cite{sl_zinn84}) metallicity scale (Rejkuba et al.~\cite{sl_rejkuba06}), reflects the good agreement between the Dartmouth metallicity scale and the Zinn \& West (\cite{sl_zinn84}) metallicity scale (e.g., Dotter et al.~\cite{sl_dotter10}; Lianou et al.~\cite{sl_lianou11}). 

  \subsection{On the old age assumption and systematic biases}

   The core assumption in our MDF analysis is that of a single, old age for the isochrones in the interpolation, and thus for the underlying stellar populations. Furthermore, we assume that the RGB width is primarily due to a metallicity spread rather than an age spread (e.g., Caldwell et al.~\cite{sl_caldwell98}; Harris, Harris \& Poole \cite{sl_harris99}; Frayn \& Gilmore \cite{sl_frayn02}). The initial star formation in early-type dwarfs may have lasted as long as 3\,Gyr or even longer (Marcolini et al.~\cite{sl_marcolini08}; Hensler, Theis \& Gallagher \cite{sl_hensler04}; Ikuta \& Arimoto \cite{sl_ikuta02}), but any adopted single age within the range of 10 - 13~Gyr would not significantly affect our results. The shape of the MDFs does not strongly depend on the choice of the age for ages older than 10~Gyr, while the choice of a $\sim$10~Gyr isochrone leads to a maximum metallicity difference on a star-by-star basis of less than 0.2~dex, occurring at the metal-poor end (Lianou et al.~\cite{sl_lianou10}). Moreover, Rejkuba et al.~(\cite{sl_rejkuba11}) present a detailed modelling of the contribution of the old AGB stars to the shape of the MDF for a halo field of NGC5128. They find that the shape of the MDF is not affected by the contribution of the old AGB stars, while they also estimate that the induced metal-poor bias is of the order of 0.1-0.2~dex per 3-4~Gyr as the age shifts from 12~Gyr to 8~Gyr.

   By using any constant age for the isochrones within the age range of about 10~Gyr to 13~Gyr, we assume that the intermediate-age (within 1~Gyr to 10~Gyr) stellar content of these dwarfs is low, although stars on the RGB may be as young as $\sim$2~Gyr (e.g., Salaris et al.~\cite{sl_salaris02}). On the other hand, the observations show that our studied dwarfs contain stars with a wide spread in age. This suggests that the choice of a single old age should be regarded as an assumption that allows us to make an initial estimate of the stellar metallicities, while the derived metallicities can only be considered as lower metal-poor limits to their true metallicities. 

   In order to estimate by how much we are biasing the metallicities towards the metal-poor part due to the old single age assumption, we can assume isochrones of 6.5~Gyr and derive the metallicities anew. The initially arbitrary choice of 6.5~Gyr stems from the stellar mass-weighted mean ages, $\tau$, of LG dIrrs, as listed in Orban et al.~(\cite{sl_orban08}). LG dIrrs have mass-weighted mean ages as low as 6.2~Gyr (Leo A; Orban et al.~\cite{sl_orban08}), while the five LG transition-type dwarfs have ages ranging from 7.8~Gyr (PegDIG) to 12~Gyr (DDO210). Thus, we expect that the assumption of a 6.5~Gyr dominant population will place upper limits to the true metallicities of our studied dwarfs. We use 
\begin{table}
\begin{minipage}[t]{\columnwidth}
\caption[]{Derived properties based on 6.5~Gyr Dartmouth isochrones. }
\label{table7}
\centering
\renewcommand{\footnoterule}{}
\begin{tabular}{l c c c}
\hline\hline
  Galaxy    &$\langle$[Fe/H]$\rangle\pm\sigma$    &$\langle$[Fe/H]$\rangle_{w}\pm\sigma$     &[Fe/H]$_{med}$\\
            &(dex)              &(dex)            &(dex)      \\
   (1)      &(2)                &(3)                   &(4)        \\
\hline                                                              
 Scl-dE1      &$-1.8\pm0.3$     &$-1.7\pm0.2$      &$-1.8$         \\
 ESO294-G010  &$-1.5\pm0.4$     &$-1.4\pm0.1$      &$-1.5$         \\
 ESO540-G032  &$-1.6\pm0.3$     &$-1.4\pm0.1$      &$-1.6$         \\
 ESO540-G030  &$-1.7\pm0.3$     &$-1.6\pm0.1$      &$-1.7$         \\
 ESO410-G005  &$-1.6\pm0.3$     &$-1.5\pm0.2$      &$-1.6$         \\
\hline
\end{tabular}
\end{minipage}
\end{table}
%
the exact same analysis as for the 12.5~Gyr isochrone set. The results for the \emph{mean} metallicities are listed in Table~\ref{table7} and indicate that using a younger isochrone leads to metallicities that are higher by 0.2~dex to 0.3~dex on average. These values are comparable to the 1-$\sigma$ dispersion listed in Table~\ref{table5}. Assuming that the dominant age of our studied dwarfs ranges from 6.5~Gyr to 13~Gyr, their average photometric metallicities will lie in between the ones computed in these extremes. We further note that the presence of an age spread points to the need for either deeper photometric observations reaching the old MS turn-offs or spectroscopic observations in order to break the age-metallicity degeneracy and to estimate the true metallicities. 

   Other than age and the age-metallicity degeneracy, the systematics that may affect the photometric metallicities are the adopted distance modulus, the reddening, and the $\alpha$-elements enhancement assumption (e.g., Lianou et al.~\cite{sl_lianou11}). Varying the distance modulus by 1\% leads to a difference in the [Fe/H] photometric metallicities of 10\%, where an increase of the distance modulus leads to more metal-poor values. Likewise, an increase of the reddening leads to more metal-poor values, with a variation of the reddening by 16\% leading to a difference of the derived metallicities of 12\%. The solar-scaled assumption for the isochrones influences the photometric metallicities such that an increase of the $\alpha$-enhancement to $+$0.2~dex lowers the median metallicity by 0.1~dex.  

   \subsection{Metallicity gradients and age-metallicity degeneracy}

   Dwarf galaxies in the Local, Centaurus A, and M81 Groups show a diversity in terms of the presence or absence of metallicity gradients (e.g., Harbeck et al.~\cite{sl_harbeck01}; Crnojevic et al.~\cite{sl_crnojevic10}; Lianou et al.~\cite{sl_lianou10}). Results from simulated isolated dwarf galaxies suggest that galaxy rotation seems to play a role in understanding the diversity of the observed metallicity gradients, with the latter preserved when external processes act on initially gas-rich dwarfs converting them to gas-poor dwarfs (Schroyen et al.~\cite{sl_schroyen11}). Pasetto et al.~(\cite{sl_pasetto10}), in their study on the effect of star formation processes in shaping the dark matter profile of an isolated dSph, conclude that there is a link between the formation of radial metallicity gradients and the change of the stellar density profile, with the latter changing due to a redistribution of the baryonic material caused by star formation processes and feedback and within a timescale compatible with the reshaping of the dark matter profile. According to Marcolini et al.~(\cite{sl_marcolini08}), metallicity gradients in a dSph can be explained through supernovae of type II (SNe\,II) or of type Ia (SNe\,Ia), with the former leading to a homogenisation of the metallicity by driving the metals outwards, and the latter producing localised iron-rich inhomogeneities.

  Our dwarf galaxy sample contains stellar populations with a wide range of ages and fractions of intermediate-age (between 1~Gyr to 10~Gyr) stars. This holds also for Scl-dE1, which has a fraction of intermediate-age stars of 26\% (Weisz et al.~\cite{sl_weisz11a}), which is higher than that of the Milky Way satellite Sculptor dSph (14\%; Orban et al.~\cite{sl_orban08}), and thus Scl-dE1 may also suffer from an age-metallicity degeneracy.  Old (say, 12 Gyr), metal-poor stars on the RGB occupy the same locus in the CMD as more metal-rich, but younger ($\sim$7 Gyr) stars. This well-known age-metallicity degeneracy can lead to over- and underestimates of individual stars' metallicities, if the age distribution of the underlying population is not known (Lianou et al.~\cite{sl_lianou11}). When also considering that in the case of a metallicity gradient, the metal-rich and young stars are usually observed towards the center (e.g., de Boer et al.~\cite{sl_deboer11}), the effect of an age-metallicity degeneracy will work against uncovering existing metallicity gradients.

 \begin{figure}
    \centering
      \includegraphics[width=4cm,clip]{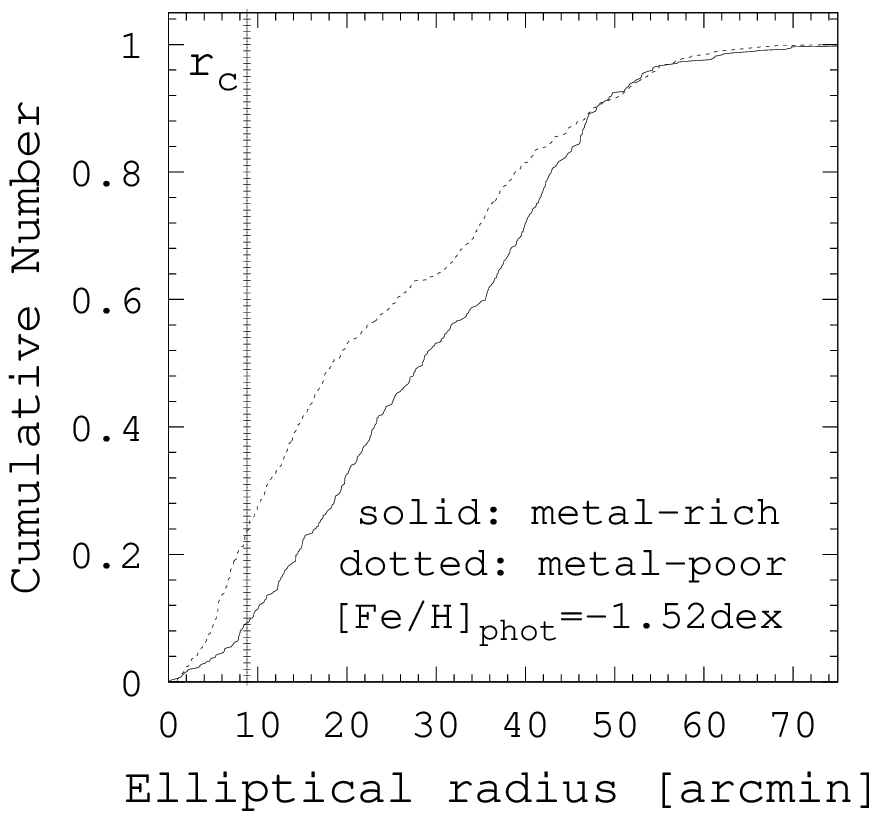}
      \includegraphics[width=4cm,clip]{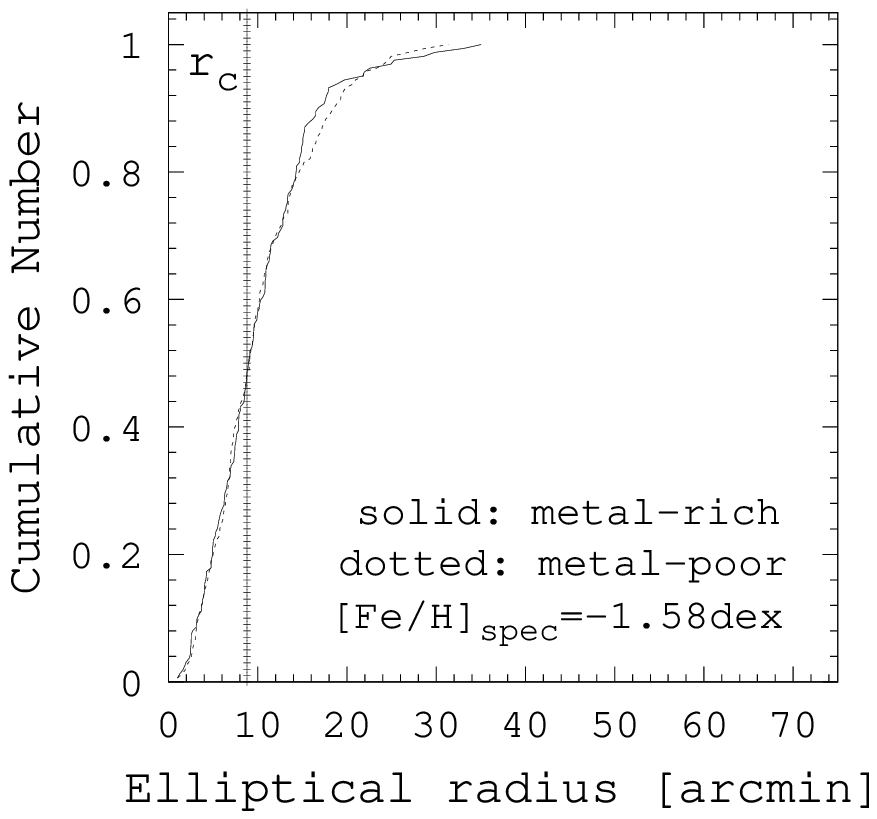}
        \caption{Left panel: Cumulative metallicity distribution functions based on photometric metallicities for the Carina dSph. Right panel: The same as in the left panel, but based on spectroscopic metallicities. The cumulative distributions are separated according to values higher or lower than the median metallicity, indicated within each panel. In both cases, the vertical line indicates the location of the core radius, r$_{c}$, from Irwin \& Hatzidimitriou (\cite{sl_irwin95}). }
       \label{sl_figure13}
\end{figure}
%
   In order to see the effect of the age-metallicity degeneracy on the derivation of metallicity gradients, we use observations of the Carina dSph, which shows a pronounced age-metallicity degeneracy and a mild spectroscopically derived metallicity gradient (Koch et al.~\cite{sl_koch06}; and references therein). We use the photometric metallicities derived in Lianou et al.~(\cite{sl_lianou11}), and the spectroscopic metallicities from Koch et al.~(\cite{sl_koch06}), with the aim to compare their cumulative metallicity distributions. We briefly mention that the photometric metallicities of Carina are derived using the same interpolation scheme as that presented in this work, while the spectroscopic metallicities are based on low-resolution spectra. We account for the elliptical shape of Carina using the elliptical radius and adopting the structural parameters from Irwin \& Hatzidimitriou (\cite{sl_irwin95}). The derived cumulative metallicity distributions are shown in Fig.~\ref{sl_figure13}. While the spectroscopic metallicity gradient (right panel) is only mildly present, there is a pronounced photometric metallicity gradient (left panel). The photometric metallicity gradient goes in the opposite direction, i.e., the metal-poor stars are more centrally concentrated compared to the metal-rich stars. We note that Harbeck et al.~(\cite{sl_harbeck01}) find instead a strong age gradient in Carina's populations, which would be consistent with an age-metallicity degeneracy affecting the photometric metallicity gradients, as discussed above. Therefore, considering the actual spatial location of the stars within the galaxy, a potential presence of metallicity gradients will be lost with the photometric metallicities. In light of the age-metallicity degeneracy effect on uncovering metallicity gradients, when using photometric metallicities, we do not go on further with exploring their potential presence in our dwarf galaxy sample. 
%

\section{Morphology-distance relation}

  The environment has an impact on the evolution of dwarf galaxies, as imprinted for instance in the morphology-distance relation (e.g., Binggeli, Tarenghi \& Sandage \cite{sl_binggeli90}), or on the HI properties of individual galaxies within a group or cluster environment, such as in the LG or the Virgo cluster. Thomas et al.~(\cite{sl_thomas10}) study a large sample of SDSS galaxies and they find that the early epoch of star formation seems to be governed by internal processes while later phases of star formation activity are influenced by the environment, with the latter effect being stronger for lower galaxy masses. Weisz et al.~(\cite{sl_weisz11a}) reach the same conclusion regarding the SFHs of Local Volume dwarfs, suggesting that while their mean cumulative SFHs are similar, it is the recent star formation that is different among their studied galaxies. Bouchard et al.~(\cite{sl_bouchard09}) find a strong level of correlation between the luminosity density $\rho_{L}$, which represents a measure of the environment, and several indicators, including star formation rate and blue absolute luminosity, of nearby dwarf galaxies.

 The Sculptor Group offers a different environment than that of the LG or other nearby groups to study the potential influence of the environment on the evolution of dwarf galaxies. It forms an extended, little evolved filament of mainly gas-rich galaxies more akin to the loose CVn group (Karachentsev et al.~\cite{sl_karachentsev03b}) than the LG, Cen A, or M81 groups. According to Karachentsev et al.~(\cite{sl_karachentsev03,sl_karachentsev04}), ESO410-G005 and ESO294-G010 may be considered companions of NGC\,55 located at the near side of the Sculptor Group at a distance of 2.1$\pm$0.1~Mpc (Tanaka et al.~\cite{sl_tanaka11}; Dalcanton et al.~\cite{sl_dalcanton09}), while ESO540-G030, ESO540-G032, and Scl-dE1 may be companions of NGC\,253 at 3.44$\pm$0.26~Mpc (Dalcanton et al.~\cite{sl_dalcanton09}). We note that since the Sculptor group is a low-density cloud of galaxies distributed along a filament and covering a distance from $\sim$1.5\,Mpc to $\sim$4.5\,Mpc, assigning a dwarf member to the brightest nearby galaxy may be regarded as only approximate, as also pointed out in Bouchard et al.~(\cite{sl_bouchard09}). 

  Grcevich \& Putman (\cite{sl_grcevich09}) confirm the existence of a morphology-distance relation for LG dwarf galaxies, in terms of HI depleted dwarfs being detected within $\sim$270~kpc from the Andromeda or the Milky Way galaxy. We place our studied Sculptor group dwarfs in the plot of HI mass as a function of galactocentric distance from the closest bright galaxy. The data regarding the HI masses are adopted from Bouchard et al.~(\cite{sl_bouchard05}) for the Sculptor group dwarfs, and from Grcevich \& Putman (\cite{sl_grcevich09}) for the brightest of the LG dwarfs. We select to plot those LG dwarfs with M${_V}$ higher than $-$11~mag, as well as the LG transition-type dwarfs. We compute the deprojected distances of our studied dwarfs from their main disturber, i.e., the neighbouring galaxy producing the maximum tidal action on a given galaxy, following the method described in, e.g., Karachentsev et al.~(\cite{sl_karachentsev03}), and using the formula: $R^{2}=D^2+D^{2}_{MD}-2 D D_{MD}cos\Theta$. Here, $D_{MD}$ is the distance in Mpc of the main disturber, $D$ is the distance in Mpc of the dwarf under consideration, and $\Theta$ is the angular separation of each dwarf from their main disturber, calculated using NASA/IPAC Extragalactic Database (NED). The derived deprojected distances $R=R_{disturber}$ are listed in column (11) of Table~\ref{table2}, along with their uncertainties, which were computed taking into account the uncertainties in the distances of the dwarf galaxy under study and its main disturber. Deprojected distances show the current locations of the dwarfs within their environment, while their actual orbits remain unknown (e.g., Bellazzini, Fusi Pecci \& Ferraro \cite{sl_bellazzini96}).

 \begin{figure}
    \centering
      \includegraphics[width=8cm,clip]{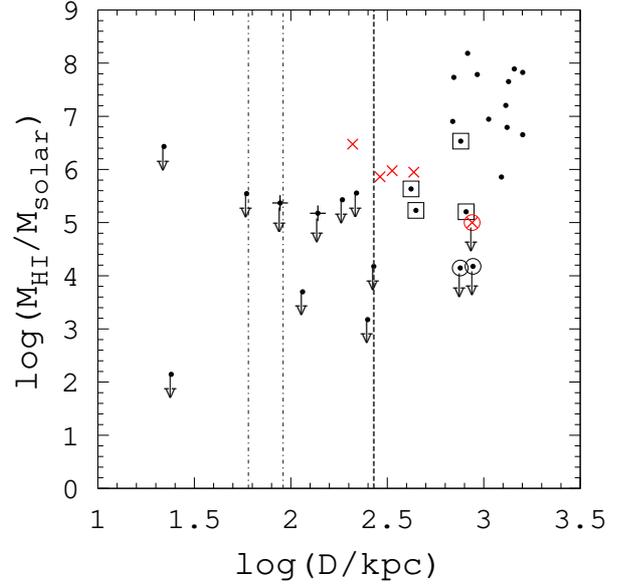}
        \caption{HI mass as a function of galactocentric distances for the brightest LG dwarfs, shown with black dots and adopted from Grcevich \& Putman (\cite{sl_grcevich09}), and Sculptor group dwarfs of our sample shown in red crosses. The HI masses for the Sculptor group dwarfs are adopted from Bouchard et al.~(\cite{sl_bouchard05}). The Tucana and Cetus dSphs are indicated with the black circled dots, while Scl-dE1 is marked with the red circled-cross. The Sculptor and Fornax dSphs of the LG are shown with the black plus signs. The LG transition-type dwarfs, LGS3, Phoenix, Leo~T, PegDIG, are shown with the squared-dots. Arrows indicate HI upper limits. The vertical dashed line indicates the distance of 270~kpc, while the two vertical dashed-dotted inner lines indicate the 60~kpc and 90~kpc distances.}
       \label{sl_figure14}
\end{figure}
%
   The main disturbers of our studied dwarfs, NGC\,55 and NGC\,253, are less massive than the Galaxy or Andromeda (Puche \& Carignan \cite{sl_puche91a,sl_puche91b}), therefore, their virial radii are smaller than 270~kpc, the radius within which the majority of the LG dSphs are located (Grcevich \& Putman \cite{sl_grcevich09}). We estimate the virial radii of NGC\,55 and NGC\,253 to $\sim$60~kpc and $\sim$90~kpc, respectively, using the dynamical mass estimates in Puche \& Carignan (\cite{sl_puche91a,sl_puche91b}), and we plot them in Fig.~\ref{sl_figure14} with the two vertical dashed-dotted lines. Fig.~\ref{sl_figure14} shows that the studied Sculptor group dwarfs lie beyond the virial radii of their nearest massive galaxies. In order to establish whether the transition-type dwarfs in our sample follow a morphology-distance relation, more observations are needed in order to uncover the low-luminosity, gas deficient dwarfs and position them in the morphology-distance diagram.

   Interestingly, as in the case of the LG dSphs Tucana and Cetus, Scl-dE1 seems also isolated at a distance of $\sim$870$\pm$260~kpc from its main disturber, NGC253. Tucana is a dSph dominated by old stellar populations (e.g., Monelli et al.~\cite{sl_monelli10a}) and taking also into account its isolation makes it an outlier galaxy in the morphology-distance relation, as is the case for the second isolated LG dSph Cetus (e.g., Monelli et al.~\cite{sl_monelli10b}). Therefore, in the low-density environment of the Sculptor group there are dwarf galaxies with properties similar to those of the dwarf galaxies in the LG. 

\section{Discussion and conclusions}

   Two early-type dwarfs have been detected in the Sculptor group thus far: Scl-dE1 and NGC59. In the LG, there are more than 40 early-type dwarfs (e.g., Kalirai et al.~\cite{sl_kalirai10}), but in this count the faintest of the dwarfs are included. Therefore, the low number of early-type dwarfs in the Sculptor group may be due to current detection limitations and surveying area coverage, since for example in the M81 group there were recently about 10 new members identified (e.g., Chiboucas et al.~\cite{sl_chiboucas09}). There are five transition-type dwarfs in the LG, and four in the Sculptor, Centaurus A, and M81 groups (Bouchard et al.~\cite{sl_bouchard09}; Weisz et al.~\cite{sl_weisz11b}; Huchtmeier \& Skillman \cite{sl_huchtmeier98}; Boyce et al.~\cite{sl_boyce01}; Karachentsev et al.~\cite{sl_karachentsev07}; but see also Makarova et al.~\cite{sl_makarova10} for the latter group). These numbers of dwarfs in the nearby groups are limited to "classical" bright satellites, given that the ultra-faint dwarf galaxies recently uncovered in large numbers in the LG are too faint to be detected in these nearby groups. In total, the Centaurus A group contains about 60 dwarfs of all types, the Sculptor group contains about 58 dwarfs of all types, while the M81 group contains about 50 dwarf galaxies of all types, with the latter down to a limiting magnitude M$_{r \arcmin}$ = $-$10 (e.g., Karachentsev et al.~\cite{sl_karachentsev04}; Bouchard et al.~\cite{sl_bouchard09}; Chiboucas et al.~\cite{sl_chiboucas09}). The Sculptor group has, currently, the largest fraction of late to early type dwarfs, as compared to, in order of increasing fractions of late to early type dwarfs, the LG, the Centaurus A group, and the M81 group (Chiboucas et al.~\cite{sl_chiboucas09}). 

   In our study, we have looked at the stellar populations and early chemical enrichment of one early-type and four transition-type dwarfs in the Sculptor group. The four transition-type dwarfs show recent star formation and all studied dwarfs show star formation at intermediate ages. Given the presence of intermediate-age populations, our derived photometric metallicities are biased towards more metal-poor values due to the age-metallicity degeneracy. Therefore, with our analysis of the photometric metallicities we only place a lower limit on their {\em mean} metallicities. We further estimate the bias induced by the age-metallicity degeneracy along the red giant branch and due to our old single age assumption for the underlying stellar populations, by considering a single intermediate age for the whole population and deriving the photometric metallicities anew under this assumption. The true {\em mean} metallicities will lie in between these two extreme values, depending on the details of the star formation and chemical evolution histories. The lower metallicity limits that we estimate range from $-$1.5~dex for ESO294-G010 to $-$1.9~dex for Scl-dE1, while a potential upper limit to these metallicities may be as high as 0.1~dex to 0.3~dex towards more metal-rich values. In spite of the wide age spread present in our dwarf sample, a metallicity spread is also required in order to justify the stellar content. 
Given the presence of an age-metallicity degeneracy, we investigate its effect on the derivation of metallicity gradients. We find that the metal-poor bias in the photometric metallicities due to the presence of intermediate-age stars may also lead to an erroneous inference of the potential presence of metallicity gradients. 

  We further examine the spatial distribution of stellar populations belonging to different stellar evolutionary phases and tracing different ages: less than 100~Myr for the MS stars; between 400~Myr to 900~Myr for the VC stars; around 1.5~Gyr to 2~Gyr for the luminous AGB stars, while older ages cannot be excluded; older than 1.5~Gyr for the RGB stars. We find that the young MS stars are, in all cases and whenever detected, more centrally concentrated with respect to the intermediate-age and old stellar populations traced by the red giant stars. The luminous AGB stars have low numbers and are following either the distribution of the VC stars, in the case of ESO294-G010 and ESO410-G005, or the distribution of the MS stars, in the case of ESO540-G030 and ESO540-G032, while for Scl-dE1 the luminous AGB stars are differently distributed than the old and intermediate-age stars in the RGB. In the case of ESO294-G010 and ESO410-G005, the detected VC stars are more spatially extended in the former dwarf, and show the same spatial distribution as the MS stars, in the latter dwarf. The detected stellar population gradients may be suggestive of metallicity gradients, assuming that an age-metallicity relation holds. Indeed, in the case of ESO294-G010 and ESO410-G005, comparing the Z metallicity of the RC and RGB stars (using Padova isochrones), the former lie in the metal-rich part of the latter's MDF, thus the detected age gradient may be indicative of a metallicity gradient as well, as is also the case for the Sculptor dSph in the LG that shows a clear relation between age and metallicity gradients to hold (e.g., de Boer et al.~\cite{sl_deboer11}).

  It is intriguing that the three transition-type dwarfs ESO294-G010, ESO410-G005, and ESO540-G030 show a stellar spatial distribution of their young component off-centred and similar to the one that the LG dIrrs show. Whether the Sculptor group dwarf galaxies follow a morphology-distance relation needs to be confirmed with more data regarding the lowest luminosity gas-deficient dwarfs, yet to be detected, and examine their location within the Sculptor group. It is interesting to note that Scl-dE1, the only dSph thus far detected in the Sculptor group, is an outlier to a potential  morphology-distance relation, as are the Tucana and Cetus dSphs in the LG. 

\begin{acknowledgements} The authors would like to thank an anonymous referee for the thoughtful comments. We are grateful to Aaron Dotter for prompt technical support on Dartmouth isochrones, Leo Girardi for clarifications on Padova isochrones, and Andrew Dolphin for DOLPHOT output specifics. SL thanks Prof. Sandra Klevansky for invaluable support, as well as the University of Heidelberg for partial grant support. AK thanks the Deutsche Forschungsgemeinschaft for funding from Emmy-Noether grant Ko 4161/1.
  
  This research has made use of the NASA/IPAC Extragalactic Database (NED) which is operated by the Jet Propulsion Laboratory, California Institute of Technology, under contract with the National Aeronautics and Space Administration, as well as of NASA's Astrophysics Data System Bibliographic Services, of SAOImage DS9 developed by Smithsonian Astrophysical Observatory, of Aladin, and of IRAF. IRAF is distributed by the National Optical Astronomy Observatories, which are operated by the Association of Universities for Research in Astronomy, Inc., under cooperative agreement with the National Science Foundation.
\end{acknowledgements}

\begin{appendix}

\section{Comparison of DOLPHOT and DAOPHOT photometries}

  In this section, we compare the DOLPHOT (Dolphin \cite{sl_dolphin00}) photometry with DAOPHOT (Stetson \cite{sl_stetson87,sl_stetson94}) photometry (Da Costa et al.~\cite{sl_dacosta09,sl_dacosta10}; Rejkuba et al.~in prep.). DOLPHOT shares some common concepts in the photometry reduction as those of DAOPHOT, but these two packages have also significant differences, as outlined in Dolphin (\cite{sl_dolphin00}). The results of the comparison between the magnitudes of the stellar sources detected with 
 \begin{figure*}
  \centering
    \includegraphics[width=7cm,clip]{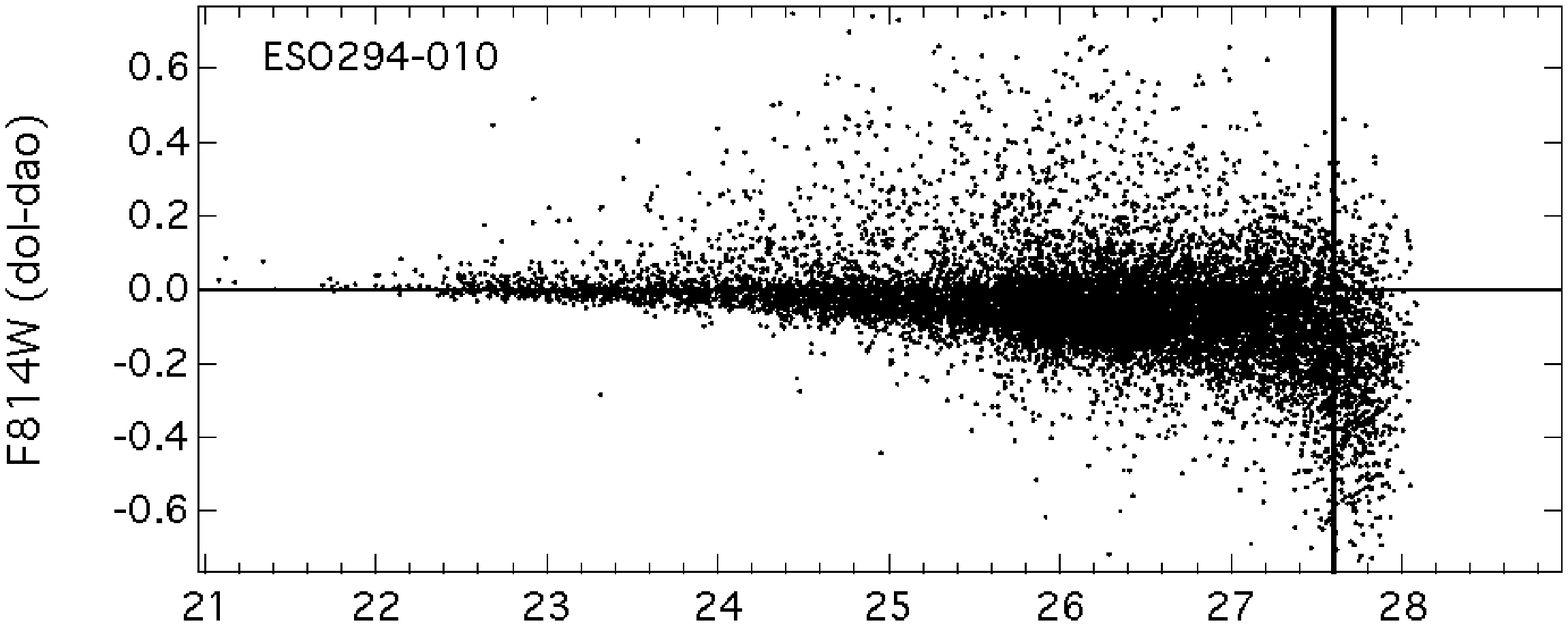}
    \includegraphics[width=7cm,clip]{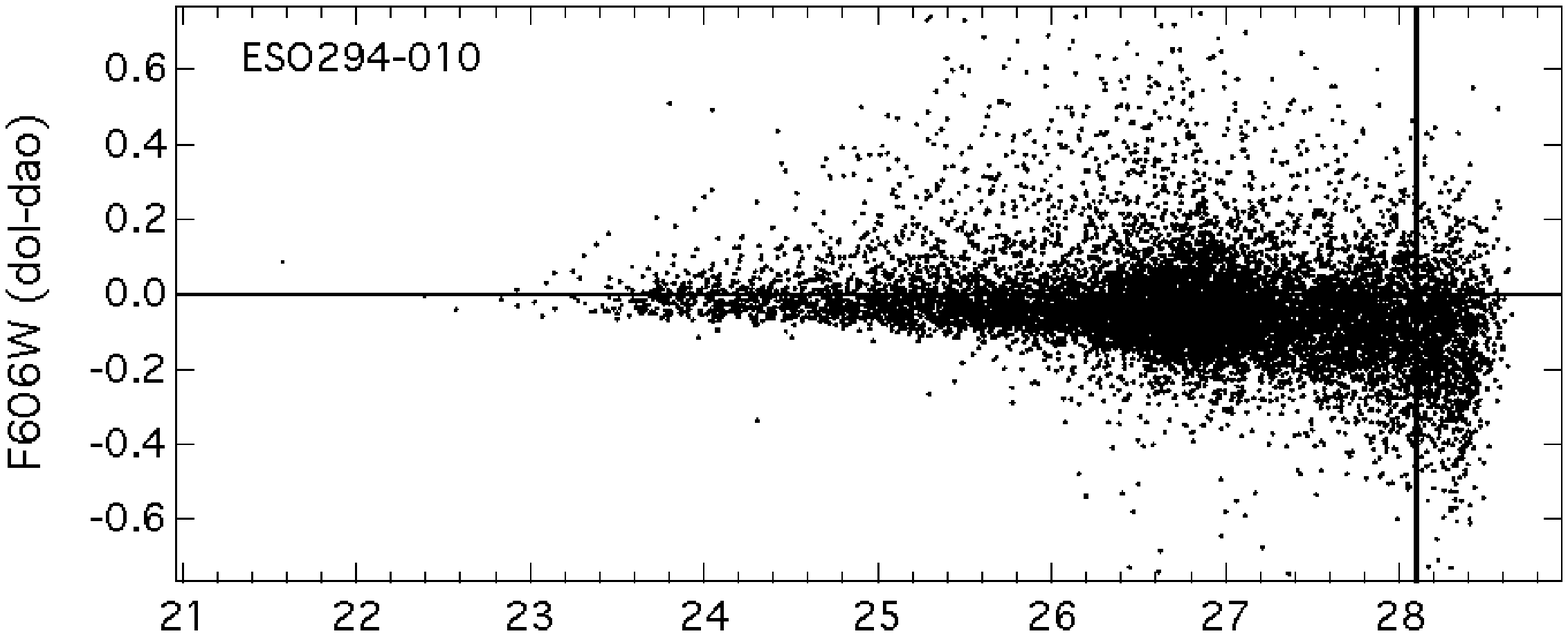}
    \includegraphics[width=7cm,clip]{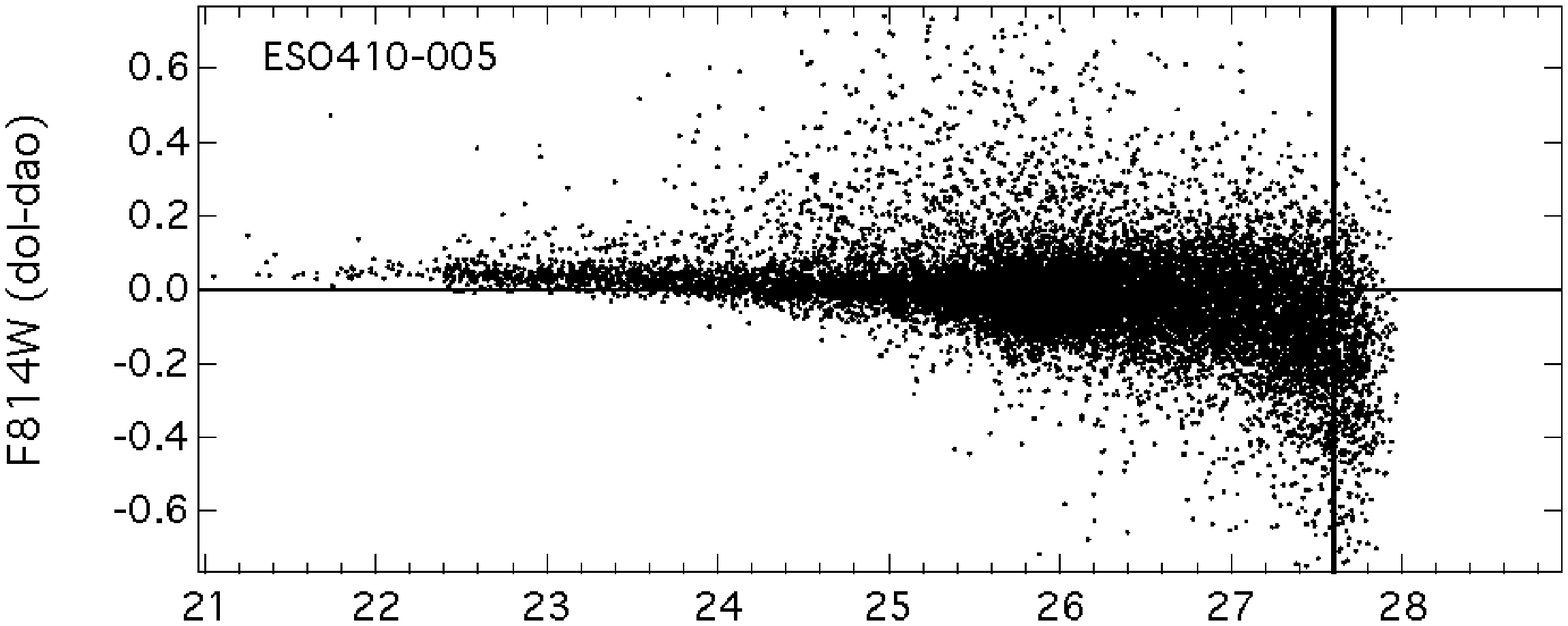}
    \includegraphics[width=7cm,clip]{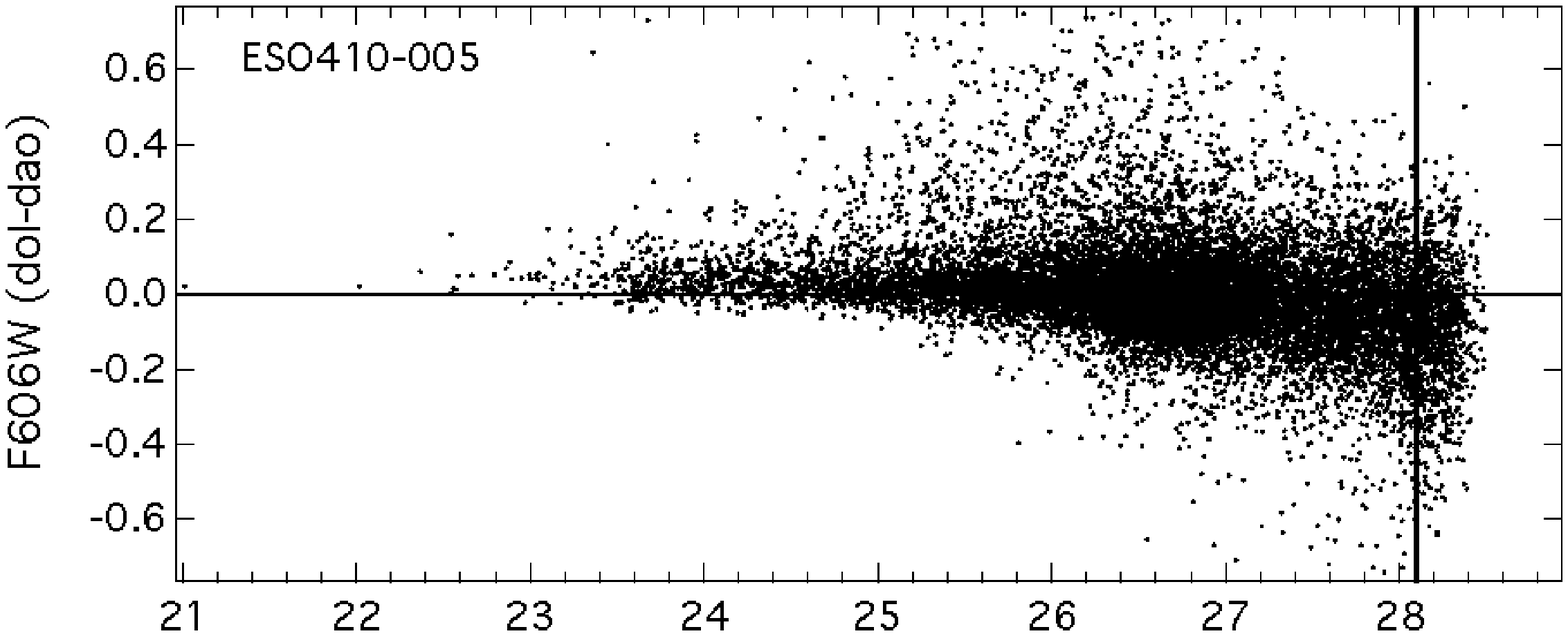}
    \includegraphics[width=7cm,clip]{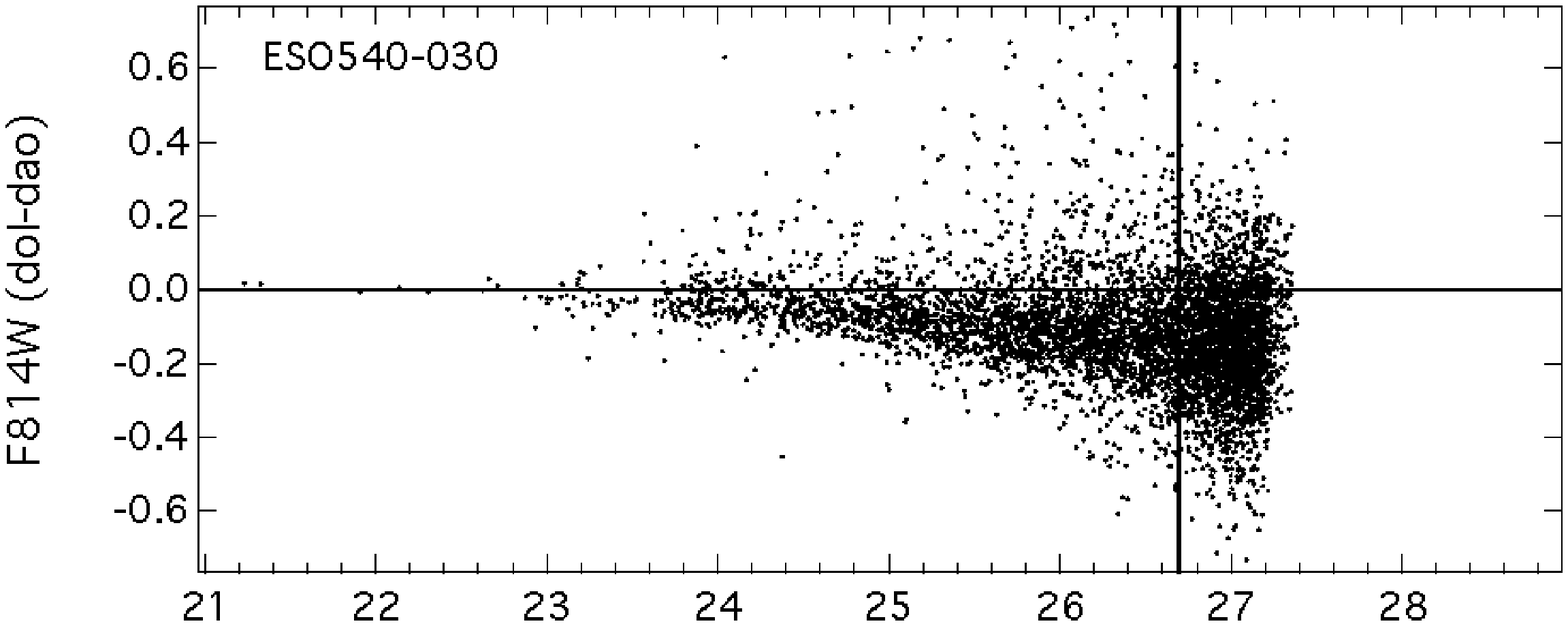}
    \includegraphics[width=7cm,clip]{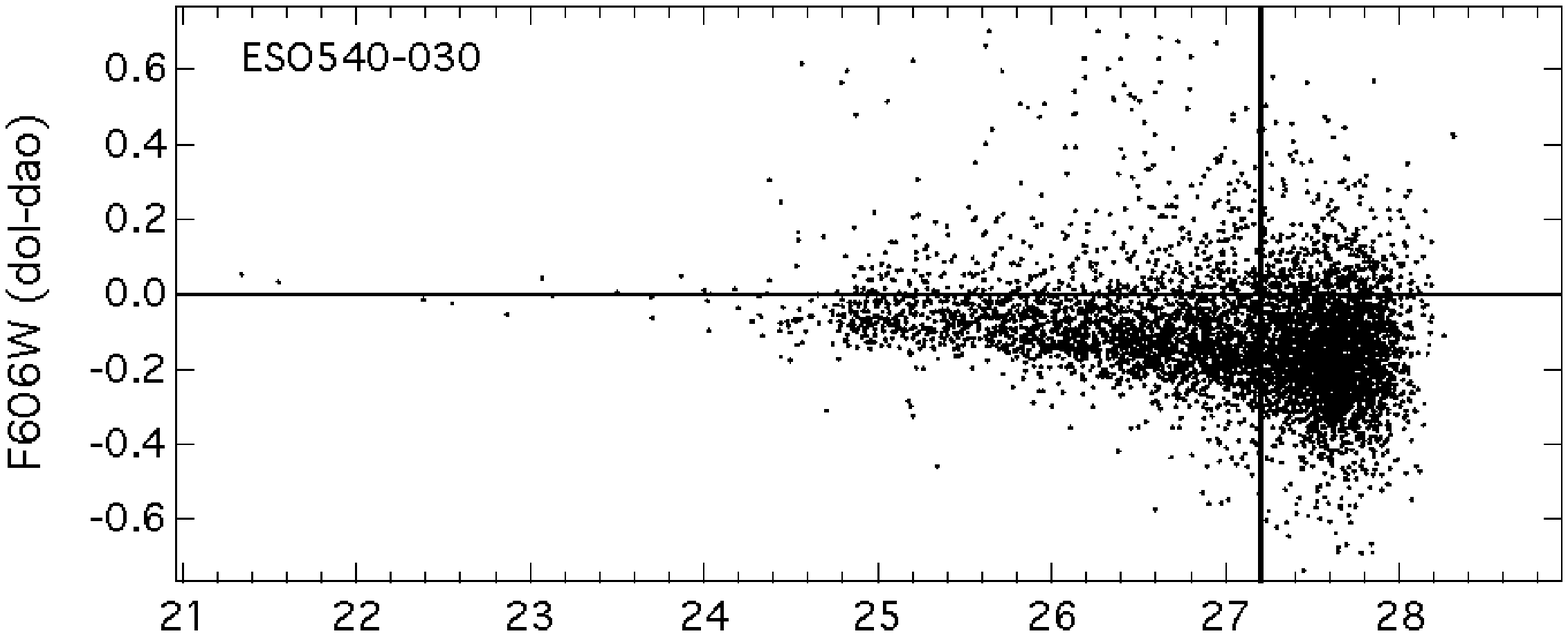}
    \includegraphics[width=7cm,clip]{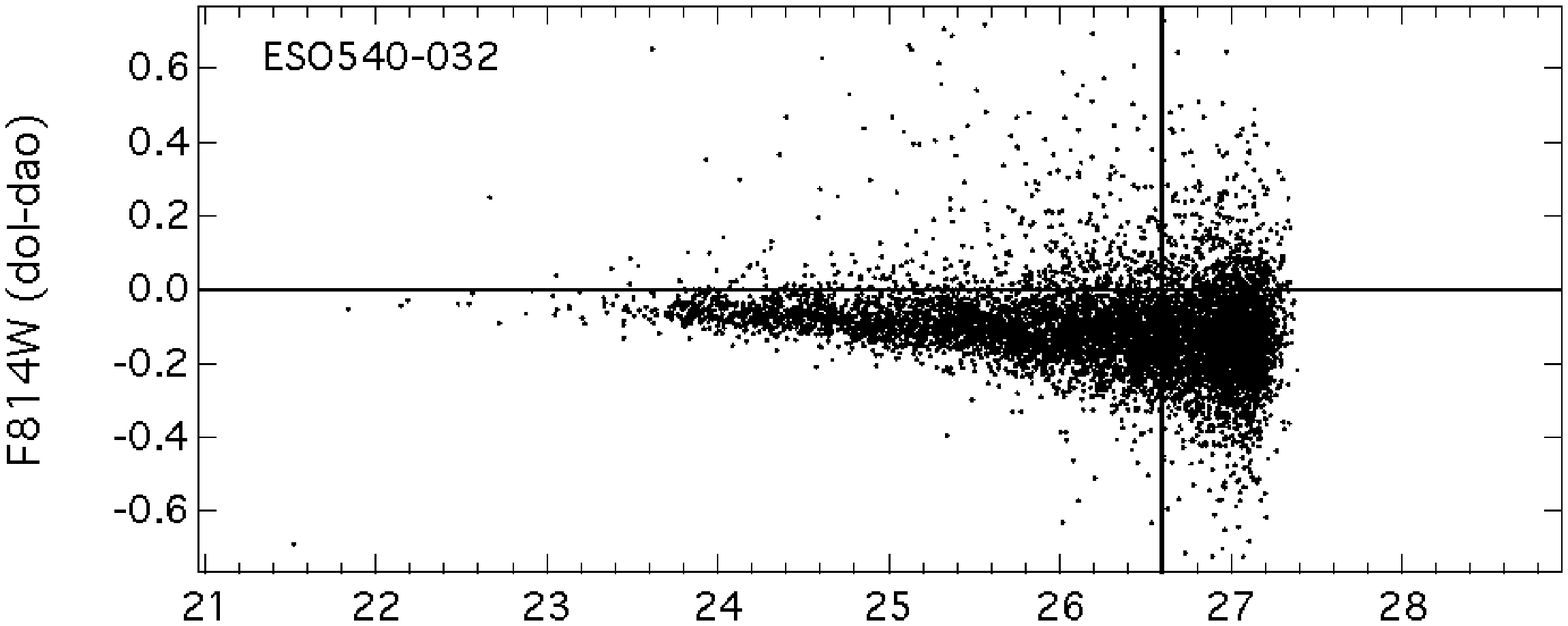}
    \includegraphics[width=7cm,clip]{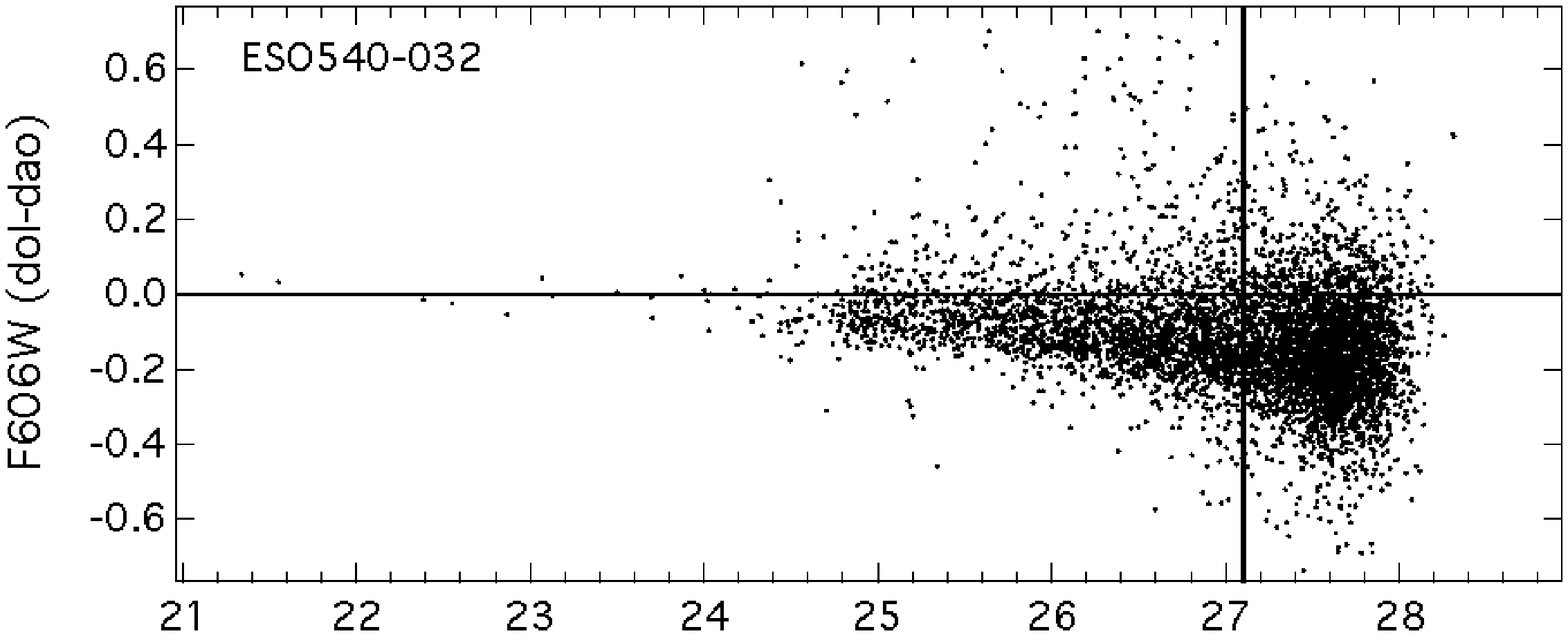}
    \includegraphics[width=7cm,clip]{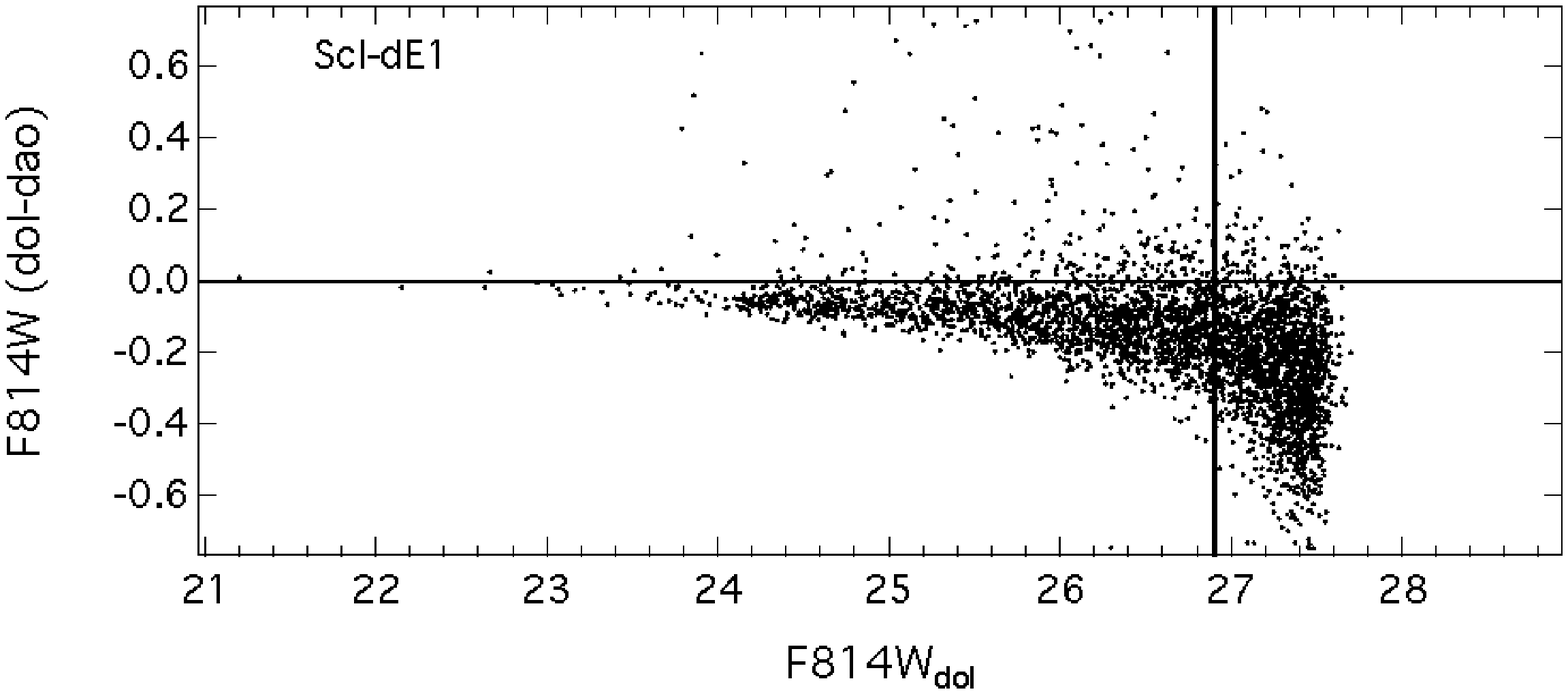}
    \includegraphics[width=7cm,clip]{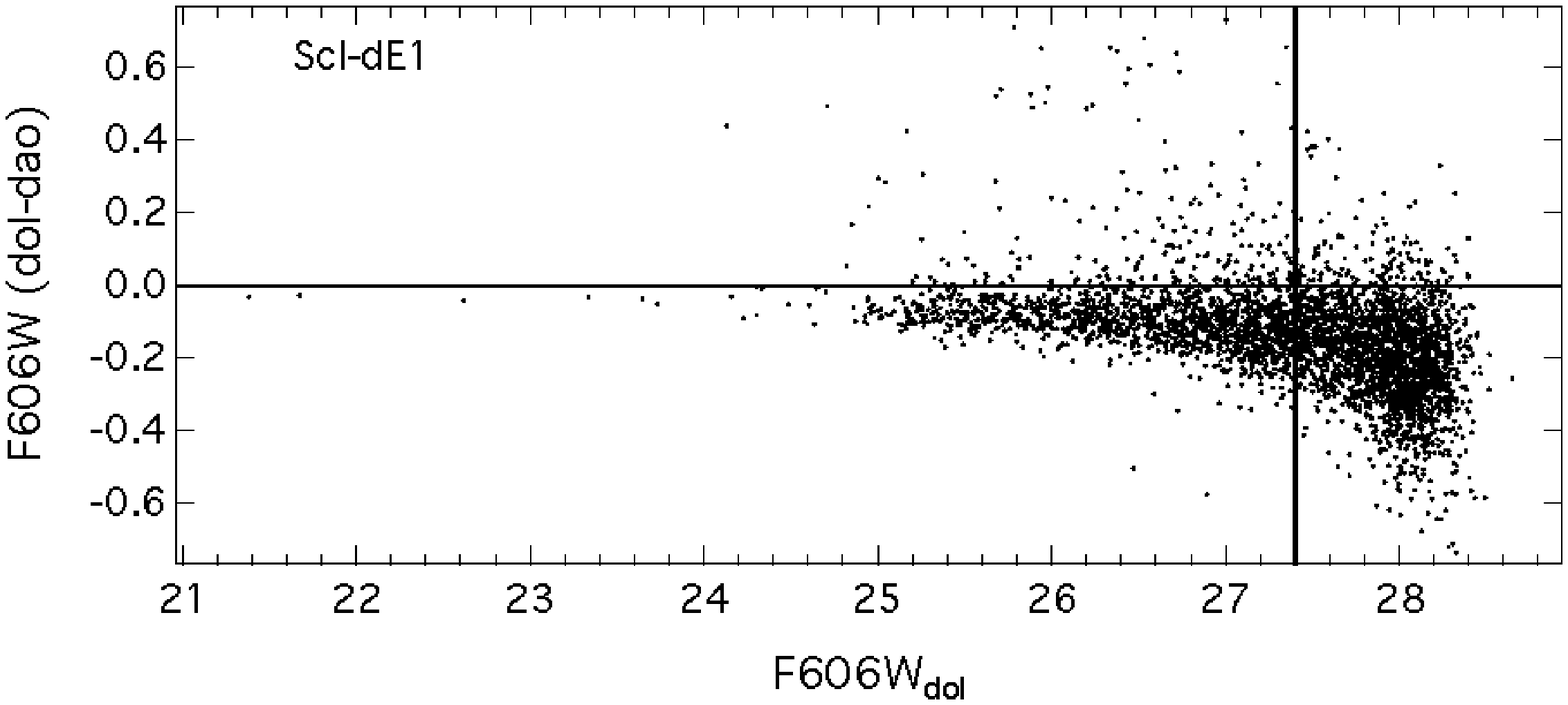}
  \caption{Comparison between the photometry obtained using DOLPHOT and DAOPHOT. The panels show the magnitudes obtained with DOLPHOT minus the magnitudes obtained with DAOPHOT as a function of the DOLPHOT magnitude, in the case of the F814W-band magnitudes (left panels) and F606W-band magnitudes (right panels). The vertical line shows the magnitude that corresponds to the 50\% completeness limit, while the horizontal line indicates the zero-magnitude level.}
      \label{sl_AppA}
\end{figure*}
%
each photometry package are shown in Fig.~\ref{sl_AppA} -- left panels for the F814W filter and right panels for the F606W filter. We note that we have applied the same quality cuts in the star selection for both photometry outputs as those we used in order to construct Fig.~\ref{sl_figure02}. 

We list the mean differences between the DOLPHOT and DAOPHOT magnitudes and colors 
%
\begin{table}
     \begin{minipage}[t]{\columnwidth}
      \caption{Differences between DOLPHOT and DAOPHOT.}
      \label{tableAppA}
      \centering
      \renewcommand{\footnoterule}{}
      \begin{tabular}{l c c c }
\hline\hline
    Galaxy      	&F814W$_{dif}$         &F606W$_{dif}$            &Color$_{dif}$              \\ 
    (1)                 &(2)                  &(3)                      &(4)                        \\ 
\hline 
    ESO540-G030         &$-$0.08$\pm$0.15    &$-$0.08$\pm$0.15        &$-$0.02$\pm$0.12             \\ 
    ESO540-G032         &$-$0.09$\pm$0.12    &$-$0.08$\pm$0.12        &$+$0.00$\pm$0.10             \\
    ESO294-G010   	&$-$0.04$\pm$0.14    &$-$0.03$\pm$0.15        &$+$0.02$\pm$0.13             \\
    ESO410-G005  	&$-$0.01$\pm$0.14    &$+$0.01$\pm$0.15        &$+$0.04$\pm$0.12             \\ 
    Scl-dE1	   	&$-$0.08$\pm$0.14    &$-$0.07$\pm$0.14        &$+$0.02$\pm$0.11             \\
\hline
\end{tabular} 
\end{minipage}
\end{table}
%
in Table~\ref{tableAppA}, along with their standard deviation. These differences are in the sense of DOLPHOT minus DAOPHOT magnitudes or colors, and were derived using all stars brighter than the magnitude corresponding to the 50\% completeness limit for each galaxy. In all dwarfs, there is a zero point offset, similar to what has been reported in Monelli et al.~(\cite{sl_monelli10b}). These zero-point differences do not affect our photometric metallicity analysis, when considering that the colors of the RGB stars are even less affected than the color differences listed in the last column of Table~\ref{tableAppA}. We use these zero-point differences as an additional source of uncertainty when we compute the TRGB magnitude uncertainties.

\section{Comparison between Padova and Dartmouth RGB metallicities}

We derive the photometric metallicities using Padova isochrones (Marigo et al.~\cite{sl_marigo08}; Girardi et al.~\cite{sl_girardi08,sl_girardi10}), using an isochrone set with a constant age of 12.5~Gyr, and [M/H] metallicity ranging from $-$2.3~dex to $-$0.5~dex, with a step of 0.1. The derived photometric metallicities are
\begin{table}
\begin{minipage}[t]{\columnwidth}
\caption[]{Derived properties based on Padova isochrones. }
\label{tableAppB} 
\centering
\renewcommand{\footnoterule}{}  
\begin{tabular}{l c c c}     
\hline\hline
  Galaxy    &$\langle$[M/H]$\rangle\pm\sigma$  &$\langle$[M/H]$\rangle_{w}\pm\sigma$     &[M/H]$_{med}$\\
            &(dex)              &(dex)                 &(dex)      \\
   (1)      &(2)                &(3)                   &(4)        \\
\hline                                                              
 Scl-dE1      &$-1.4\pm0.2$     &$-1.3\pm0.3$      &$-1.4$         \\
 ESO294-G010  &$-1.3\pm0.3$     &$-1.2\pm0.2$      &$-1.2$         \\
 ESO540-G032  &$-1.4\pm0.3$     &$-1.2\pm0.2$      &$-1.3$         \\
 ESO540-G030  &$-1.4\pm0.3$     &$-1.3\pm0.2$      &$-1.4$         \\
 ESO410-G005  &$-1.3\pm0.2$     &$-1.2\pm0.3$      &$-1.3$         \\
\hline
\end{tabular}
\end{minipage}
\end{table}
%
listed in Table~\ref{tableAppB}. The number ratio of the stars bluer than the bluest Padova isochrone versus all stars within the RGB selection box ranges from 0\% (for Scl-dE1) to 3\% (for ESO540-G030). We note a difference in the $\rm {mean}$ error-weighted photometric metallicities of the order of 0.6~dex to 0.3~dex than the [Fe/H] metallicities derived using Dartmouth isochrones, Table~\ref{table5}. The differences in the photometric metallicities derived using the two different isochrone sets stem from their different handling of the input physics relevant to the RGB phase, as discussed in Gallart et al.~(\cite{sl_gallart05}) and in Cassisi (\cite{sl_cassisi11}), thus the Padova-based and Dartmouth-based metallicities define a different metallicity scale. 

\section{Excluded bluest RGB stars}

 \begin{figure*}
  \centering
    \includegraphics[width=6cm,clip]{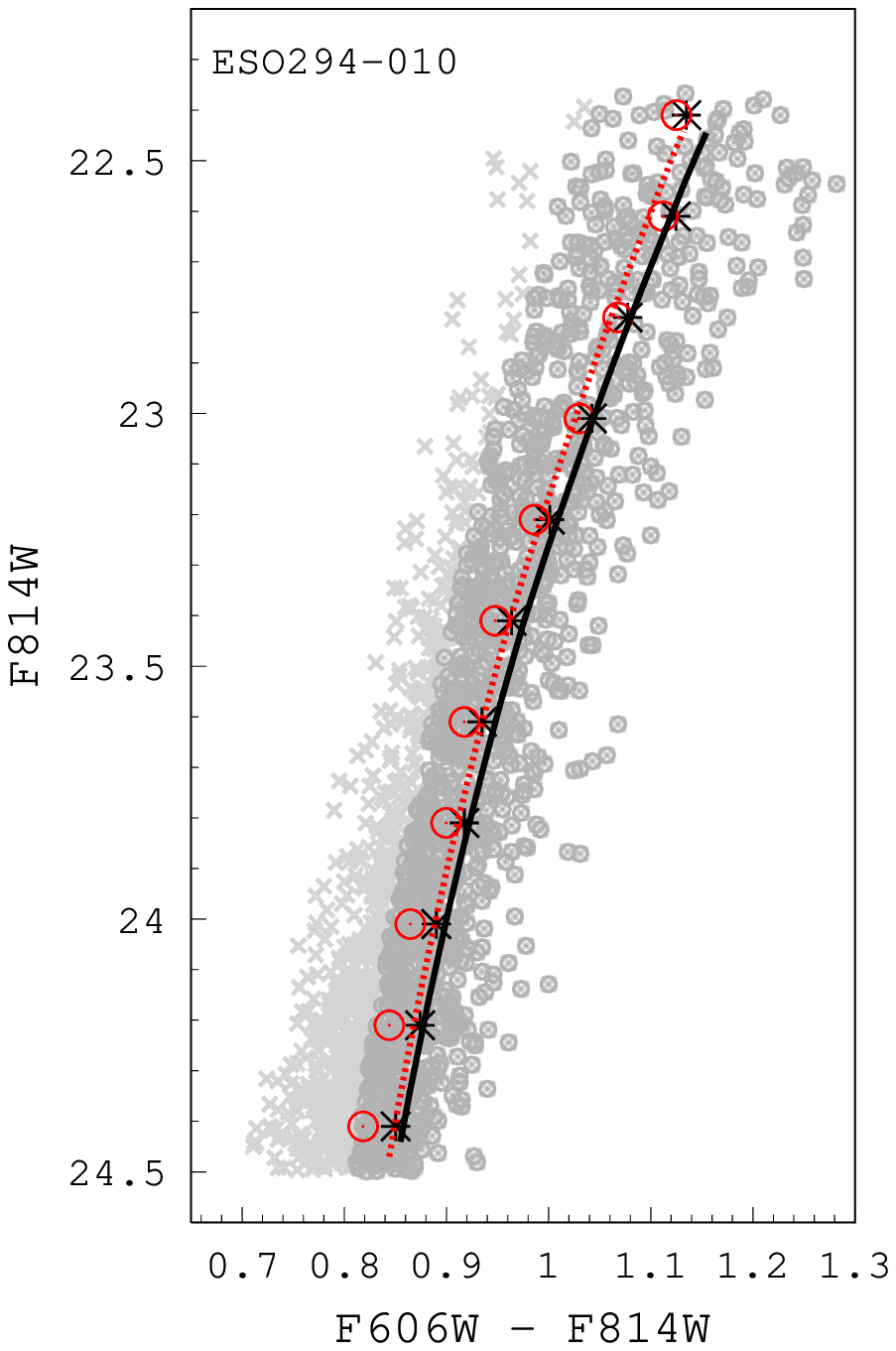}
    \includegraphics[width=6cm,clip]{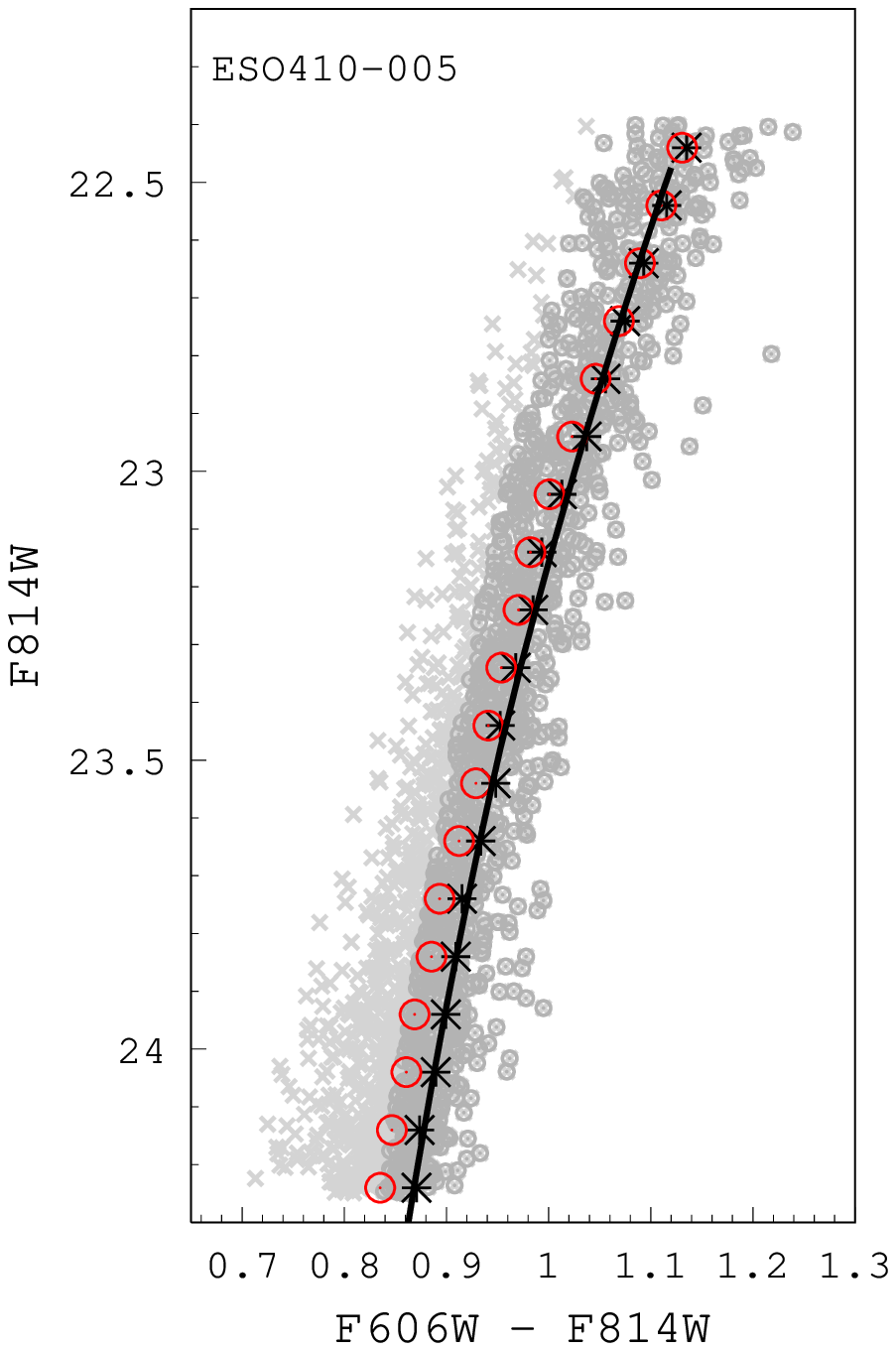}
    \includegraphics[width=6cm,clip]{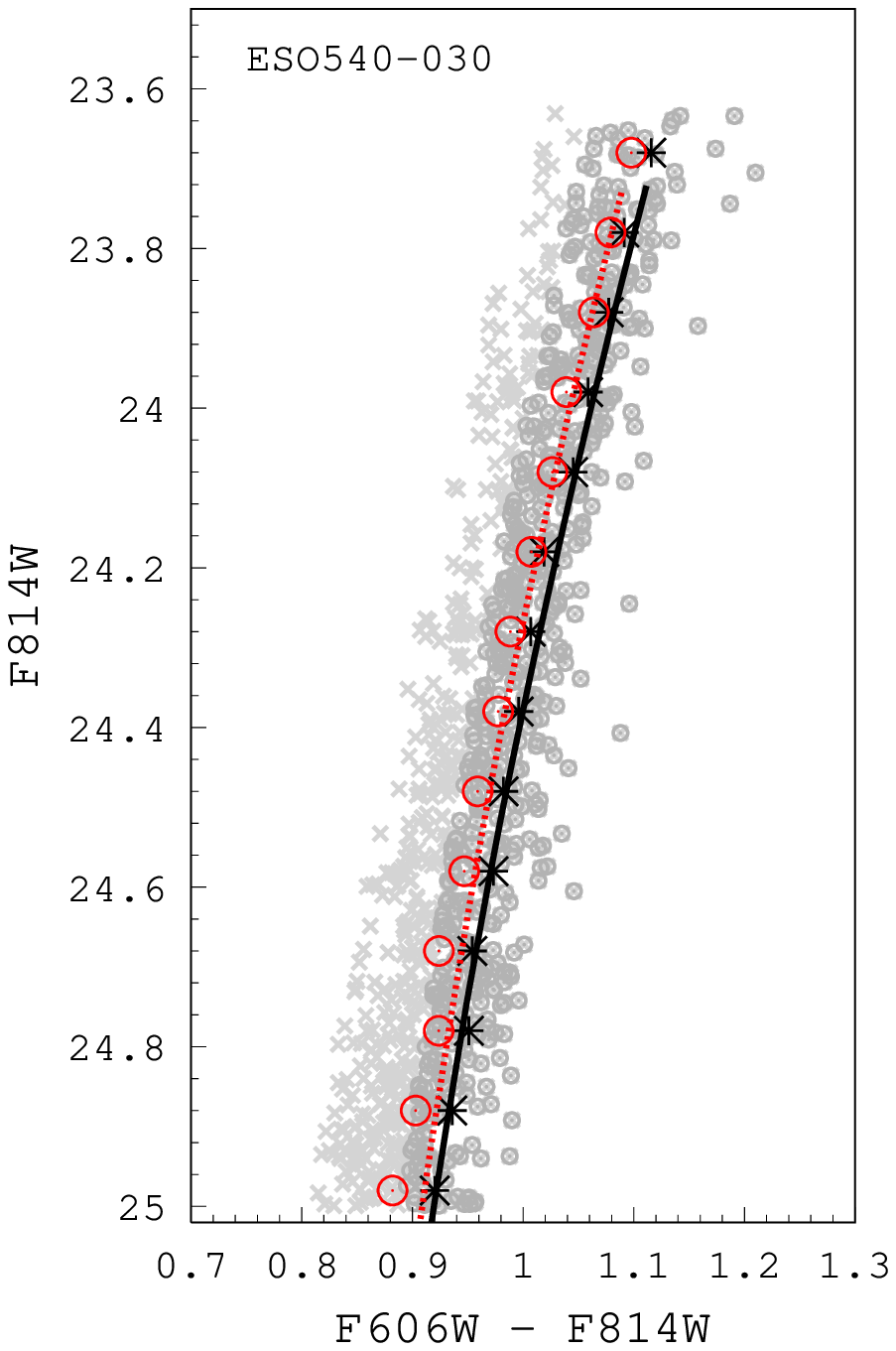}
    \includegraphics[width=6cm,clip]{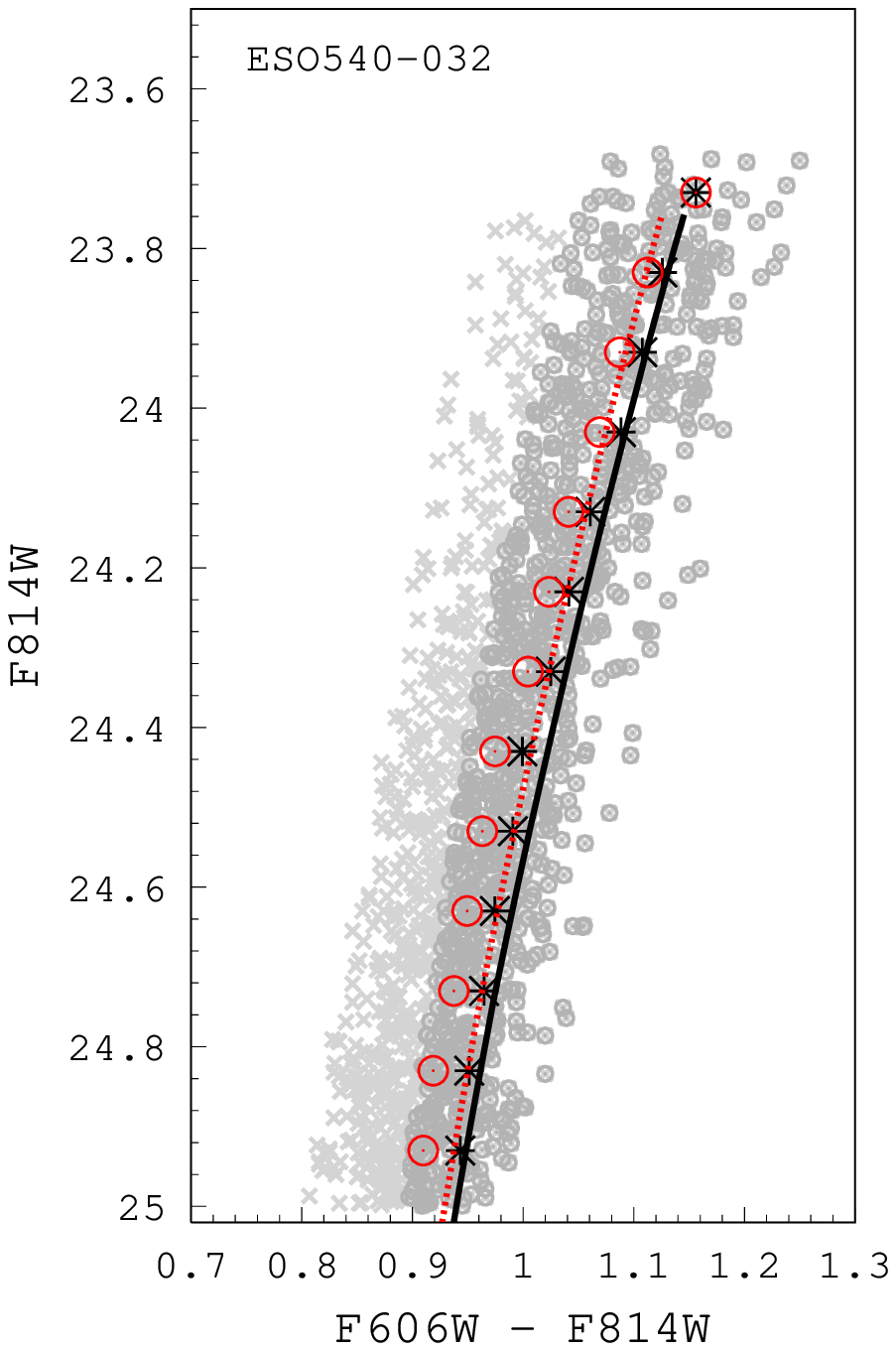}
    \includegraphics[width=6cm,clip]{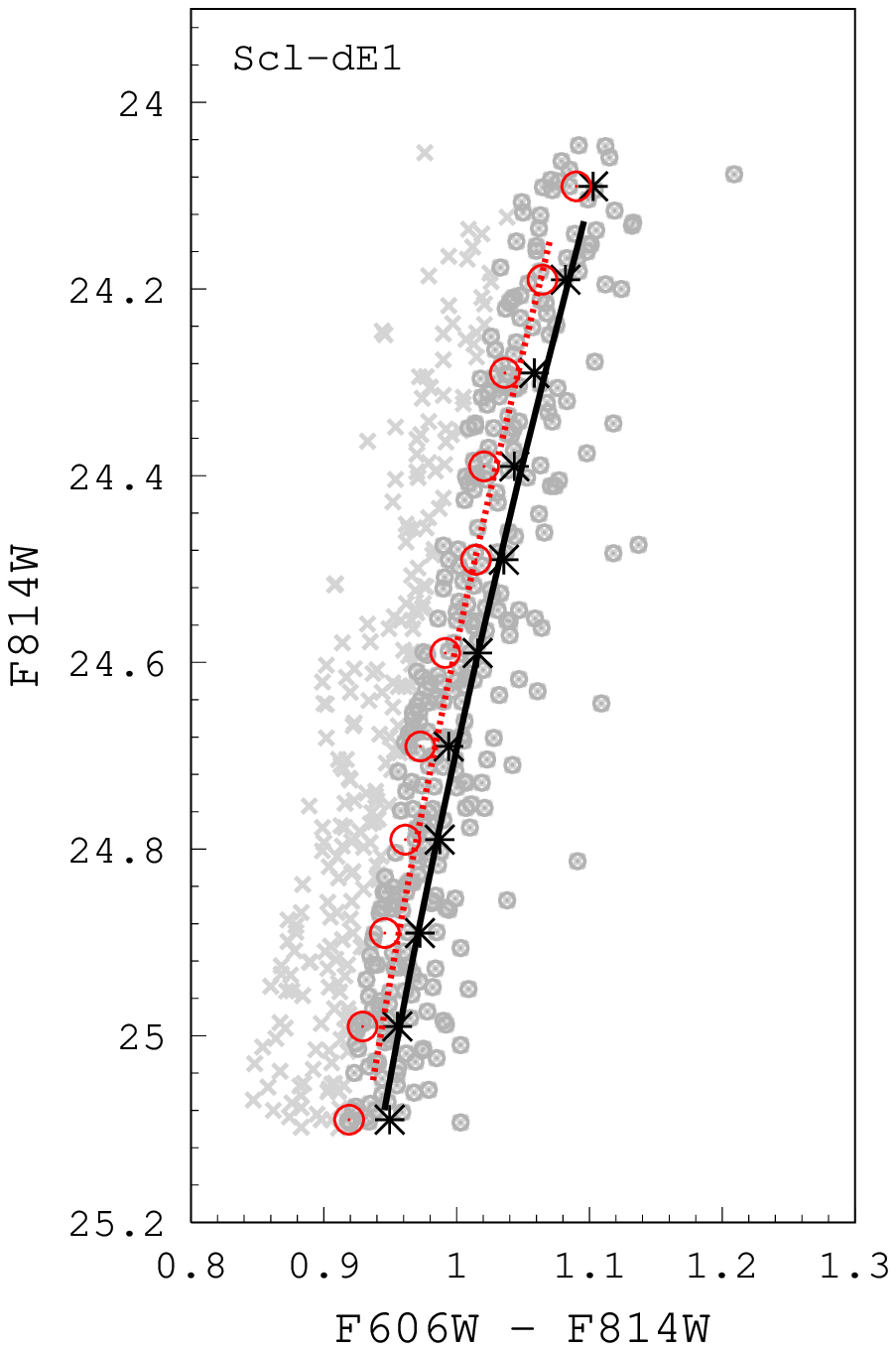}
  \caption{RGB stars shown in the lightest gray cross symbols, as well as those RGB stars that were selected for the MDFs, shown with the darker gray circled-cross symbols. The fiducial derived using all the RGB stars, RGB$_{\rm ALL}$, is outlined with the red open circles, while the fiducial derived using those RGB stars that were selected to construct the MDFs, RGB$_{\rm MDF}$, is outlined with the black asterisks. The red dotted line corresponds to the Dartmouth 12.5~Gyr isochrone that best matches the RGB$_{\rm MDF}$ fiducial, while the black solid line corresponds to the one that best matches the RGB$_{\rm MDF}$ fiducial. From left to right and from top to bottom panels, the red dotted lines correspond to a metallicity of $-$1.55~dex, $-$1.75~dex, $-$1.85~dex, $-$1.65~dex, and $-$1.9~dex, respectively, while the black solid lines to a metallicity of $-$1.65~dex, $-$1.75~dex, $-$2~dex, $-$1.75~dex, and $-$2.1~dex, respectively. } 
      \label{sl_appc}
\end{figure*}
%
  In this section, we estimate by how much the exclusion of the RGB stars that lie bluer than the most metal-poor isochrone we use in the interpolation is biasing the {\em mean} metallicity estimates. In order to do so, we compute the fiducial of those RGB stars that were selected to construct the MDFs (i.e., excluding the stars that lie bluer than the most blue isochrone used in the interpolation), as well as the fiducial of all the RGB stars within the selection box (i.e., including the stars that lie bluer from the most blue isochrone used in the interpolation). We refer to the former as the RGB$_{\rm MDF}$ fiducial, and to the latter as the RGB$_{\rm ALL}$ fiducial. For each dwarf, we show both fiducials in Fig.~\ref{sl_appc}, outlined with the black asterisks for the RGB$_{\rm MDF}$ fiducial, and with the red open circles for the RGB$_{\rm ALL}$ fiducial. Additionally, we overplot all the RGB stars in the selection box with the lightest gray cross symbols, as well as the RGB stars that were selected to construct the MDFs with the darker gray circled-cross symbols. 

  Subsequently, we overplot those two Dartmouth isochrones that best match the two fiducials, assuming an old isochrone age of 12.5~Gyr. We require a good match between the isochrone and the fiducial in the bright part of the RGB, since towards fainter magnitudes there is a larger deviation between them due to the increased influence of the excluded stars that lie bluer than the most blue isochrone used in the interpolation. 

  The comparison of the two RGB$_{\rm ALL}$ and RGB$_{\rm MDF}$ fiducials, as well as their {\em mean} metallicity estimates inferred from the best matched Dartmouth isochrone, yields a maximum bias in metallicity of 0.2~dex, occurring for Scl-dE1. The metallicity bias is in the sense of deriving a more metal-poor metallicity when including the stars that lie bluer than the most blue isochrone used in the interpolation. For ESO294-G010 and ESO540-G032, the metallicity bias is 0.1~dex, while this is 0.15~dex for ESO540-G030, and zero for ESO410-G005. These metallicity differences between RGB$_{\rm ALL}$ and RGB$_{\rm MDF}$ fiducials are within the derived standard deviations of the {\em mean} [Fe/H] metallicities listed in Table~\ref{table5}.

\end{appendix}

\end{document}